\documentclass[sigconf]{acmart}
\usepackage{acmart-taps}

\newcommand{\techname}{Perceptual Pat}

\usepackage{fontawesome5}

\usepackage{colortbl}

\usepackage{subfig}
\usepackage{float}

\usepackage{svg}

\usepackage{xcolor}
\usepackage{tikz}
\newcommand*\bcircled[1]{\tikz[baseline=(char.base)]{
            \node[shape=circle,fill,inner sep=1pt] (char) {\textcolor{white}{#1}};}}
\newcommand*\wcircled[1]{\tikz[baseline=(char.base)]{
            \node[shape=circle,draw,inner sep=1pt] (char) {\textcolor{black}{#1}};}}
\newcommand{\rev}[1]{{#1}} % print text black (option 1)
\newcommand{\revt}[1]{{#1}} % print text black (option 1)
\usepackage{enumitem}

\copyrightyear{2023}
\acmYear{2023}
\setcopyright{rightsretained}
\acmConference[CHI '23]{Proceedings of the 2023 CHI Conference on Human Factors in Computing Systems}{April 23--28, 2023}{Hamburg, Germany}
\acmBooktitle{Proceedings of the 2023 CHI Conference on Human Factors in Computing Systems (CHI '23), April 23--28, 2023, Hamburg, Germany}\acmDOI{10.1145/3544548.3580974}
\acmISBN{978-1-4503-9421-5/23/04}

\begin{document}

%% ---------------------------------------------------------------------
%% Title, author(s), and affiliations
%% ---------------------------------------------------------------------

% Paper Title
\title[\techname]{\techname{}: A Virtual Human Visual System for Iterative Visualization Design}

%% Authors
\author{Sungbok Shin}
\affiliation{%
  \institution{University of Maryland}
  \city{College Park}
  \state{Maryland}
  \country{USA}
}
\email{sbshin90@cs.umd.edu}

\author{Sanghyun Hong}
\affiliation{%
  \institution{Oregon State University}
  \city{Corvallis}
  \state{Oregon}
  \country{USA}
}
\email{sanghyun.hong@oregonstate.edu}

\author{Niklas Elmqvist}
\affiliation{%
  \institution{University of Maryland}
  \city{College Park}
  \state{Maryland}
  \country{USA}
}
\email{elm@umd.edu}

%% Short header
\renewcommand{\shortauthors}{Shin et al.}

%% ---------------------------------------------------------------------
%% Abstract (max 150 words)
%% ---------------------------------------------------------------------
\begin{abstract}
    Designing a visualization is often a process of iterative refinement where the designer improves a chart over time by adding features, improving encodings, and fixing mistakes.
    However, effective design requires external critique and evaluation. 
    Unfortunately, such critique is not always available on short notice and evaluation can be costly.
    To address this need, we present \techname{}, an extensible suite of AI and computer vision techniques that forms a virtual human visual system for supporting iterative visualization design.
    The system analyzes snapshots of a visualization using an extensible set of filters---including gaze maps, text recognition, color analysis, etc---and generates a report summarizing the findings.
    The web-based Pat Design Lab provides a version tracking system that enables the designer to track improvements over time.
    We validate \techname{} using a longitudinal qualitative study involving \rev{4} professional visualization designers that used the tool over a few days to design a new visualization.
\end{abstract}



%% Keywords. 

%% ---------------------------------------------------------------------
%% Teaser
%% ---------------------------------------------------------------------
\begin{teaserfigure}
  \includegraphics[width=\textwidth]{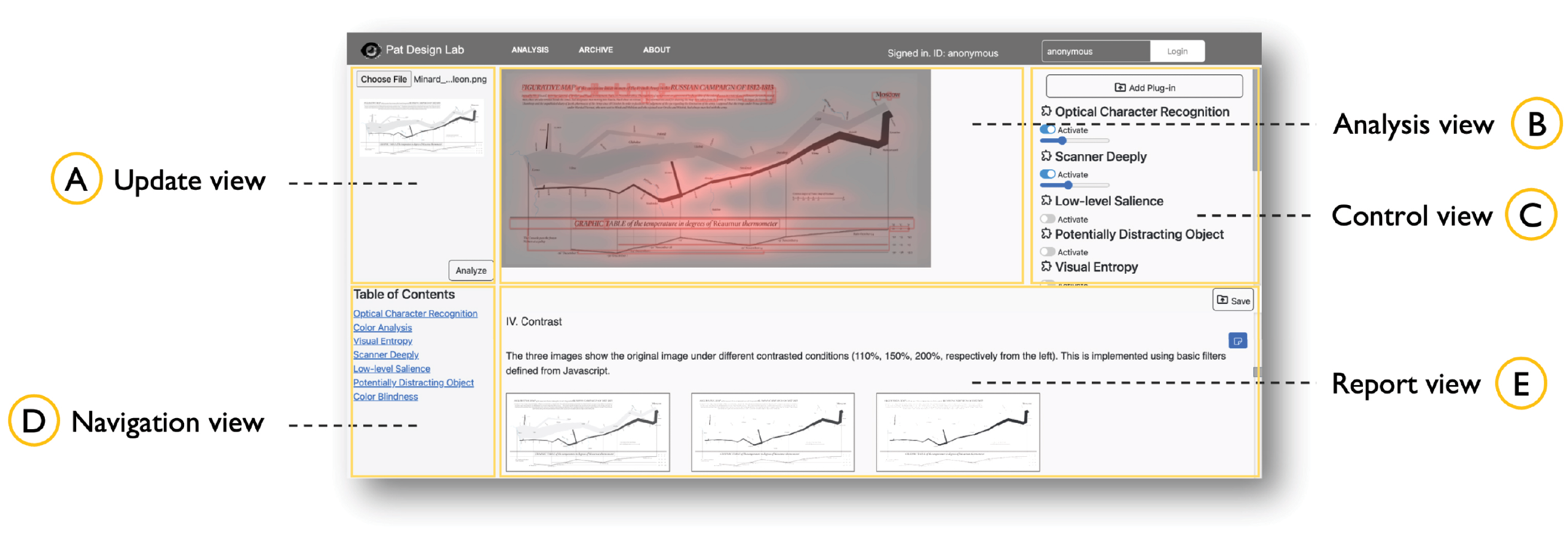}
  %\resizebox{\linewidth}{!}{\framebox{Teaser}}
  \caption{\textbf{Pat Design Lab.}
  Applying \techname{} to Charles Minard's famous Napoleon's March to Russia diagram.
  The Pat Design Lab is the web-based interface to the \techname{} suite of perceptually-inspired image filters that can be used by a designer to receive rapid and inexpensive feedback on a visualization being designed.
  } % Pat Design Lab -- table, icon that looks like a robot? human, scientist? human visual system, but it's virtual. It's a simulated virtual system. 
  \Description{This teaser shows a screenshot of the Pat Design Lab applied to Charles Minard's famous Napoleon's March to Russia diagram. We broke up the system into 5 parts. These are update view, analysis view, control view, navigation view, and report view.}
  \label{fig:teaser}
\end{teaserfigure}

%% ---------------------------------------------------------------------
%% Title
%% ---------------------------------------------------------------------
\maketitle

%% ---------------------------------------------------------------------
%% INTRODUCTION
%% ---------------------------------------------------------------------
\section{Introduction}

Visualization design, just like any other design discipline, is inherently iterative~\cite{munzner14visualization}.
During the course of designing a novel visualization, a designer may go through dozens or even hundreds of ideas, sketches, and prototypes before settling on a final composition.
However, while experienced designers have developed the ability to view their own work through the eyes of their intended audience, it is clear that all designers would benefit from having access to an objective and unbiased audience to use as a sounding board for each design iteration. 
Is this label readable?
Are the peaks and troughs in this line-series chart salient?
Would a non-expert recognize this chart type? 
Being able to answer such questions at the drop of a hat would be invaluable, but fellow designers are not always available to provide feedback and empirical evaluation is unfortunately so costly in terms of time and money that it is impractical. 

That is, until Pat came along. 
Pat is always game to take a look at a new version of your visualization, regardless of the time of day.
Pat will tell you honestly what he thinks of your visualization design. 
In fact, Pat will be precise and give you a detailed breakdown of the colors, saliency, and legibility of your design.
And, what's more, Pat doesn't sleep or take breaks.
In fact, Pat can be invoked with a mere click of your mouse button.

In this paper, we present \textsc{Perceptual Pat}: an extensible suite of image processing, computer vision, and machine learning models that taken together forms a virtual human visual system suitable for supporting visualization design and evaluation.
While we obviously do not intend for Perceptual Pat to be a drop-in replacement for a real human by a long shot, the Pat suite provides a collection of perceptually-inspired image filters that combine to yield a comprehensive picture of what a person would see when viewing a visualization.
Examples of such filters include the virtual gaze maps, such as the Scanner Deeply virtual eyetracker~\cite{shin23scannerdeeply}, color vision deficiency filters, text legibility and optical character recognition (OCR) scanning, color analysis, etc.
Inspired by the virtual human known as ``Virtual Jack''~\cite{badler93simulating} (or just ``Jack''), which has long been a staple in ergonomics design, Perceptual Pat is a complement and not a replacement for human testing.
Instead, the purpose of Pat is to provide easy access to multiple rounds of quick and cheap feedback before a design is evaluated in a focus group, expert review, or usability study.

At its core, the Perceptual Pat suite is held together by the \textsc{Pat Design Lab}, a web-based software system for managing iterative visualization design.
Using the Design Lab, a designer can upload evolving versions of a design over time, each time running the Perceptual Pat tests and receiving an interactive report in response.
The Design Lab allows for analyzing and studying the output of the tests as overlays display on top of the visualization itself. 
The tool also has functionality for designers to add their own annotation and notes to the report.
Finally, different reports can be interactively compared using the tool, enabling the designer to see how their design has evolved over time. 
Furthermore, the Pat suite is based on a flexible plugin architecture, enabling the Design Lab to be easily extended with third-party image filters.

\revt{We validated the Perceptual Pat suite and the Pat Design Lab in a longitudinal user study involving 4 expert visualization designers using the tool for their own design project.}
We first met with our participants in individual meetings to introduce the tool and its functionality, and then again a week later at the conclusion of their design project.
During this time, the participants were asked to design and refine a new visualization from scratch using the tool.
Our analysis of the resulting reports and user annotations indicate that a majority of causes for change in the design were attributed to our system.
During the exit interviews, participants acknowledged the effectiveness of the Perceptual Pat in detecting problems within visualizations, as well as its convenience in providing design feedback. 

The contribution of this paper are as follows: 
\begin{itemize}
    \item The concept of a \textit{virtual human visual system} (VHVS) as a suite of models providing feedback on a visualization image indicative of human perception; 
    \item The Perceptual Pat implementation of a virtual human visual system for iterative visualization design;
    \item The Pat Design Lab for supporting iterative evaluation of visualization design using the Perceptual Pat suite;
    \item Qualitative findings from a longitudinal study involving four expert designers using our prototype implementation in their own design projects; and
    \item \rev{Findings from external evaluators who assessed the design process and outcomes from the longitudinal study.}
\end{itemize}

%% -------------------------------------------------------------
%% BACKGROUND
%% -------------------------------------------------------------
\section{Background}
\label{sec:background}

% \rev{Here we provide an overview of the background for our work on Perceptual Pat suite. 
% We first explain the role of feedback in visualization design. 
% We review works in vision science and visualization that shows how understanding perception can also support visualization design.
% Then, we present examples of how feedback is presented to help visualization design. 
% At last, we present our motivation for the design of Perceptual Pat.}

Here we provide an overview of the background for our work on the Perceptual Pat suite by first explaining how visualization is a design discipline and reviewing systems that scaffold the design process.
We then review work in vision science and visualization that shows how understanding perception can also support visualization design.

\subsection[The Role of Feedback in Visualizations]{\rev{The Role of Feedback in Visualizations}}

Data visualization is still a largely empirical research field, and iterative design is thus a key component in authoring a new visualization~\cite{munzner14visualization}.
It has also entailed a focus within the field on design heuristics and rules of thumb, such as Edward Tufte's reviews of effective visualizations~\cite{Tufte1983}, visualization textbooks with a design emphasis~\cite{munzner14visualization}, and many blog posts and practical handbooks drawn from practice.

All design benefits from external feedback.
There are two main mechanisms for feedback: user and usage feedback vs.\ peer and supervisor feedback. 
The former is more common in academia and focuses on validation, often through empirical evaluation.
Munzner present a nested model for visual design and validation~\cite{munzner2009} that incorporates primarily the former through various forms of validation.
She presents four design layers for creating visualizations: (1) domain problem characterization, (2) data/operation abstraction design, (3) encoding/interaction technique design, and finally \rev{(4) algorithm design.}
Sedlmair et al.~\cite{Sedlmair2012} develop a nine-stage design study framework (consisting of three higher level category), and provide practical guidance for conducting design studies.
The three high-level categories are: (1) personal (precondition), (2) inward-facing (core), and (3) outward-facing (analysis).
\rev{Peer feedback, on the other hand, is more common in industry and design practice, where critique is critical when authoring visualizations intended for mass consumption.}
It has also reached academia within the general interaction design community.
Bardzell et al.~\cite{DBLP:journals/interactions/BardzellBL10, DBLP:journals/iwc/Bardzell11} talk about the importance of criticisms in designing interfaces, and assert that they raise people's perceptual ability, which over time constitutes a heightened sensibility or competence.

\paragraph{\rev{Comparison.}}

\rev{Obtaining external feedback requires time and money and is not always accessible.
In contrast, Perceptual Pat provides external feedback using automated methods, and is thus quick, always available, and virtually cost-free.}

\subsection[Facilitating the Chart Design Process]{Facilitating the Chart Design Process}

\rev{External feedback helps to improve chart design, but there exist many different ways to provide such feedback}.
One method is based on practical visualization recommendation.
Building on seminal automatic visualization work by Jock Mackinlay~\cite{Mackinlay1986}, Tableau's Show Me~\cite{mackinlay07showme} feature recommends charts based on data types as well as best practices. 
\rev{Since then, various types of recommendation tools have been developed, such as those based on data properties~\cite{key12vizdeck, wongsuphasawat16voyager}, perceptual principles~\cite{wongsuphasawat16voyager}, expert feedback~\cite{luo18deepeye}, large-scale dataset-visualization pairs~\cite{hu19vizml}, and design knowledge~\cite{moritz19draco}.}

\rev{Another method is to directly aid iterative design for visualization.}
The data visualization saliency (DVS)~\cite{matzen18visualsaliency} model is a general-purpose saliency heatmap generator for visualizations.
This saliency map \rev{enables designers to understand a viewer's attention on a chart.}
Recently, there is also a rising interest in techniques that provide automated design feedback on visualizations using a linting or sanity check metaphor.
Examples are tools for detecting chart construction errors~\cite{hopkins20visualint}, visualization mirages~\cite{mcnutt20mirage}, and deceptively-designed line charts~\cite{fan22linechartdeception}.
\rev{Finally, VizLinter~\cite{chen22vizlinter} even provides solutions to help chart designers.}

\paragraph{\rev{Comparison.}}

Our Perceptual Pat suite is similar to many of these techniques, \rev{but it} does not recommend visualizations to be authored, leaving \rev{authoring} entirely in the hands of the designer.
Instead, our approach is to provide a toolbox of different filters that together can give the user multiple lenses through which to view the visualization artifact they are designing.
This is akin to a supertool~\cite{Shneiderman2022} augmenting the capabilities of the designer a hundredfold.

\subsection[Understanding Human Perception for Visualizations]{\rev{Understanding Human Perception for Visualization}}

\rev{Understanding how humans perceive charts is vital to supporting design, and has been an active area of interest within vision science.}
These efforts began as early as the end of the 19th Century in work done by the so-called ``Berlin School'' of experimental psychology.
This eventually led to the development of \textit{Gestalt psychology}~\cite{Koffka1922}, a theory of mind based on a holistic view of human visual perception where the sum of the perceived ``gestalt'' is qualitatively different than its component parts, and in effect has an identity of its own.

Experimental work within vision science has also spent significant effort collecting empirical data on how humans perceive charts.
In 1926, Eells et al.~\cite{Eells1926} study how people viewed statistical graphics.
\rev{Croxton et al.~\cite{Croxton1927}} compare bar charts with pie charts in 1927 and study the effectiveness other shapes for comparison in 1932~\cite{Croxton1932}.
In 1954, Peterson et al.~\cite{Peterson1954} perform experiments for eight different statistical graphics. 
These findings, and many more, were summarized in Cleveland and McGill's seminal 1984 paper~\cite{Cleveland1984} on graphical perception of statistical graphics.

Empirical work on sophisticated visualization mechanics has continued at a rapid pace; an exhaustive survey is beyond the scope of this paper.
Bateman~\cite{DBLP:conf/chi/BatemanMGGMB10} study the impact of visual embellishment compared to minimalistic chart design, finding memorability improvements.
Chalbi et al.~\cite{DBLP:journals/tvcg/ChalbiRPCREC20} extend the original Gestalt laws for dynamic graphical properties.
Michal and Franconeri~\cite{michal17visual} present findings on the order readers follow when viewing visualizations (e.g., looking at the tallest bar when reading bar charts). 
Many of these detailed empirical findings have been condensed into specific design rules in a recent paper by Franconeri et al.~\cite{franconeri21visualdatacommunication}, listing both visualizations that succeed in effectively communicating data through visualizations and those that fail to do so because of illusion and misperception.

\rev{With the introduction of high-performance eye tracking devices, researchers use eye movement data~\cite{bylinskii19differentmodels} to understand human perception. 
This provides us with new knowledge about what causes confusion in charts~\cite{lalle16confusion}, visual patterns that benefit recall~\cite{borkin16recall}, visual saliency as a measure of attention~\cite{riche13metricssaliency}, and assessment methods for visualization proficiency~\cite{toker14untrainedvisusers}. 
It is even possible to infer one's personality with eye movement patterns during chart reading~\cite{toker13infoviseyetrackingchar}.}

\paragraph{\rev{Comparison.}}

\rev{All this work are candidates for inclusion} into our pragmatic take on a virtual human visual system.

\subsection[Modeling Human Perception]{\rev{Modeling Human Perception}}

Our work is inspired by the concept of \textit{Jack}, a human simulation system devised by Badler et al.~\cite{badler93simulating} in 1999. 
Jack is an abstracted version of a human body, with special focus on the body's physical properties and its movements.
In general, simulating humans as virtual agents provides a virtual experience that can be used to detect and prevent many of problems in a preemptive manner during early design. 

We propose an approach for modeling not the physical body of a human, but its visual system \rev{for the purposes of supporting visualization design}.
In 1993, Lohse propose the first simulation of a human perceptual system based on a cognitive modeling approach drawn from both experimental and theoretical findings up to that date~\cite{Lohse1993}.
More recently, Haehn et al.~\cite{haehn19understandingpercwithcnns} study the use of convolutional neural networks (CNNs) as a possible candidate model for a human perceptual system.
While they were able to replicate several of Cleveland and McGill's seminal graphical perception results, they ultimately decide that CNNs are currently not a good model for human perception.
Finally, Shin et al.~\cite{shin23scannerdeeply} use crowdsourced eyetracking data to build a bespoke deep learning model that simulates human eye movement to generate gazemaps on any uploaded data visualization presented to it.

\paragraph{\rev{Comparison.}}

\rev{Perceptual Pat is different from all of this prior work because it integrates and synthesizes many models based on image processing and computer vision.}
In fact, many of the aforementioned tools have already been integrated into the Perceptual Pat suite.
\rev{However, we are aware of no design-oriented virtual human visual system similar to ours.}

%% -------------------------------------------------------------
%% DESIGN
%% -------------------------------------------------------------
\section{Design: Virtual Human Visual System}

We informally define the concept of a \textit{virtual human visual system} (VHVS) as a simulated human perceptual system based on computer software filters that use imagery as their main input channel and outputs information about the expected human visual response to this input.
In our implementation, we think of these image filters as \textit{perceptually-inspired} to indicate the pragmatic and practical approach we adopt in this paper; instead of attempting a high fidelity simulation of the human visual system, our goal is to provide actionable information to a designer iterating on a visualization artifact.

\begin{figure}[tbh]
    \centering
    \includegraphics[width=\linewidth]{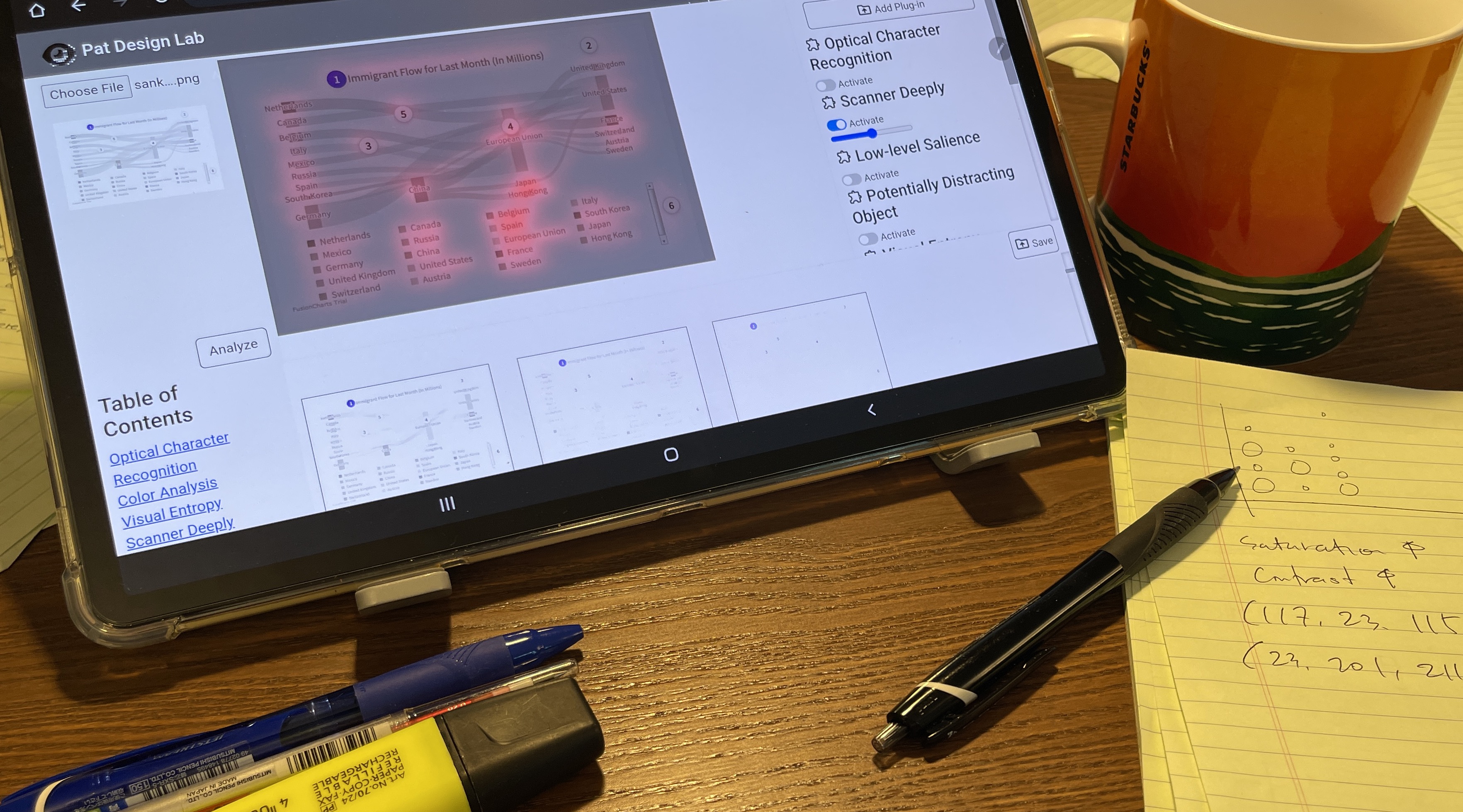}
    \caption{\textbf{The Pat Design Lab in action.}
    \rev{Example of a designer} using the Pat Design Lab during their chart design process.}
    \label{fig:perceptual_pat_ex}
    \Description{Photograph of a desk with an Apple iPad showing the Perceptual Pat interface being used to analyze a Sankey diagram. There are various pens and highlighters on the desk as well as a legal pad and a coffee cup.}
\end{figure}

\subsection{Motivation}

There are many potential reasons for designing a virtual human visual system (VHVS), ranging from vision science---e.g., the ability to completely simulate the human visual system at high fidelity---to more practical applications for specific design and evaluation situations.
Our goal in this paper is the latter: while the overall VHVS goal may seem lofty, we are primarily interested in providing pragmatic and actionable feedback to designers through a suite of image filters ``inspired by'' human perception \rev{(see Fig.~\ref{fig:perceptual_pat_ex} for a demonstration of its intended use)}.
In other words, we do not purport to faithfully model the human visual system, \rev{but rather to pragmatically support iterative visualization design}.

% Nevertheless, there is much to gain by creating such a pragmatic VHVS for applied design.
% Design in a vacuum is difficult, so most designers---even experienced ones---benefit from external feedback on their work in order to iterate towards an acceptable end product.
% Since visualization is a design discipline~\cite{munzner14visualization}, this also applies to visualization: receiving external feedback is the best way to quickly converge on a design.
% For this reason, most visualization textbooks and design processes emphasize the need for an iterative workflow that progressively fixates towards final product. 
% Thus, the goal of our ``pragmatic'' VHVS is to provide rapid and actionable feedback on perceptual aspects of a visual design in order to promote improvement over time.

\subsection{Scope}

The scope of our VHVS implementation is to serve as a \textit{supertool}~\cite{Shneiderman2022} that uses AI and computer vision to augment, amplify, and extend the capabilities of a human designer during iterative visualization refinement. 
\rev{While our intended user is a visualization designer of any skill level, we note that our approach is not currently to suggest fixes to identified concerns.
This means that while the feedback is useful to anyone, an expert designer will often be in a better situation to address it because of their greater experience.}
We organize the external design feedback into two main types:

\begin{itemize}

    \item\textbf{Design feedback and critique:} This form of external feedback involves receiving criticism---preferably constructive, i.e., focused on improvement, rather than merely pointing out flaws---from peer designers or supervisors. 
    Designers often work in teams or at least as part of organizations with multiple designers, so critique is intrinsic to design~\cite{DBLP:journals/iwc/Bardzell11}. 
    However, receiving feedback from peers can be time-consuming because (a) the process is laborious in itself, and (b) peers are often not immediately available because of their own commitments.
    Furthermore, feedback from an uninitiated peer can often be general and not sufficiently detailed.
    Nevertheless, design feedback remains important, not only for designers working alone on a project, but also for teams who can benefit from an outside and unbiased critical eye on their work.
    
    \item\textbf{Empirical evaluation:} Given that visualization is a primarily empirical discipline, evaluation involving human participants is essential~\cite{DBLP:conf/avi/Plaisant04}.
    This is true for both academia as well as industry and practice~\cite{DBLP:journals/tvcg/Isenberg2013, DBLP:journals/ivs/SedlmairIBB11}.
    However, even small-scale empirical evaluation studies are costly in terms of time, money, and preparation---they certainly cannot provide answers to specific questions designers have about their visual design at short notice.
    For this reason, empirical evaluation is conducted at a time granularity of days or, more likely, weeks.
    
\end{itemize}

\subsection{Design Requirements}
\label{subsec:design-requirements}

Because of the automated nature of the VHVS that we are proposing, it is important to note that we are not replacing but \textit{augmenting} human feedback.
This means that the fidelity requirements are lower because our automated feedback is meant for guidance rather than enforcing specific perceptual or design rules.
We summarize our design requirements from the motivation and scope above as follows: 

\begin{itemize}
    \item[DR1]\textbf{Rapid:} To facilitate a conversation between the designer and the VHVS, feedback must be rapid, preferably yielding output in less than a minute or two (the faster, the better).
    
    \item[DR2]\textbf{Inexpensive:} Similarly, to promote frequent iteration, the feedback cannot require costly investment, or ideally any investment at all.
    This likely precludes crowdsourced critique, which still incurs some cost.
    
    \item[DR3]\textbf{Automated:} The feedback should be mostly automated and not require intricate configuration or setup; ideally, the user should be able to submit the current state of their visualization artifact.
    
    \item[DR4]\textbf{Progressive:} In recognition of the iterative refinement commonplace in visualization design, the feedback should track the evolution of an artifact over time.
    
    \item[DR5]\textbf{Constructive:} The feedback provided should be organized to help the designer improve their artifact.
\end{itemize}

\subsection{Practical Perceptual Feedback}
\label{subsec:practical-feedback}

Focusing on pragmatic, actionable, and practical perceptual feedback on a visualization artifact frees us from having to design a fully comprehensive and accurate model of the human visual system.
Instead we choose types of feedback that will aid the designer in making improvements to a visualization that will \rev{directly} benefit viewers.
We summarize the main types of such feedback below. 
Note that this is a suggestion and not an exhaustive list.

\begin{itemize}
    \item[\faSun] \textbf{Visual saliency:} Saliency is a measure of the patterns and parts of a visual scene that attracts a viewer's eyes~\cite{riche13metricssaliency}, and has long been a core part of vision science~\cite{itti98saliency, harel06gbvs, bruce04saliencyIM}.
    However, visual saliency is also a highly practical and pragmatic aspect of visualization design in helping the designer determine which parts of a visualization will attract the viewer and in what order, and, analogously, which parts will not.
    Receiving unbiased and objective feedback on the visual saliency can certainly help a designer iterate on their visualization to ensure that the saliency is consistent with their intent.
    
    \begin{itemize}
    
        \item\textit{Eyetracking:} Eye trackers empirically measure saliency by collecting eye movement data from the real world~\cite{kim12eyetracker}. 
        Eyetracking technology is now becoming cheaper and increasingly available, enabling designers to use themselves as test subjects. 
        However, such hardware solutions are beyond the scope of our work.

        \rev{\item\textit{Virtual eye trackers:} A virtual eye tracker is trained on eye movement data to generate artificial gaze maps, enabling the designer to conduct a virtual eye tracking experiment.
        Large-scale eye movement data can be used to build eye movement prediction models, such as the work by Itti et al.~\cite{itti98saliency}, the CAT2000 benchmark~\cite{mit-saliency-benchmark}, and SALICON~\cite{jiang15salicon}.}
        While gaze prediction models can be used for this purpose, the Scanner Deeply~\cite{shin23scannerdeeply} tracker is specifically trained on visualization images.

%        \item\textit{Gaze prediction:} Large-scale eye movement data can be used to build eye movement prediction models, such as the work by Itti et al.~\cite{itti98saliency}, the CAT2000 benchmark~\cite{mit-saliency-benchmark}, and SALICON~\cite{jiang15salicon}.
%        This information can in turn be used to predict how a user will perceive a data visualization.

        \item\textit{Information theory:} Information-theoretic approaches to visualization~\cite{DBLP:journals/tvcg/ChenJ10} measure the ratio of raw data that are communicated using a visual representation.
        Calculating the local entropy across a visualization image can thus be seen as a theoretical measure of its visual saliency; its information content.
    \end{itemize}

    \item[\faPalette]\textbf{Color perception:} Color is a basic building block of any image, data visualization in particular~\cite{munzner14visualization, Rhyne2016}.
    Effective use of color is therefore a key factor in any design projects involving data visualization.
    
    \begin{itemize}
        \item\textit{Color statistics:} Understanding dominant color schemes, color distribution, and color scale choices is a useful mechanism for any visualization designer.
        
        \item\textit{Color choice:} Visualization practice stipulates using a limited number of distinguishable and easily named colors~\cite{DBLP:conf/visualization/Healey96}, potentially as a function of the mark used~\cite{szafir18modelingcolordifference}. 
        
        \item\textit{Opponent processing:} Opponent process theory casts color perception as balances between three pairs of colors~\cite{Hering1964, Hurvich1957}; while the exact constituent colors are disputed, these are often held to be red vs.\ green, blue vs.\ yellow, and black vs.\ white.
        Pragmatic visualization would promote avoiding color combinations that involve both parts drawn from one of these pairs.

    \end{itemize}
    
    \item[\faFont]\textbf{Text:} Most visualizations incorporate text in some form, much of it central to understanding the scale, extents, names, and details of the visualized data.
    Textual characters are obviously visual objects that are modeled by other vision models, but because of their special meaning in visualization, text is worth studying on its own.
    We propose several specialized forms of text identification and classification feedback.

    \begin{itemize}
        \item\textit{Legends:} Some visualizations require legends to enable deciphering color allocations or color scales. 
        Identifying the legend, or notifying the designer that none is present, would therefore be useful feedback.
        
        \item\textit{Labels:} Axis labels, titles, and tick marks are central to comprehending a visualization. 
        
        % \item\textit{Legibility:} Sometimes labels and titles in a chart may be difficult to read, for example due to small font size, poor font color choice, or detailed chart backgrounds.
        % Automatically recognizing text may yield useful feedback to this effect to the designer.
        
        \item\textit{\rev{Textual content:}} Disregarding the visual representation of the text, what about \rev{its actual content?
        Feedback on spelling, grammar, and meaning} can help the designer here.
    \end{itemize}
    
    \item[\faChartBar]\textbf{Visual representation:} Some perceptual feedback may be specific to the visualization technique being used.
    While some of this type of feedback may stray into visualization linting~\cite{chen22vizlinter, hopkins20visualint} for finding chart design and construction errors, we here focus our feedback on perceptual aspects.
    
    \begin{itemize}
        \item\textit{Chart type:} Merely using an automatic classifier to identify the chart type can be useful feedback to a designer.
        If the designer is working on a non-standard visualization and the classifier does not recognize it, this may be indicative to change representation, or to better signpost the representation.
        Alternatively, if the designer is using a standard chart type and it is not recognized (or incorrectly so), this may be a signal that they need to improve and standardize their visual design.

        \item\textit{Data extraction:} Taking chart recognition a step further would be to use automatic methods to recover the data from a visualization, essentially reverse-engineering the visualization~\cite{DBLP:journals/cgf/ChoiJPCE19, Poco2017, Savva2011}.
        This would enable the designer to determine if the visual encoding is lossy by recovering symbolic data that has been encoded.
        
        \item\textit{Visual embellishment (``chart junk''):} Visual embellishments to charts---sometimes referred to as ``chart junk''~\cite{Tufte1983}---while potentially beneficial to the memorability of a chart~\cite{DBLP:conf/chi/BatemanMGGMB10}, may detract from the chart or even distract the viewer compared to a more minimalistic visual representation~\cite{Cleveland1985}. 
        Automated object recognition can inform the designer about any visual embellishments and their potential for distraction.
        
    \end{itemize}
    
    \item[\faEye]\textbf{Vision science:} Beyond the visual saliency discussed above, there are many useful metrics from vision science that we may pragmatically adopt for visualization design feedback. 
    As stated above, since our goal is not to accurately model the visual system, we can instead choose concepts that lend themselves to iterative design. 
    
    \begin{itemize}
        \item\textit{Preattentiveness:} Preattentive features~\cite{Treisman1980, Treisman1985, Wolfe2019} are those that guide the viewer's attention so that they ``pop out'' in a visual scene, and that cannot be decomposed into simpler features.
        Automatically detecting and highlighting preattentive features in a visualization would be highly useful, because they can help the designer confirm conscious design choices and discover---and likely address---inadvertent ones.
        
        \item\textit{Ensemble processing:} How do people estimate characteristics from a group of visual objects, such as marks in a scatterplot? 
        \textit{Ensemble processing}~\cite{Alvarez2011} models how the visual system computes averages of visual features in a group with even complex shapes and configurations.
        Implementing an ensemble processing filter could help designers understand how groups of visual marks would likely be interpreted by the viewer.
        
        \item\textit{Shape identification:} Recognizing and identifying shapes in a visualization artifact may be another confirmatory piece of feedback for a designer.
        If shapes are not correctly identified, perhaps due to scale or overplotting, the designer may use this feedback to make revisions to their artifact.
        
        \item\textit{Image segmentation:} More of a computer vision than a vision science tool, \textit{image segmentation}~\cite{DBLP:journals/pami/MinaeeBPPKT22, Shapiro2001} is the process of partitioning an image into segments or regions based on image content. 
        These segments may help a designer to understand about the fundamental structure of the visualization artifact being designed.
        
        \item\textit{Moving object tracking: } While our approach in this paper is based on static screenshots of a visualization, a dynamic animation may yield further perceptual information about a visualization.
        In particular, providing feedback on moving objects may be useful given human perceptual limits on tracking multiple objects~\cite{Cavanagh2005, Pylyshyn1988}.
        It could also be used to understand temporal aspects of an animation~\cite{DBLP:conf/chi/DragicevicBJEF11}, which can aid perception.

    \end{itemize}
    
    \item[\faAccessibleIcon]\textbf{Accessibility}: The accessibility of visualizations has recently become an area of burgeoning research within the community~\cite{DBLP:journals/cgf/ChoiJPCE19, DBLP:journals/tvcg/ChunduryPRTLE22}, and many visualization designers are in dire need of assistance in making their charts accessible. 
    Here we review some specific forms of feedback that can help.
    
    \begin{itemize}
        \item\textit{\rev{Color vision deficiency:}} An estimated 300 million people in the world suffer from some form of color vision deficiency where their ability to distinguish colors is diminished. 
        There are already websites and native tools to show what a visualization tool looks like depending on the form of color vision deficiency; integrating such feedback into a framework would help further.
        \rev{In addition, Angerbauer et al.~\cite{DBLP:conf/chi/AngerbauerRCOPM22} present several findings drawn from a large-scale crowdsourced assessment of color vision deficiency in visualization that could be operationalized for a virtual human visual system.}
        
        \item\textit{General accessibility:} The Chartability framework by Elavsky et al.~\cite{DBLP:journals/cgf/ElavskyBM22} provides heuristics for evaluating and auditing visualizations based on their accessibility to people with different disabilities, including visual, motor, vestibular, neurological, and cognitive ones. 
        However, the framework must currently be applied manually during an accessibility audit.
        Operationalizing these heuristics into an automated model would enable integrating it into a framework such as ours.
        
    \end{itemize}
    
\end{itemize}

%% -------------------------------------------------------------
%% TECHNIQUE
%% -------------------------------------------------------------
\section{The \techname{} Suite}

\techname{} is an implementation for a virtual human visual system designed for iterative design of data visualization artifacts.
In this section, we describe its system architecture, components, and implementation.

\begin{figure*}[tbh]
    \centering
    \includegraphics[width=\linewidth]{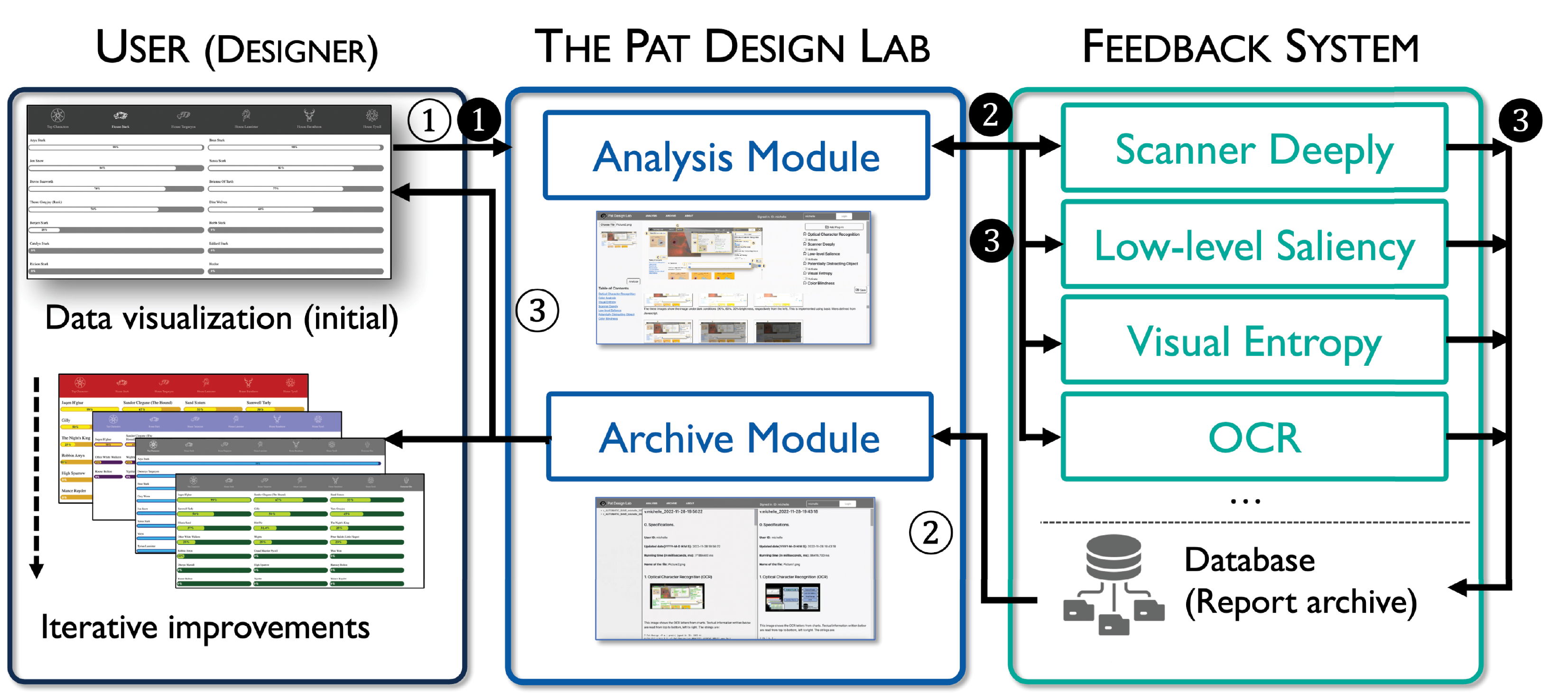}
    \caption{\textbf{\rev{System overview.}}
    The \techname{} suite is comprised of a user interface (The Pat Design Lab) and a feedback system.
    The user interface receives a data visualization (chart) from a user and automatically performs an analysis of the chart.
    The feedback system consists of diverse visual analysis components and uses them to generate a report containing design feedback.
    \rev{(We used Perceptual Pat to iteratively refine this figure by providing a freelance graphic designer with the original version and a PDL report.)}
    }
    \Description{This figure is a diagram showing an overview of our system. 
    We present this in three parts.
    The first is a user part, the second is the Pat Design Lab (PDL) part, and the third is the feedback system part.
    The user part describes the role the designer is doing during the design process. 
    They conduct iterative designs to improve the visualization.
    The Pat Design Lab presents two modules, the analysis module, and the archive module.
    The analysis module provides analysis from various machine-learning-steered tools that analyze the visualization. 
    The design lab also provides the archive module, where the designer can refer to past versions of their design. 
    The feedback system provides analysis to the analysis module.
    The feedback system consists of a list of image filters and a database for storing reports.}
    \label{fig:system-overview}
\end{figure*}

\begin{figure*}[tbh]
    \centering
    \includegraphics[width=\linewidth]{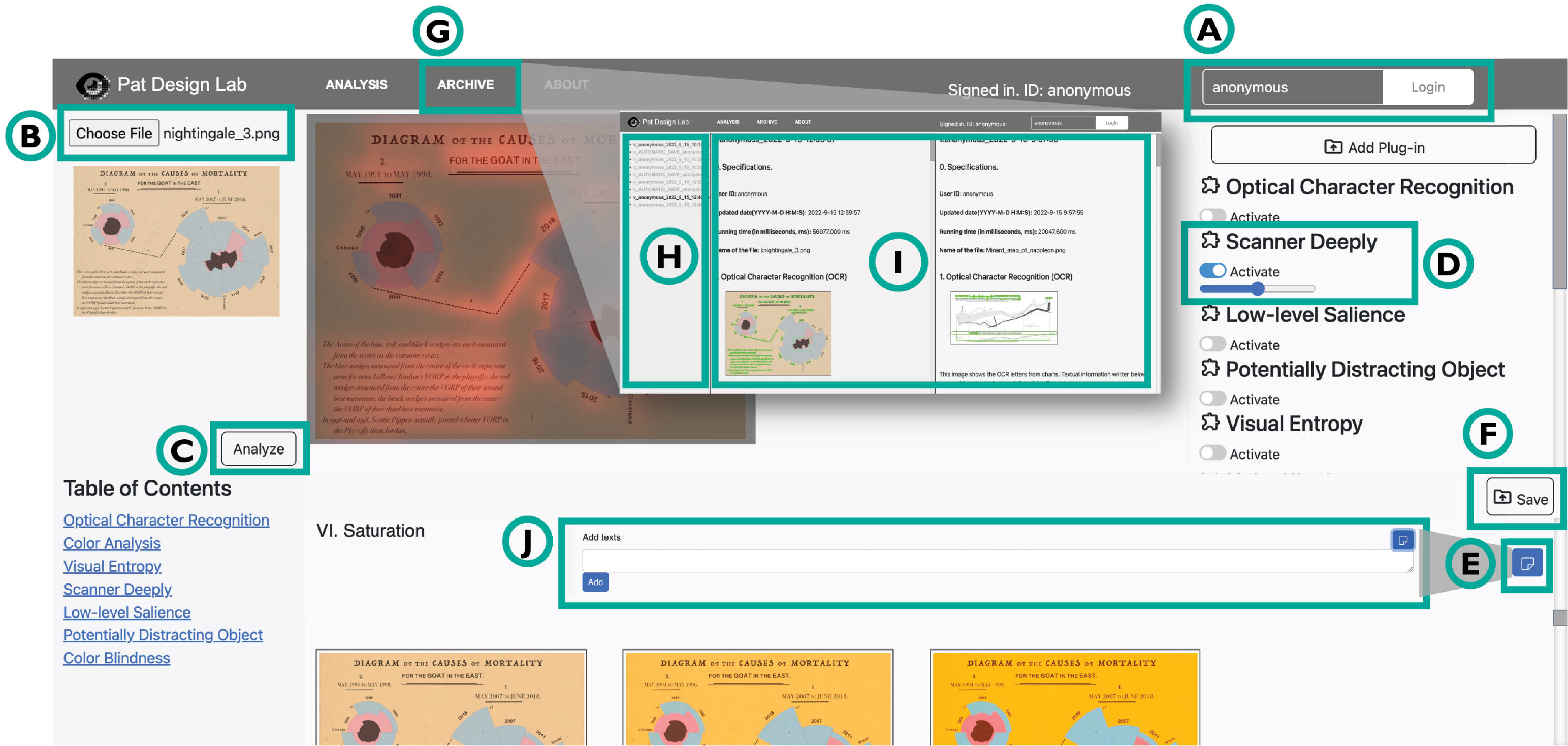}
    \caption{\textbf{Designing visualizations with the Pat Design Lab.}
    \rev{The user can log in to the system by signing in on (A). 
    Then, they can upload images on (B), and start the analysis process by clicking the `analyze' button in (C).
    They can choose filters in the control view, and at the same time control the opacity of the filter in (D). 
    To take notes in the report, they need to click the blue button in (E). 
    Then, they can add texts in (J) and click the button `add.'
    Users have to manually press the save button in (F) to keep record of them report with notes. 
    They can check her past versions of the design by clicking the hyperlink `archive' at the navigation bar (G).
    In the archive view, they can select a version of her interest in (H), and the most-recently chosen version will appear at the left window, pushing the old one into the right window (I). }}
    \Description{This image shows various interface tools deployed for the Pat Design Lab.
    The user can log in to the system by signing in. 
    The sign in interface is at the top rightmost part of the interface.
    Then, they can upload images by clicking the button `choose file', and start the analysis process by clicking the `analyze' button, which is below the `choose file' button.
    After the designer clicks `Analyze,' an enlarged version of the image shows at the center of the interface.
    They can choose which filters to show on top of the large image from the control view, and at the same time control the opacity of the filter. 
    A report is shown below the large image. 
    To take notes in the report, there is a blue button with a `document' icon below the title of each section of the report, and they need to click it. 
    Then, they can add texts and click the button `add.'
    The designers have to manually press the save button to keep record of them report with notes. 
    They can check past versions of the design by clicking the hyperlink `archive' at the navigation bar, which is situated at the top of the interface.
    The archive view consists of three parts. 
    The first is the list view, where past versions are listed.
    Then, there two window views that show two archived versions of the design.
    In the list view, they can select a version of their interest, and the most-recently chosen version will appear at the left window view, pushing the old one into the right.}
    \label{fig:interface-overview}
\end{figure*}

\begin{figure*}[tbh]
    \centering
    \includegraphics[width=\linewidth]{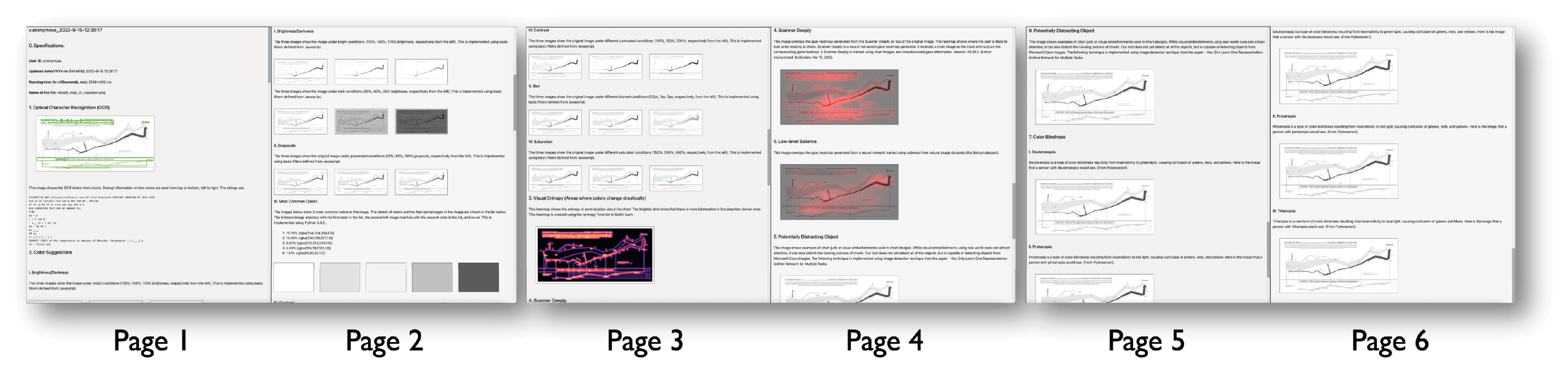}
    \caption{\textbf{\rev{Report generated by the Pat Design Lab.}}
    The report contains information about filters used for the analysis of the updated chart images. 
    It consists of 8 sections.
    These include (1) the chart's specifications, (2) OCR, (3) visual entropy, (4) a Scanner Deeply, (5) low-level salience, (6) potentially distracting objects (Chartjunk~\cite{DBLP:conf/chi/BatemanMGGMB10}), and (7) \rev{color-vision deficiency.}}
    \Description{The report contains information about filters used for the analysis of the updated chart images. 
    It consists of 8 sections. 
    These include the chart's basic properties, results from OCR filter, visual entropy filter, a Scanner Deeply filter, low-level salience filter, potentially distracting object filter, also known as Chartjunk, and color vision deficiency filter.}
    \label{fig:pat-report}
\end{figure*}

\subsection{System Overview}
\label{subsec:system-overview}

The Perceptual Pat suite is designed to meet the requirements 
in Sec~\ref{subsec:design-requirements} (see Figure~\ref{fig:system-overview} for an overview; \rev{the black \bcircled{0} and white \wcircled{0} circles refer to the steps in the figure}).
The suite consists of two components: a web-based user interface (the Pat Design Lab) and a feedback system. 
The user interface (PDL) contains analysis and archive modules---see the following subsection for more details.
\bcircled{1}
The analysis module receives a chart image from a user
in a graphics format (e.g., \verb|.png|, \verb|.jpeg|, or \verb|.jpg|).
\bcircled{2}
This action automatically triggers the feedback system to generate a report containing the results of the diverse visual analysis we develop in Sec~\ref{subsec:components} (\textbf{DR1}, \textbf{DR2} and \textbf{DR3}).
\bcircled{3}
The report will be shown to the user, and is also stored in the database for future reference.
%We provide a feature that a user can add annotations to the report.
%By clicking the `save' button, the notes will be stored along with the report.

The archive module is designed to support the iterative chart design process of a user (\textbf{DR4} and \textbf{DR5}).
\wcircled{1}
It allows a user to retrieve the report that they generated before or to compare two or more reports at the same time.
The user will have a list of report names shown in chronological order.
By clicking those report names, the user interface requests all reports from the feedback system.
\wcircled{2}
The feedback system will pull those reports from the database.
\wcircled{3}
Reports are shown in a single view side-by-side and are scrolled up/down together.

\subsection{The Pat Design Lab}

The Pat Design Lab (PDL) is a web-based single-page application. 
At its core, PDL consists of three functions: upload, analyze, and save. 
Users can upload a screenshot of their chart image. 
Then, they can analyze the image \rev{using the perceptual components} provided by the PDL. 
Finally, users can save their versions for comparison with past designs. 

To initialize the Pat Design Lab, the user needs to login to the system (Fig.~\ref{fig:interface-overview}(A)). 
Once they have signed in, Pat Design Lab activates and the update view (Fig.~\ref{fig:teaser}(A)) appears. 
In the update view, they can choose a file to upload \rev{into the main interface.}
They can start the analysis process by pressing the `analyze' button in Fig.~\ref{fig:interface-overview}(C).

When PDL completes the analysis, \rev{the analysis report is displayed in the report view (Fig.~\ref{fig:teaser}(E)).}
Users can \rev{toggle filters that are overlaid on top of the visualization image}, as shown in Fig.~\ref{fig:interface-overview}. 
\rev{The filter opacity can be controlled using a slider} (Fig.~\ref{fig:interface-overview}(D)).
\rev{Multiple filters can be overlaid at once to enable studying compound effects.}

The report view in Fig.~\ref{fig:teaser}(E) provides access to the full analysis report; Fig.~\ref{fig:pat-report} shows an example.
\rev{The component names in the navigation view provides easy access to corresponding report sections~\ref{fig:teaser}(D).}
\rev{Users can also add notes about each component for each section by clicking the document button in Fig.~\ref{fig:interface-overview}(E), which displays a text input interface (Fig.~\ref{fig:interface-overview}(J)).}
They can save the version with the notes by clicking the blue-colored button with a document icon in Fig.~\ref{fig:interface-overview}(F). 

\rev{The archives tab provides access to saved reports} (Fig.~\ref{fig:interface-overview}(G)).
\rev{This page provides a full list of past reports and notes (Fig.~\ref{fig:interface-overview}(H)).}
\rev{It also shows two windows that lets users compare two reports of their choice (Fig.~\ref{fig:interface-overview}(I)).}

\begin{table*}[htb]
    \centering
    \begin{tabular}{lll}
    \toprule
    \rowcolor{gray!10} 
    \textbf{\textsc{Component Name}} &
    \textbf{\textsc{Feedback Type}} &
    \textbf{\textsc{Implementation}}\\
    \midrule
    
    Scanner Deeply & \faSun~Visual Saliency & Scanner Deeply~\cite{shin23scannerdeeply}\\
    \rowcolor{gray!10}
    Low-level Salience & \faSun~Visual Saliency & Trained SimpleNet~\cite{reddy20deepsaliencyprediction} using Salicon dataset~\cite{jiang15salicon}\\
    Visual Entropy & \faSun~Visual Saliency & SciPy library~\cite{virtanen20scipy}\\
    \rowcolor{gray!10}
    Color Suggestions & \faPalette~Color perception & CSS filters, Python Image, OpenCV libraries \\
    OCR & \faFont~Text & Google PyTesseract-OCR~\cite{kay07tesseract, PyTesseract_perf} \\
    \rowcolor{gray!10}
    Chartjunk/Visual Embellishment & \faChartBar~Visual Representation & YoloR~\cite{wang21yolor} \\
    \rev{Color Vision Deficiency} & \faAccessibleIcon~Accessibility & Python Color-blindness library~\cite{elrefaei18colorblind}\\
    \bottomrule
    \end{tabular}
    \caption{\textbf{Components composing Pat's feedback system.}
    Names, feedback type and implementation of seven components included in the suite.
    \rev{These were selected to be representative and cover the design space in Section~\ref{subsec:practical-feedback}.}
    \rev{We provide additional technical and performance details on these components in Appendix (supplemental material).}
    }
    \label{tab:filters}
    \Description{This table provides information about the tools used in implementing Perceptual Pat. The list of component names is Scanner Deeply, low-level salience, visual entropy, color suggestions, optical character recognition, chartjunk and lastly, color vision deficiency. Per tool, we provide its feedback type and its implementation source.}
\end{table*}

\subsection{Current \techname{} Components}
\label{subsec:components}

\rev{Table~\ref{tab:filters} shows the components in Pat's feedback system, drawn from Sec.~\ref{subsec:practical-feedback}.
The focus of this work is not primarily in evaluating the performance of these components, but in exploring how access to a virtual human visual system can help users during iterative design. 
For this reason, our goal was not to exhaustively cover the entire design space in Section~\ref{subsec:practical-feedback}, but to find a representative sample of components.}

\rev{Towards this end, we chose several perceptual components that cover a wide range of the feedback types in Section~\ref{subsec:practical-feedback} and are readily adaptable from existing computer vision and machine learning libraries.
Note that Pat uses a plugin system, so adding new components is straightforward.}
Below we explain how we implemented the seven components.

\paragraph{Optical Character Recognition (OCR)}

Optical character recognition \rev{is a computer vision technique~\cite{shi17ocrnn}} that detects text characters from natural images.
This technique \rev{gives the designer an ``smoke test'' of the legibility of the text; if the OCR technique fails to detect the text, there may be a legibility problem and redesign may be needed.}

\paragraph{A Scanner Deeply}

Scanner Deeply~\cite{shin23scannerdeeply} \rev{is a \textit{virtual eyetracker}; a gaze heatmap generator using a neural network model trained on more than 10K} instances of eye movement data with chart images as the visual stimuli.
\rev{The model will generate a simulated heatmap predicting where a person's attention will be directed when viewing a visualization.}

\paragraph{Low-level Salience}

This \rev{salience heatmap generator uses a neural network model trained on the Salicon dataset}~\cite{jiang15salicon}, which contains fixation maps for a natural image dataset (i.e., Microsoft COCO~\cite{lin14coco}).
\revt{The component is particularly useful in showing viewer attention for visualizations that contain natural image data, or in visualizations that are situated within the world, such as for a visualization embedded in Augmented Reality.}

%That said, it can be a useful source of reference for salience when the visualization is drawn with natural image as its background, or when it is situated within the world, such as for a visualization embedded in Augmented Reality.

\paragraph{Visual Entropy}

\rev{This components generates a heatmap showing \textit{visual entropy}; pixels whose RGB values differ from neighboring ones.}
\rev{This can potentially give designers awareness of the data distribution in the image.}

\paragraph{\rev{Color Vision Deficiency}}

\rev{This filter component provides color overlays to enable the designer to see how people suffering from different types of color vision deficiency (CVD) would see the visualization}. 
\rev{The component supports three types of CVD: (1) deuteranopia, (2) protanopia, and (3) tritanopia.}

\paragraph{Chartjunk/Visual Embellishment}

\rev{Our component for detecting chart junk is implemented using the YoloR object detection algorithm~\cite{wang21yolor,yolor_perf}, which detects real-world objects visible in the visualization image.}
\rev{The model is limited by its training, which may cause it to fail to detect all objects, but this functionality can still serve as early warning.}

\paragraph{Color Suggestions}

\rev{Our color suggestions component is implemented using CSS filters and include blur, gamma, grayscale, contrast, and saturation.}
\rev{It enables envisioning how a visualization would look under different color themes.} 

\subsection{Implementation Notes}
\label{subsec:implementation}

The Perceptual Pat suite is implemented as a \rev{client/server web application}.
\rev{The Pat Design Lab was built} using HTML, CSS, JavaScript, and JQuery.
We implemented the feedback system using the Python Flask web framework\footnote{https://palletsprojects.com/p/flask/}; \rev{the analysis components were implemented server-side in Python 3}.
We store data in MongoDB.\footnote{https://www.mongodb.com/}
\rev{During the user study, the platform was hosted} on a Ubuntu Linux server equipped with an Intel i7-12800K processor, 32GB RAM, 2 Pascal Titan RTX GPUs, and 2TB of flash memory.

%% -------------------------------------------------------------
%% METHOD
%% -------------------------------------------------------------
\section{User Study}
\label{sec:evaluation}

We conducted a user study to assess the utility of the Perceptual Pat suite and the web-based Pat Design Lab implementation. 
The study asked participants to use the Pat Design Lab in support of a visualization design task spanning between three and five days.
Rather than compare our approach to an existing tool, our study is mostly qualitative in nature and focuses on understanding how professional visualization designers might use the tool to iteratively refine a visualization artifact.
We motivate this choice by the fact that there exists no directly comparable tool that would serve as a useful baseline.

Here we describe the participants, methods, and metrics of this study.
In the next section, we report on the results.

\begin{table*}[t!]
    \centering
    \begin{tabular}{llllll}
        \toprule
        \rowcolor{gray!10} 
        \textbf{\textsc{ID}} &
        \textbf{\textsc{Gender}} &
        \textbf{\textsc{Age}} &
        \textbf{\textsc{Education}} &
        \textbf{\textsc{Job Title}} &
        \textbf{\textsc{Years in Vis}}\\
        \midrule
        P1 & Male & 28 & Master's Degree & Software Engineer & More than 3 years\\
        \rowcolor{gray!10}
        P2 & Female & 27 & Master's Degree & Ph.D. Student in Visualization & More than 3 years \\
        P3 & Female & 43 & Master's Degree & UX/UI/Visualization Designer & More than 5 years\\
        \rowcolor{gray!10}
        P4 & Female & 24 & Bachelor's Degree & Freelance Visualization Designer & More than 1 year \\
        P5 & Male$^*$ & -- & (no information) & (no information) & (no information)\\
        \bottomrule
    \end{tabular}
    \caption{\textbf{User study demographics.}
    Our participants were recruited from the Data Visualization Society Slack channel (\url{https://www.datavisualizationsociety.org/slack-community}) and community of IT Engineers in Facebook South Korea.
    $*$ = participant P5 did not finish the study.
    }
    \Description{This table has demographics information about the participants that participated in our experiment.
    Participants are given US \$ 40 as compensation.
    We provide the participants' gender, age, education status, job title, and visualization experience.
    The first participant is a male, 28 years old, holds a master's degree, is a software engineering and has more than three years of experience in visualization.
    The second participant is a female, 27 years old, holds a master's degree, and is a Ph.D. student in visualization and has more than three years of experience in visualization.
    The third participant is a female, 43 years old, holds a master's degree, is a UX/UI/Visualization designer and has more than five years of experience in visualization.
    The fourth participant is a female, 24 years old, holds a bachelor's degree, is a freelance visualization designer and has more than one year of experience in visualization.}
    \label{tab:participant}
\end{table*}

\subsection{Participants}

We recruited 5 professional visualization designers using the Data Visualization Society Slack channel.
Table~\ref{tab:participant} presents an overview of the participant demographics.
All participants were paid a total of \$40 as an Amazon gift card upon completing the study (or its equivalent in the participant's requested currency).
Unfortunately, one participant abandoned the study after a week without producing any design artifacts; their demographic data is shown in Table~\ref{tab:participant}, but in no other part of the paper.

\subsection{Apparatus}

The study was conducted online in its entirety using the participant's own computers.
We imposed no specific hardware on the study, only that the participant would use a personal computer, and not a mobile device such as a smartphone or tablet for the design task.
We recommended them to use the website in Google Chrome.
Since the Pat Design Lab is designed to use only screenshots of data visualization tools, participants were free to use their own preferred visualization tool and workflows.

\subsection{Design Task}

Participants were asked to do only one thing during the 3-5 days of the study: to design a new visualization from scratch using the Pat Design Lab as supporting software.
The visualization could be anything: use any dataset, any representation, and for any use (including merely for the purpose of this study).
They were asked to record at least 5 versions of their visualization into the Design Lab.
Ideally the five stored versions would be taken from representative stages in the design process.
Participants were instructed to read the reports, write at least one annotation, and to endeavor to use the reports to improve their designs.

As stated above, we placed no restriction on the design workflow or visualization tools used in the process---participants could use whatever tool they preferred, or even switch tools during the process.
For example, a participant could start with a pen-based sketch and then proceed to using Tableau, Excel, or Spotfire.
We only stipulated that the participant spent a total of at least \textbf{two hours} on the design process.

In summary, we made the following requirements on the visualization design process:

\begin{itemize}
    \item Resolution minimum 400$\times$300, maximum 1000$\times$1000 pixels in resolution (higher resolutions were downscaled).
    \item At least 5 versions stored into the Pat Design Lab.
    \item Artifact must include at least one chart (any form of chart), and could could include multiple charts.
    \item Artifacts may include text.
    \item Artifacts may \textbf{not} include photographs.
    \item No confidential or identifying information or data is to be included.
    \item Participants give permission for publishing images of their created artifacts in academic papers about the work.
\end{itemize}

\subsection{Procedure}

Our study was approved by our university's IRB.
The study consisted of three phases that spanned over three to five days, \rev{followed by an independent fourth phase involving external evaluators.}
We scheduled the dates for each phase during initial recruitment.
\rev{The fourth phase was independent of the preceding three and did not involve the designer participants.}
Below we describe each of the \rev{four} phases in detail.

\paragraph{Phase I: Initial Interview.}

During our initial interview (Phase I), we gave a brief introduction of the study and then asked participants to provide informed consent (signature waived due to the online format) using an online form according to stipulations from our IRB.
We then collected participants demographics using another form. 
The remainder of the initial interview consisted of training during which time the experiment administrator (the first author of this paper) demonstrated how to create an account in the Pat Design Lab, authenticate and log in, and then upload a screenshot of a visualization into the tool.
The administrator then showed how to run a Perceptual Pat analysis and interpret the resulting report, as well as how to compare two different reports.

After finishing demonstrations, the experimenter asked the participant to repeat each step, and answered any questions that the participant had. 
Then the experimenter gave the participant their design charge, including the requirements listed above.
The participant was given the design task in electronic form. 
Finally, the session ended by confirming the date and time for the exit interview (Phase III).
Each session lasted approximately 30 minutes.

\paragraph{Phase II: Individual Design Process.}

During the intervening time between the initial interview (Phase I) and the exit interview (Phase III), the design process (Phase II) consisted of the participant working on the visualization artifact they were designing.
Participants were free to spend this time in any way they wished---we only asked that they spent at least a total of \textbf{two hours} on the design work and that they used the Pat Design Lab to support the process.
They were instructed to reach out to the experimenter with any questions or problems that arose during the study; none of them did, and the system was stable during the time period of the experiment.

\paragraph{Phase III: Exit Interview.}

The exit interview (Phase III) involved asking the participant about their experiences using the Pat Design Lab, their feedback about its strengths and weaknesses, and their design process using the tool.
The experiment administrator then stepped through the participant's version history in the Pat Design Lab, one version at a time and asking about details for each version.
In case the participant had stored more than five versions, the experimenter asked the participant to identify the five most significant versions.
Each session lasted approximately 30 minutes.

\paragraph{\rev{Phase IV: External Assessment.}}

\rev{Finally, we recruited three external evaluators to objectively assess the visualization design process.
All evaluators were senior visualization faculty or researchers with experience in teaching data visualization and/or designing their own visualizations.
They gave informed consent and were then given access to the sequence of visualization versions for each of the participants.
They were then asked to provide their assessment of the quality of changes for each pair of versions (i.e., from version 1 to 2, 2 to 3, etc) as well as from the initial (version 1) to the final (version 5) design using a 1-5 Likert scale (1 = significant decline in quality, 3 = neutral, 5 = significant improvement in quality).
They were also asked to motivate their assessment using free text.}

\subsection{Data Collection}

Interviews were video and audio recorded, and the audio was transcribed for later analysis. 
Furthermore, demographics, subjective preferences, and tool rankings were collected using online forms. 
The Pat Design Lab itself was the main source of data collection. 
This data encompasses each of the visualization versions uploaded into the Pat Design Lab by the participants, including the screenshot of the visualization, the resulting Pat design report, and the annotations added by the participant.
These annotations were augmented with the spoken comments that participants made about each version during the version walkthrough in Phase III.

While we included all versions uploaded by participants in our analysis, we also asked participants to identify the five most significant versions in case they had uploaded more than five.

%% -------------------------------------------------------------
%% RESULTS
%% -------------------------------------------------------------
\section{Results}

Here we report on the results from our user study. 
First, we introduce participants' responses on interviews prior to starting the experiment. 
Then, we analyze the evolution of the visualization artifacts and the impact of \techname{} on the iterative design.
Finally, we present participants' comments after conducting the experiment.  

\subsection{Results from the Initial Interview}

Besides overall instructions and a demographics survey during the initial interview in Phase I, we also asked the participants three questions to understand how they usually get feedback for visualizations; see Table~\ref{tab:pre_study} for the questions and responses.
As is clear from the table, the majority of participants receive feedback from people who are easily reachable, such as peers, colleagues, or supervisors (P2, P3). 
If the visualization is upon a client's request, then these designers tend to get feedback directly from clients to confirm that their work is acceptable, as it is the most direct way to understand the client's intentions (P1, P4).

\begin{table*}[t!]
    \centering
    \begin{tabular}{lm{14em}m{14em}m{14em}}
        \toprule
        \rowcolor{gray!10}
        \textbf{P\#} &
        \textbf{\textsc{Q1}:} How do you currently get feedback to improve a visualization?&
        \textbf{\textsc{Q2}:} Are you familiar with any tools that provide design feedback?&
        \textbf{\textsc{Q3}:} If such a tool existed, would you find it useful?\\
        \midrule
        \textbf{\textsc{P1}} & clients & not familiar & somewhat useful \\
        \rowcolor{gray!10}
        \textbf{\textsc{P2}} & peers, friends & not familiar & very useful \\
        \textbf{\textsc{P3}} & supervisors, managers, peers~~~~ & not familiar & somewhat useful \\
        \rowcolor{gray!10}
        \textbf{\textsc{P4}} & client, peers & not familiar & somewhat useful \\
        \bottomrule
    \end{tabular}
    \caption{\label{tab:pre_study} \textbf{Initial interview with participants.}
    Reponses for Q1 are given in the participant's own order.}
    \Description{This table provides questions we asked during the initial interview with participants. We asked 5 questions, and this table answers three questions.
    The first question we asked was `How do you currently get feedback to improve a visualization?'
    The second question we asked was `Are you familiar with any tools that provide design feedback?'
    The third question we asked was `If such a tool existed, would you find it useful?'
    For the first question, P1 answered as clients, P2 answered as peers and friends, P3 answered as supervisors, managers and peers, and P4 answered as client and peers. 
    For the second question, all participants answered as `not familiar.'
    For the third question, P1, P3, and P4 answered as somewhat useful, and P2 answered as very useful.}
\end{table*}

When asked whether they were familiar with tools that provide design feedback, no participant was familiar with any such tool.
All participants felt, however, that such a tool would be useful.
As a main reason for this, all participants pointed out that receiving timely feedback is sometimes a challenge and there are only a few people who can provide feedback in visualizations.
Asking the same people over and over can be problematic.
P2 provided us another reason---that it is difficult to verbally communicate about visualizations with peers and by using a machine, she might not have such a problem. 

P3 did raise an issue about the credibility and transparency of the system. 
She felt that such a tool would be useful, but said, ``\textit{I am not sure if I can 100\% trust the feedback provided by a machine.
If the machine makes a comment, I would like to know why, in detail, it provided that comment, so that I can decide whether or not to trust it.}'' 

%That said, however, the reasons for getting the answer were all varied. 
%P1 mentioned the difficulty in receiving feedback from peers, as it is difficult to find visualization experts nearby. 
%P2 talked about the difficulty in exchanging back-and-forth visualization feedback between peers verbally. 
% Oftentimes, the design feedback that she gets from peers can be contradictory and repetitive.
% She said that if a tool could provide a clear message to the visualization, then it would be helpful.
% P3 also admitted that a feedback tool would be helpful, but raised doubts on the content of feedback a machine could provide. 
% She was curious what type of feedback can a tool provide designers, and was unsure if it was something that she could use for design
% P4 said that during the chart design process, she sometimes feels unsure whether she is `doing the right thing' and wants to see how others would think of her designs. 
% So she asks her friends, but visualization experts are scarce, and so she keeps on asking to a small range of people. 
% Such a tool, she said, can help me rely on them less. 
% P5 
% ]because it is not always easy to get feedback (we have to ask people), the comments we get from peers or supervisors can sometimes be general and too obvious that it may not be very insightful (P5) and  

\begin{figure*}[tbh]
    \setlength{\tabcolsep}{0pt}
    \begin{tabular}{lccccc}
        \hspace{0.5cm} &
        \colorbox{blue!10}{\parbox[t][0.25cm]{0.16\textwidth}{\centering Version 1}} &
        \colorbox{blue!20}{\parbox[t][0.25cm]{0.16\textwidth}{\centering Version 2}} &
        \colorbox{blue!30}{\parbox[t][0.25cm]{0.16\textwidth}{\centering Version 3}} &
        \colorbox{blue!40}{\parbox[t][0.25cm]{0.16\textwidth}{\centering Version 4}} &
        \colorbox{blue!50}{\parbox[t][0.25cm]{0.16\textwidth}{\centering \textcolor{white}{Version 5}}}
        \\
        
        P1 & 
        \includegraphics[width=0.17\textwidth]{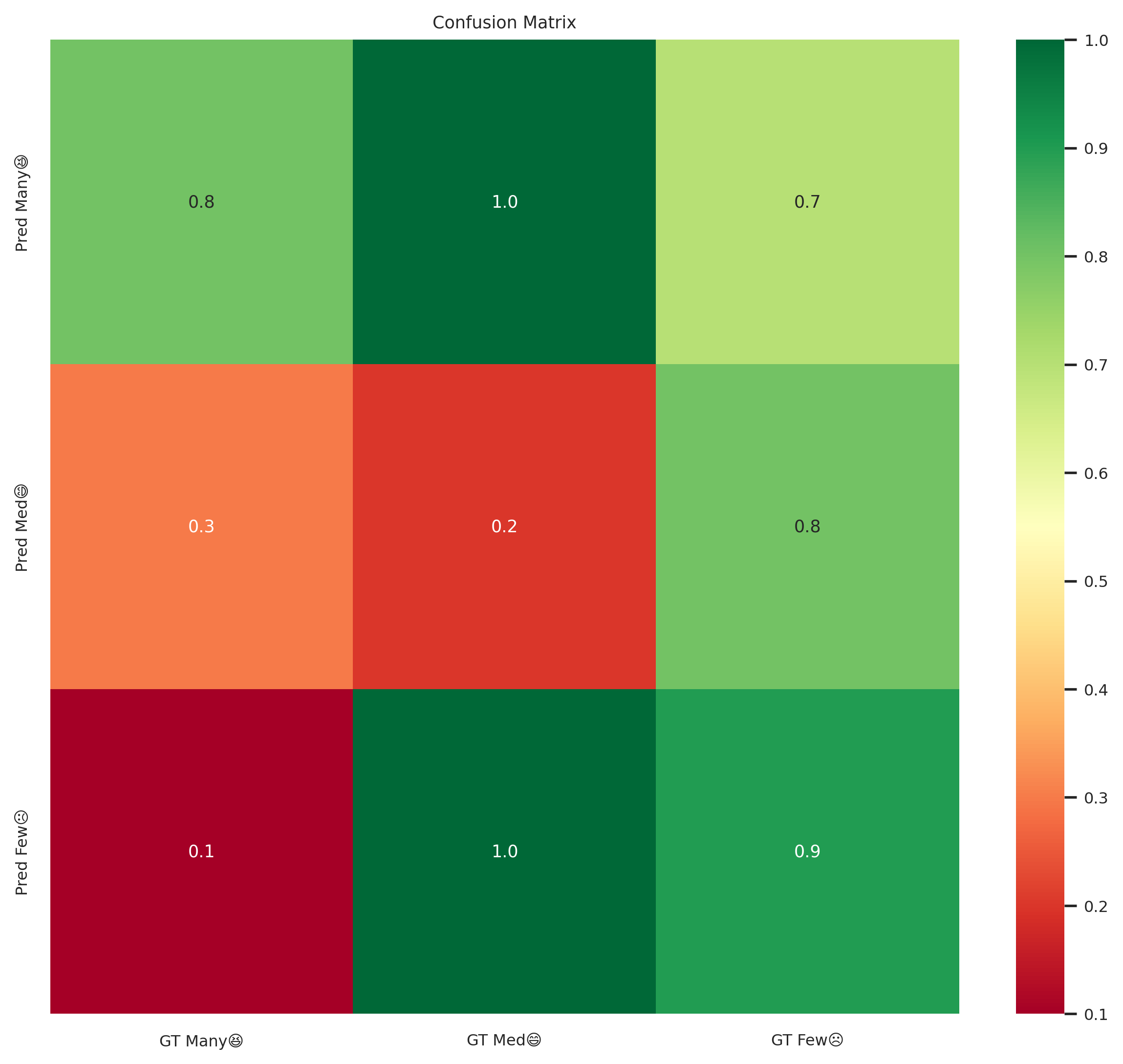} &
        %\resizebox{0.17\textwidth}{figures/P1_1.png} &
        \includegraphics[width=0.17\textwidth]{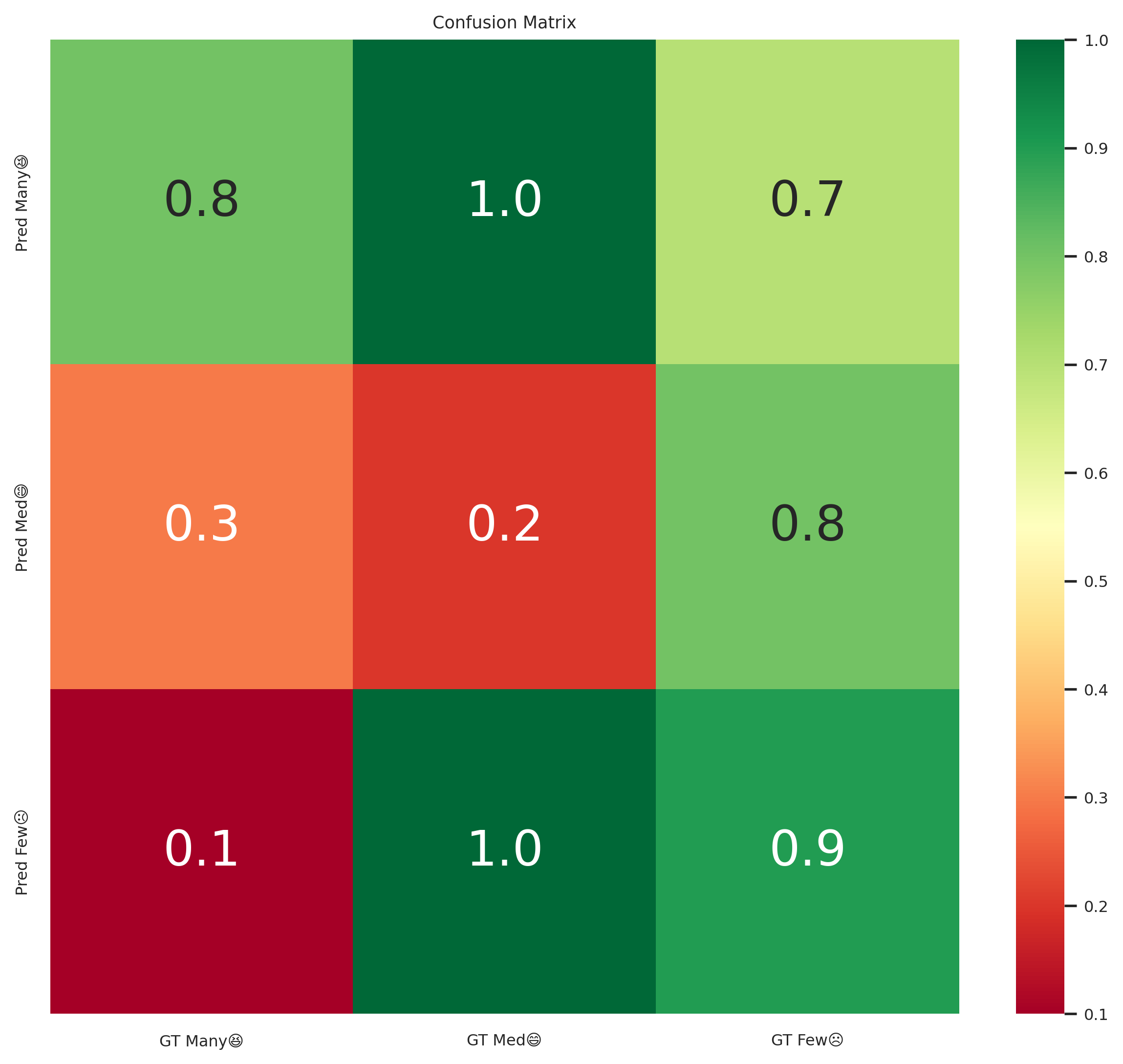} &
        %\resizebox{0.17\textwidth}{!}{\framebox{Vis}} &
        \includegraphics[width=0.17\textwidth]{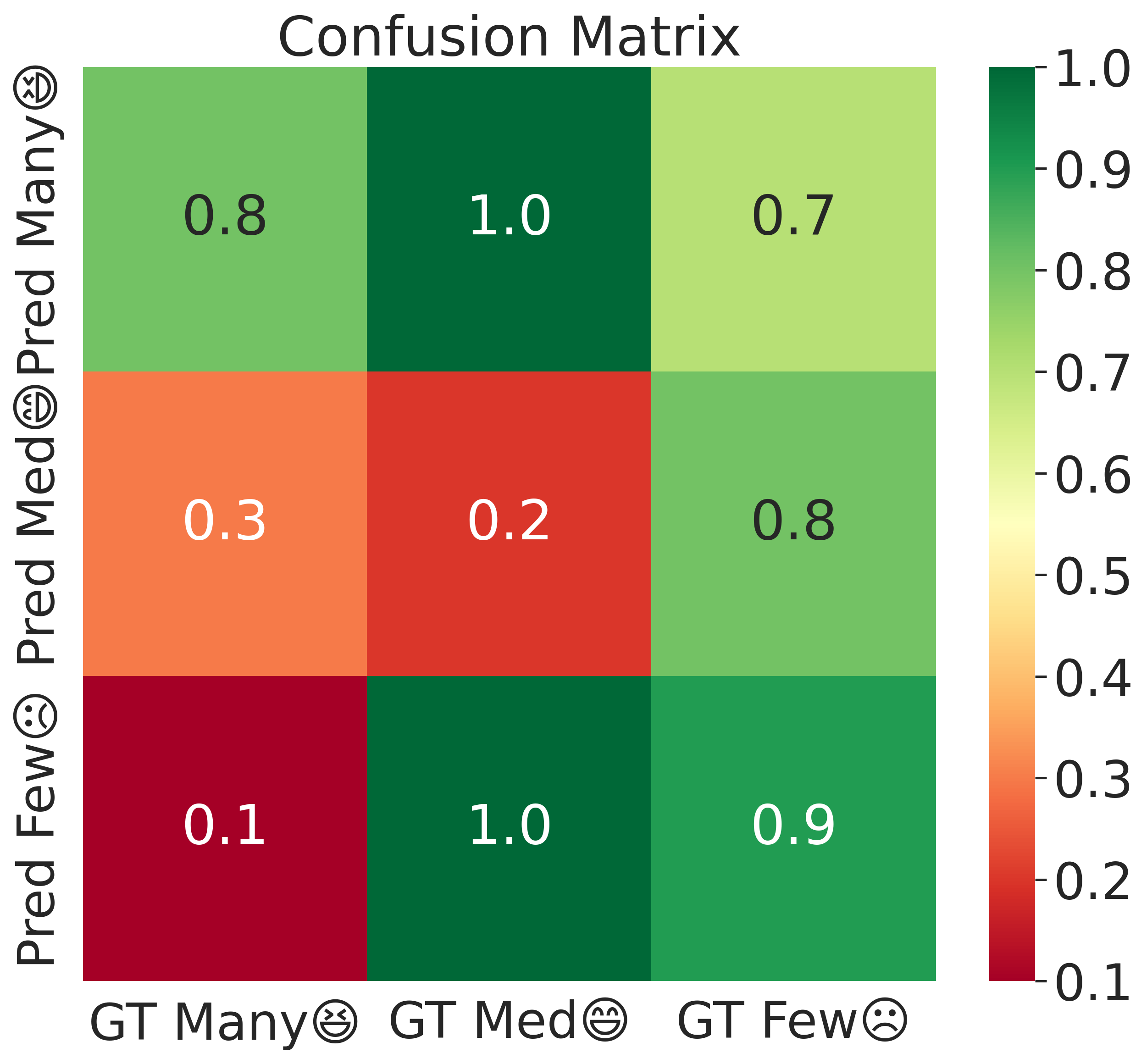} &
        %\resizebox{0.17\textwidth}{!}{\framebox{Vis}} &
        \includegraphics[width=0.17\textwidth]{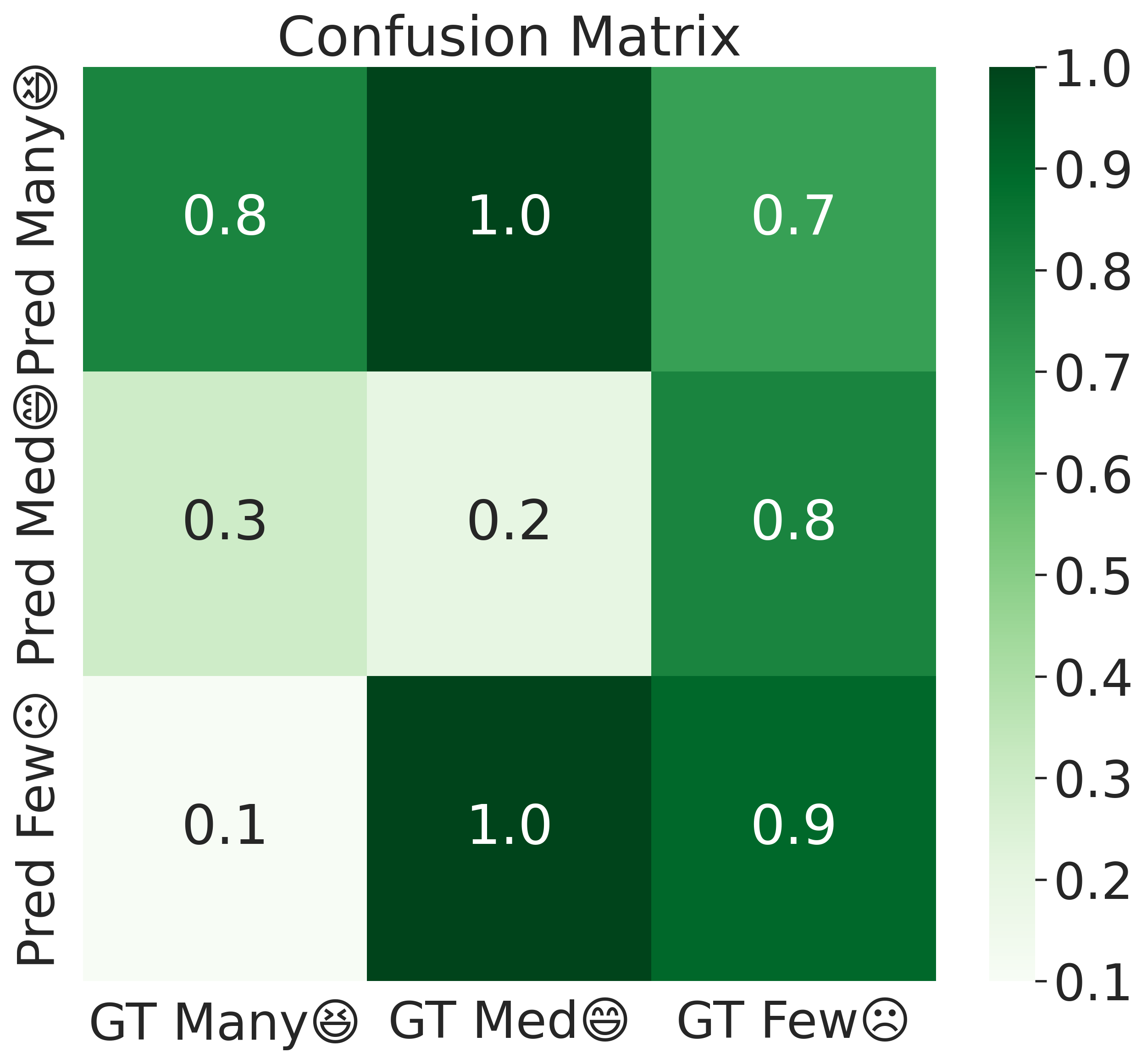} &
        %\resizebox{0.17\textwidth}{!}{\framebox{Vis}} &
        \includegraphics[width=0.17\textwidth]{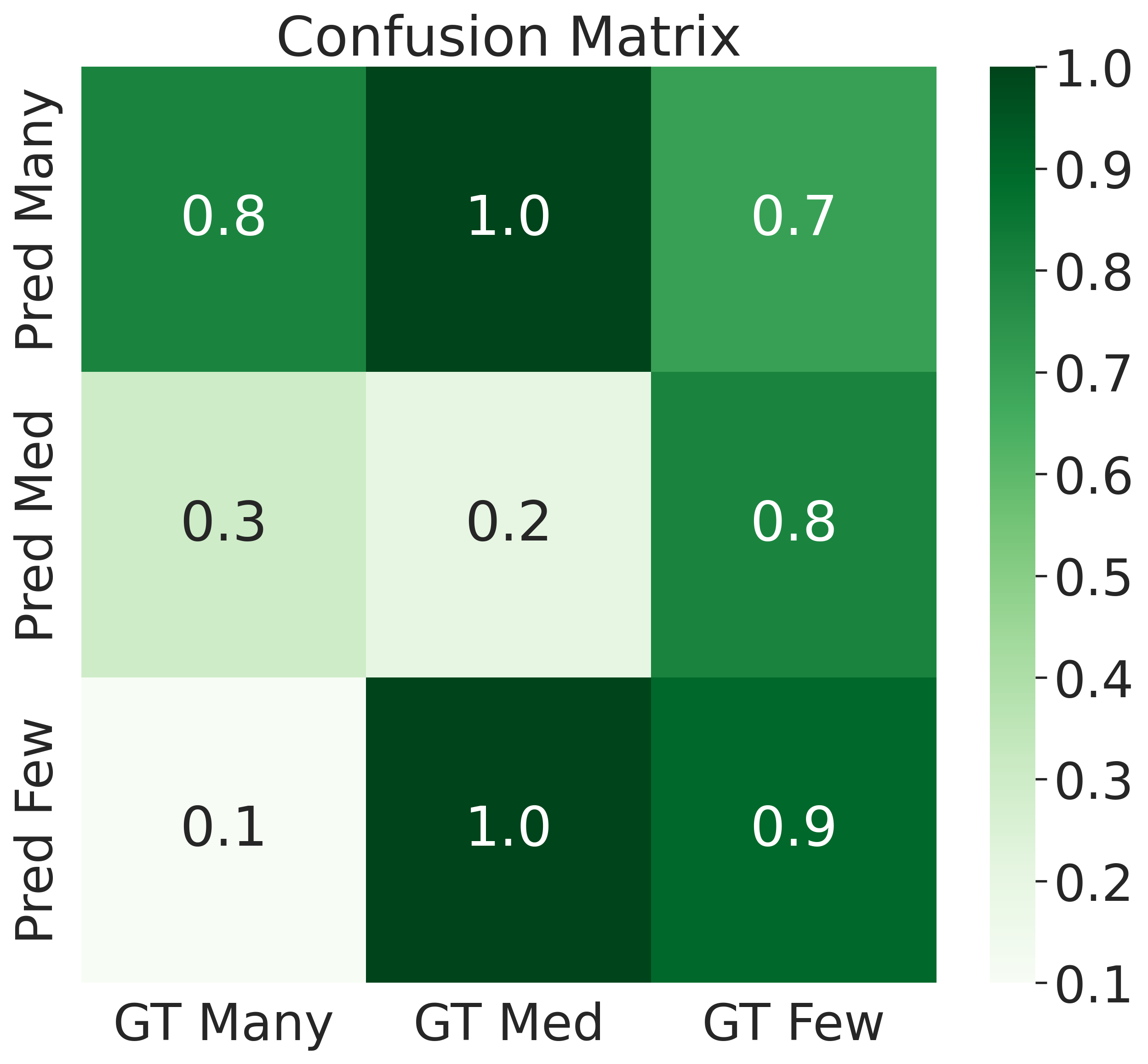} 
        %\resizebox{0.17\textwidth}{!}{\framebox{Vis}}
        \\
        \midrule

        P2 & 
        \includegraphics[width=0.17\textwidth]{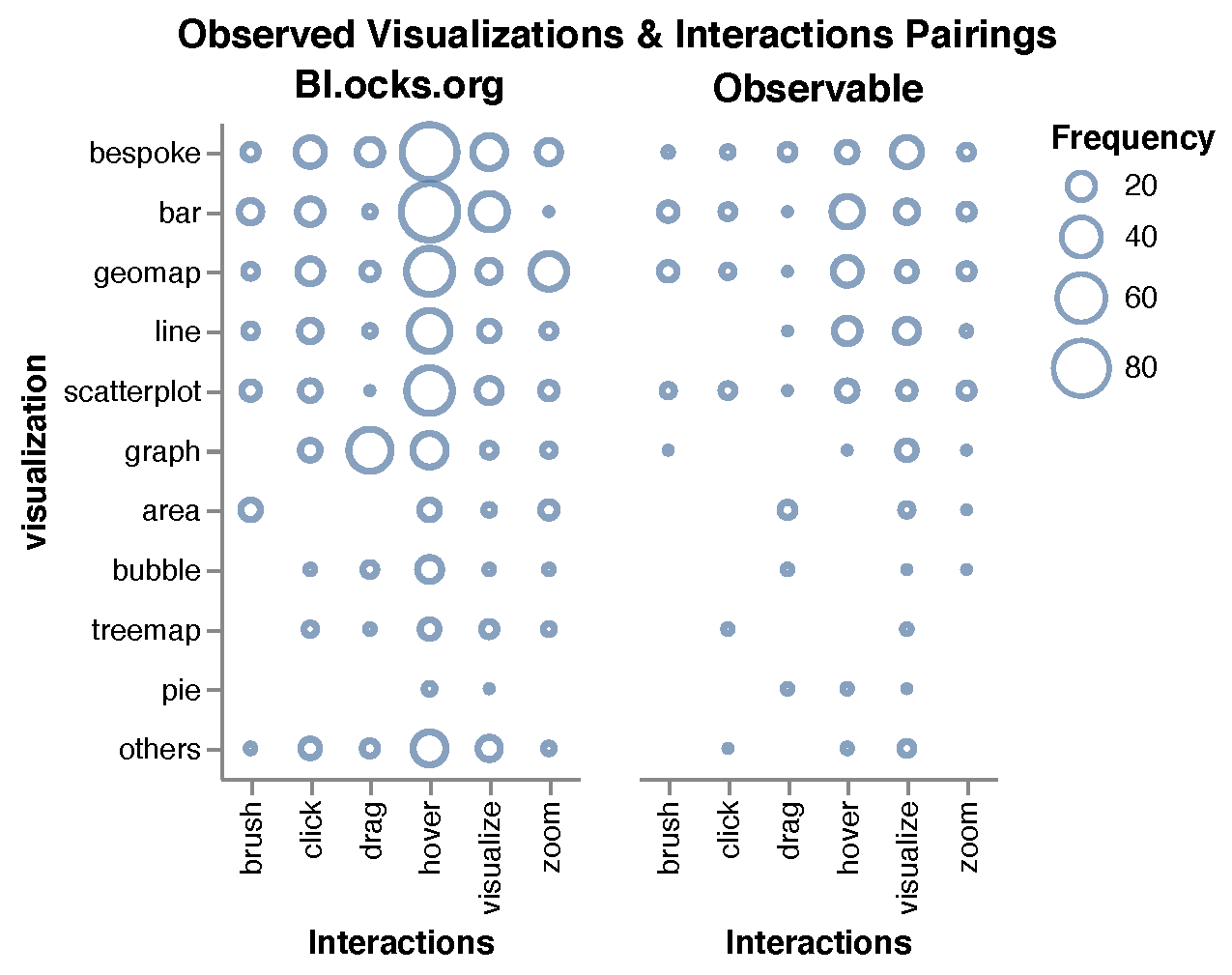} &
        \includegraphics[width=0.17\textwidth]{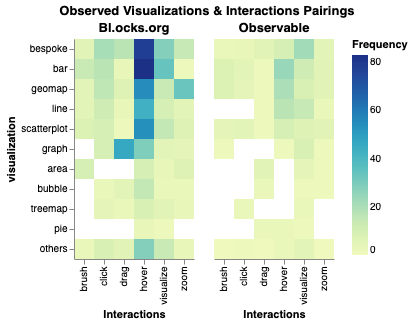} &
        \includegraphics[width=0.17\textwidth]{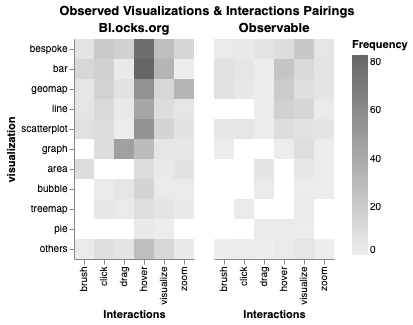} &
        \includegraphics[width=0.17\textwidth]{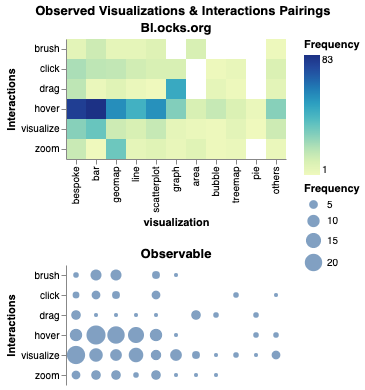} &
        \includegraphics[width=0.17\textwidth]{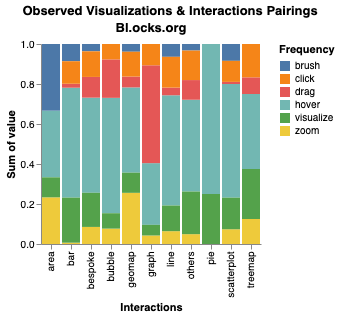} 
        \\
        \midrule

        P3 & 
        \includegraphics[width=0.17\textwidth]{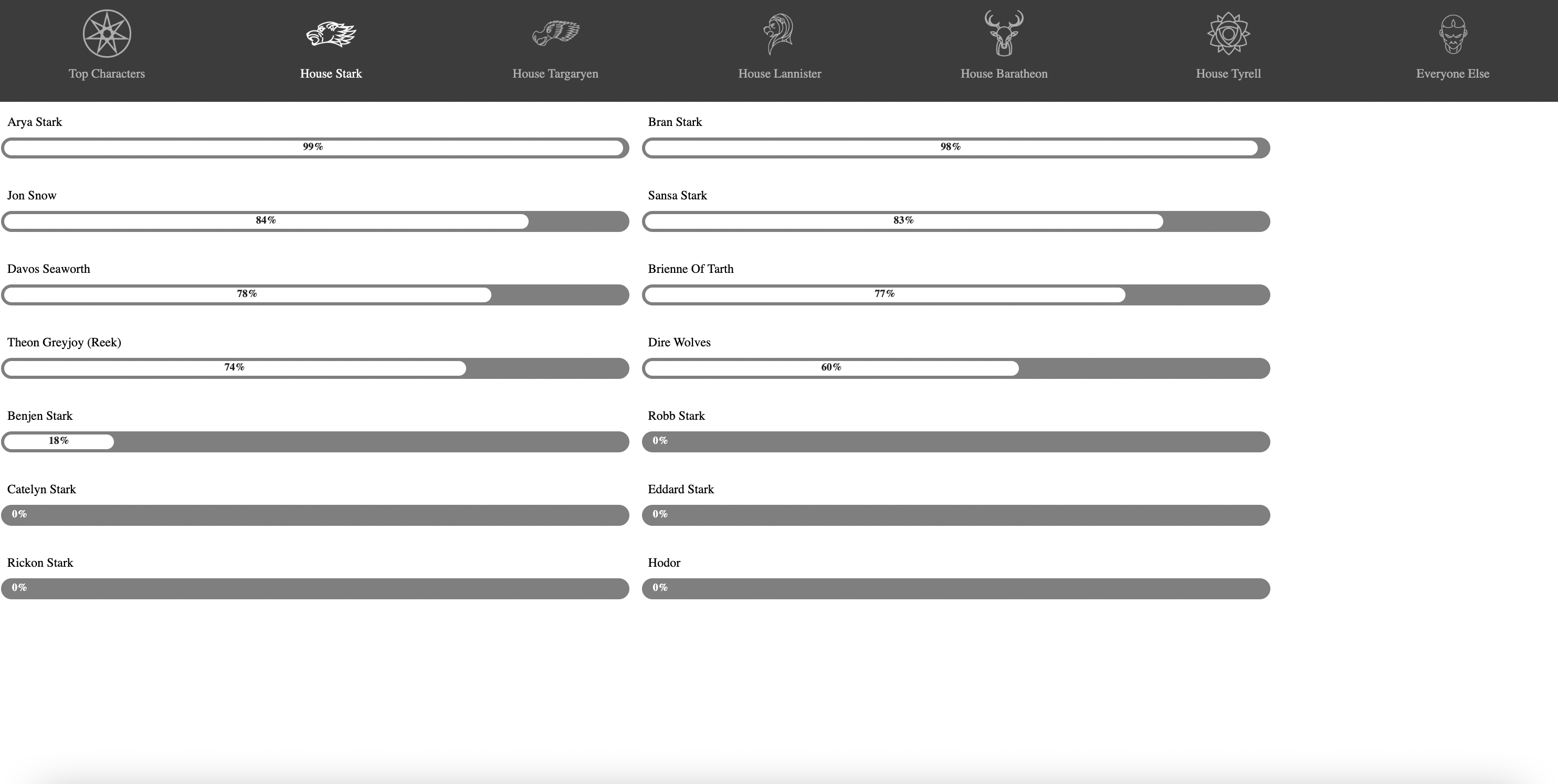} &
        \includegraphics[width=0.17\textwidth]{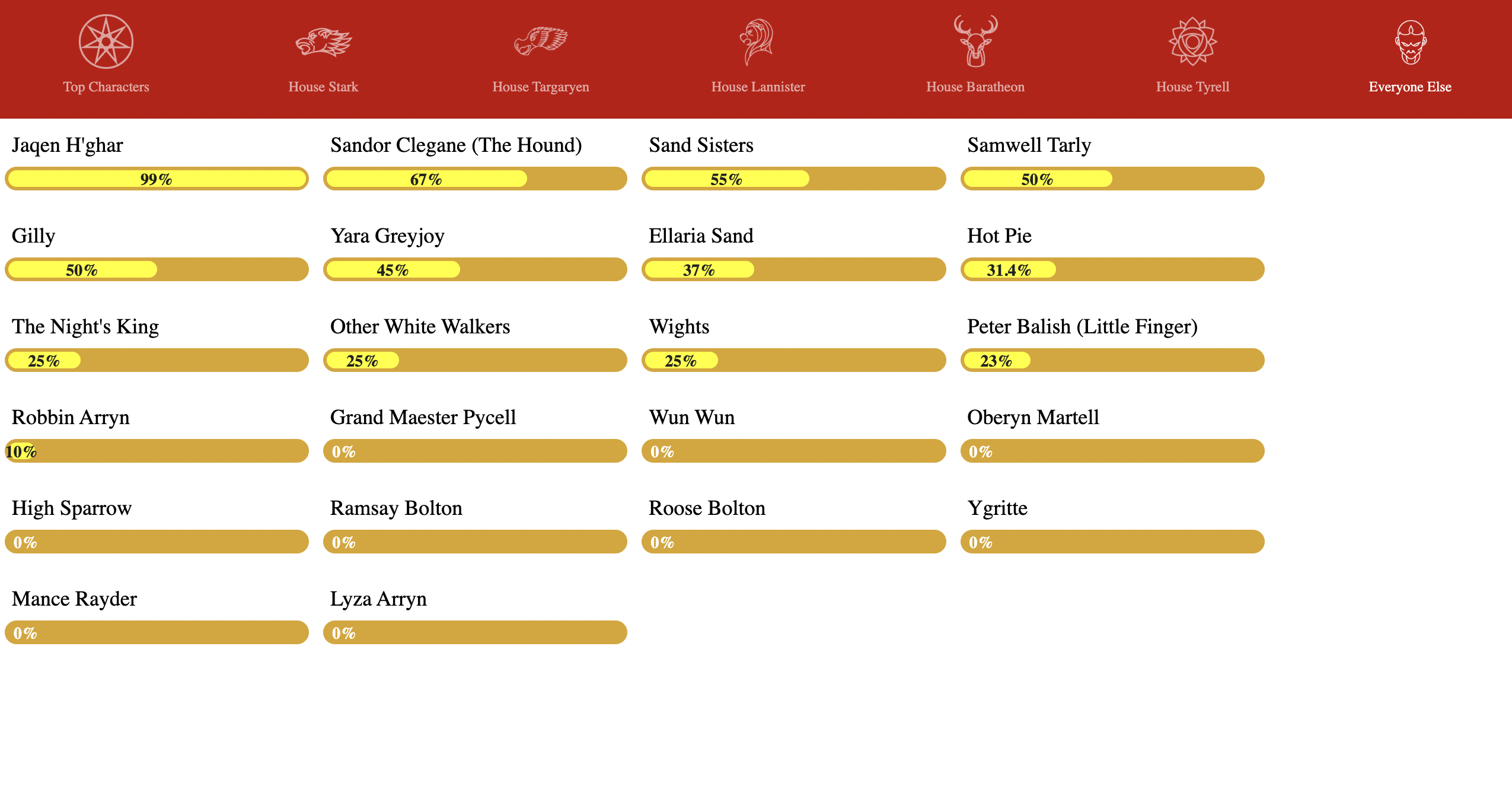} &
        \includegraphics[width=0.17\textwidth]{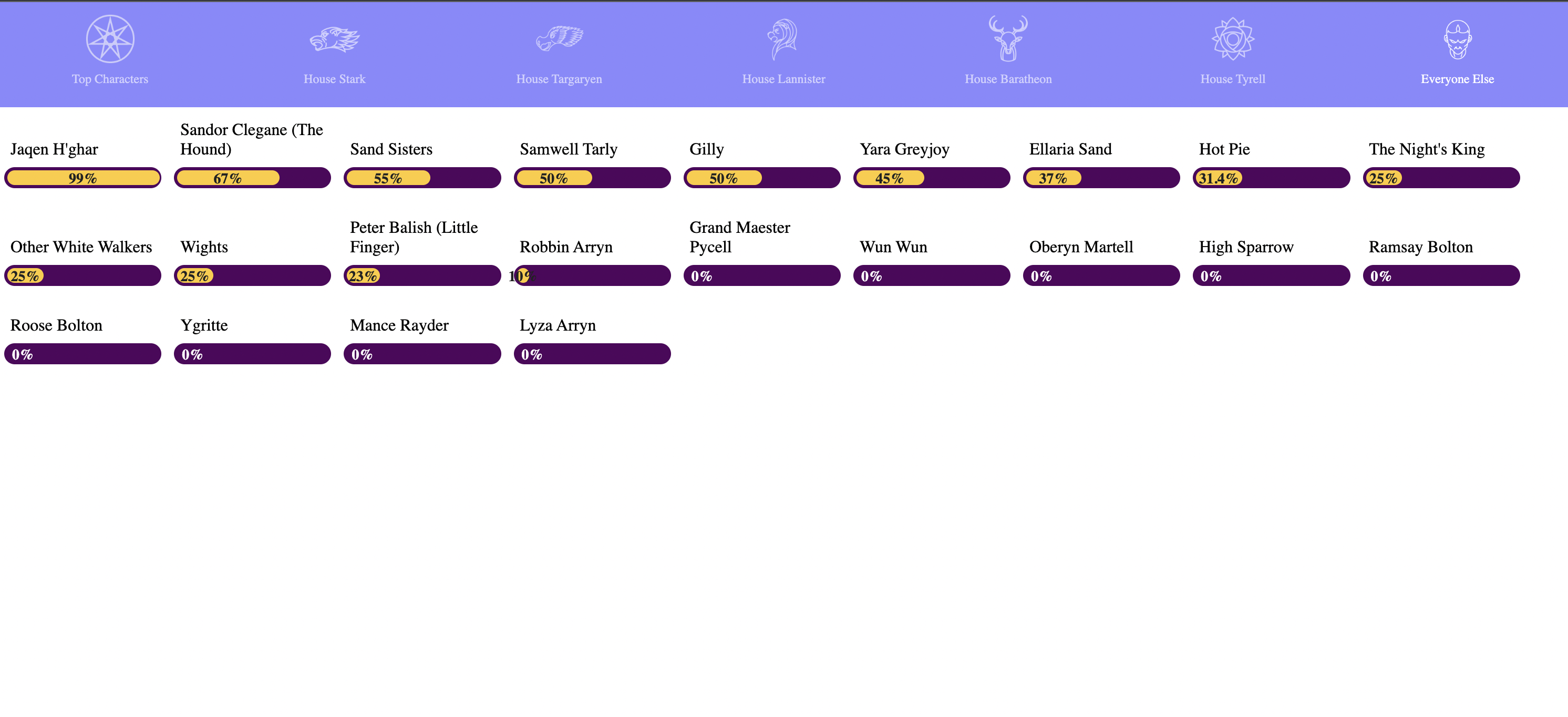} &
        \includegraphics[width=0.17\textwidth]{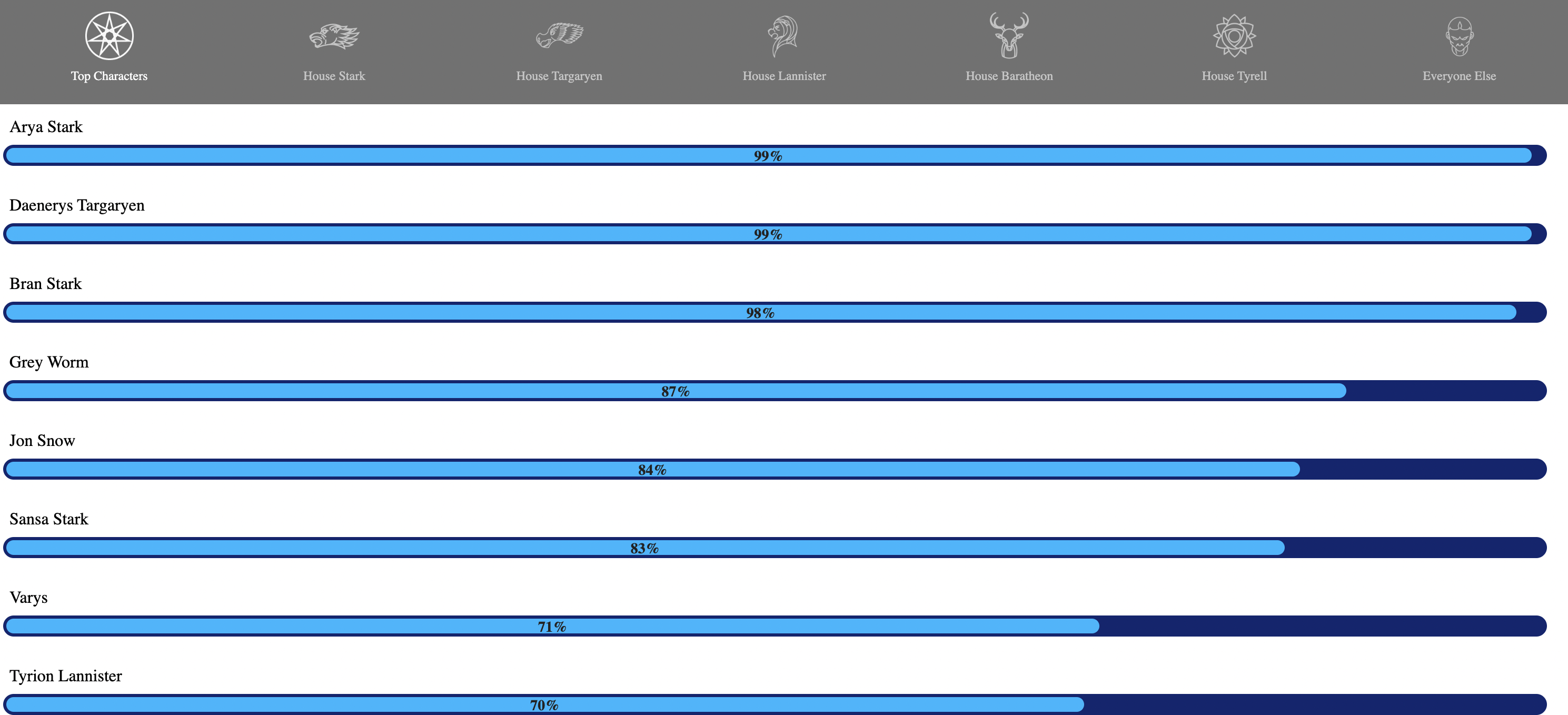} &
        \includegraphics[width=0.17\textwidth]{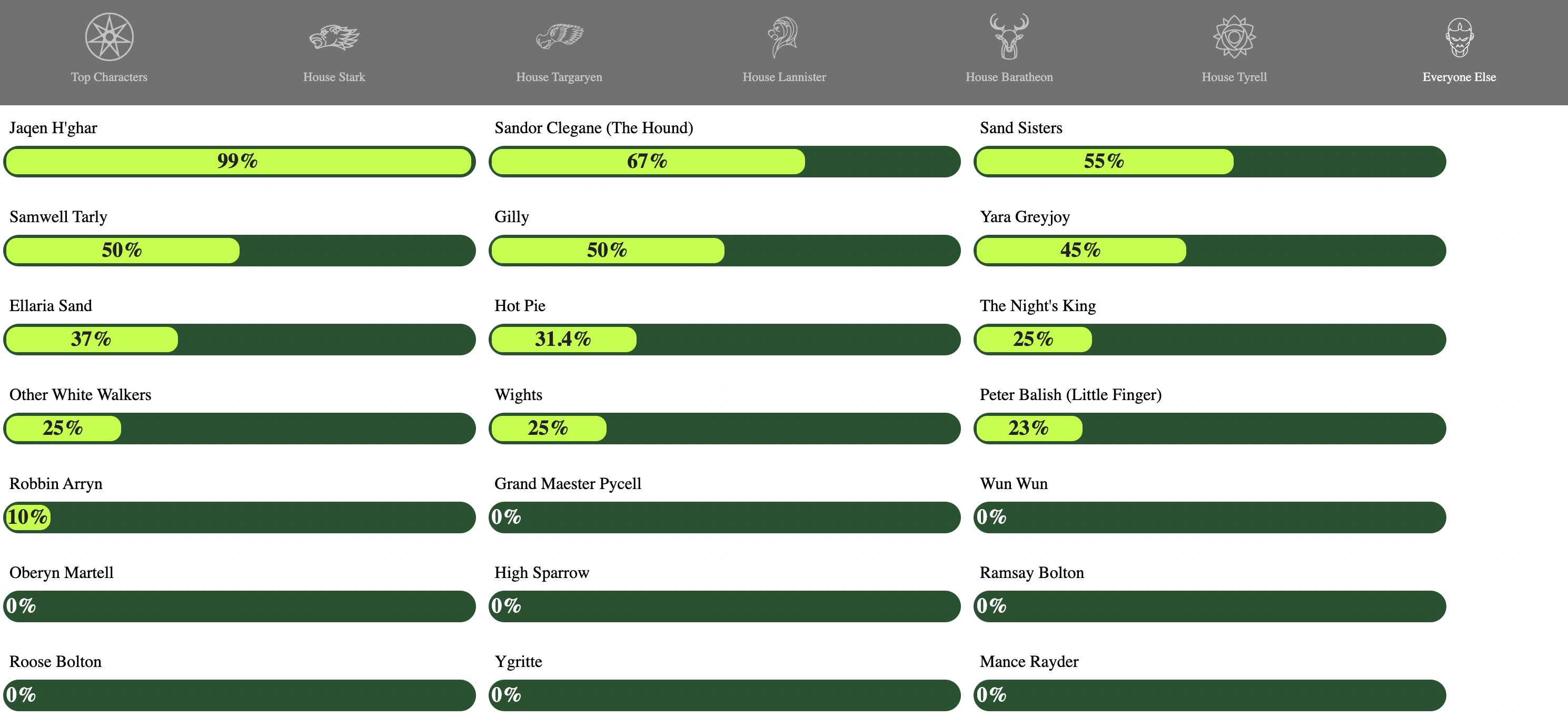}
        \\
        \midrule

        P4 & 
        \includegraphics[width=0.17\textwidth]{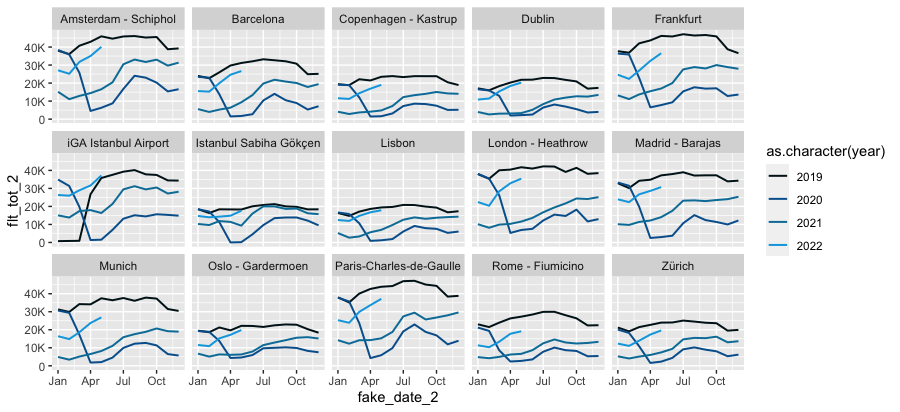} &
        \includegraphics[width=0.17\textwidth]{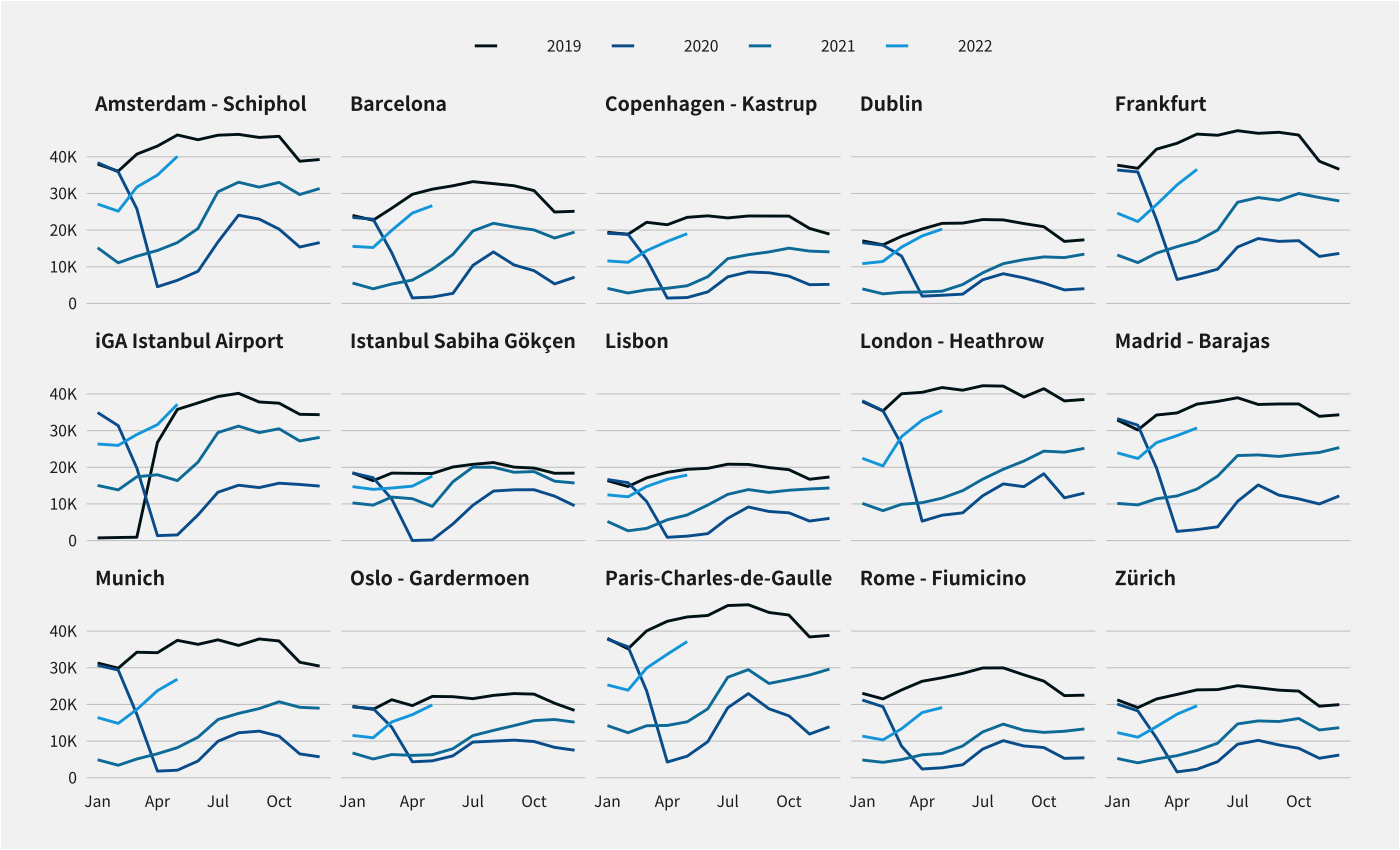} &
        \includegraphics[width=0.17\textwidth]{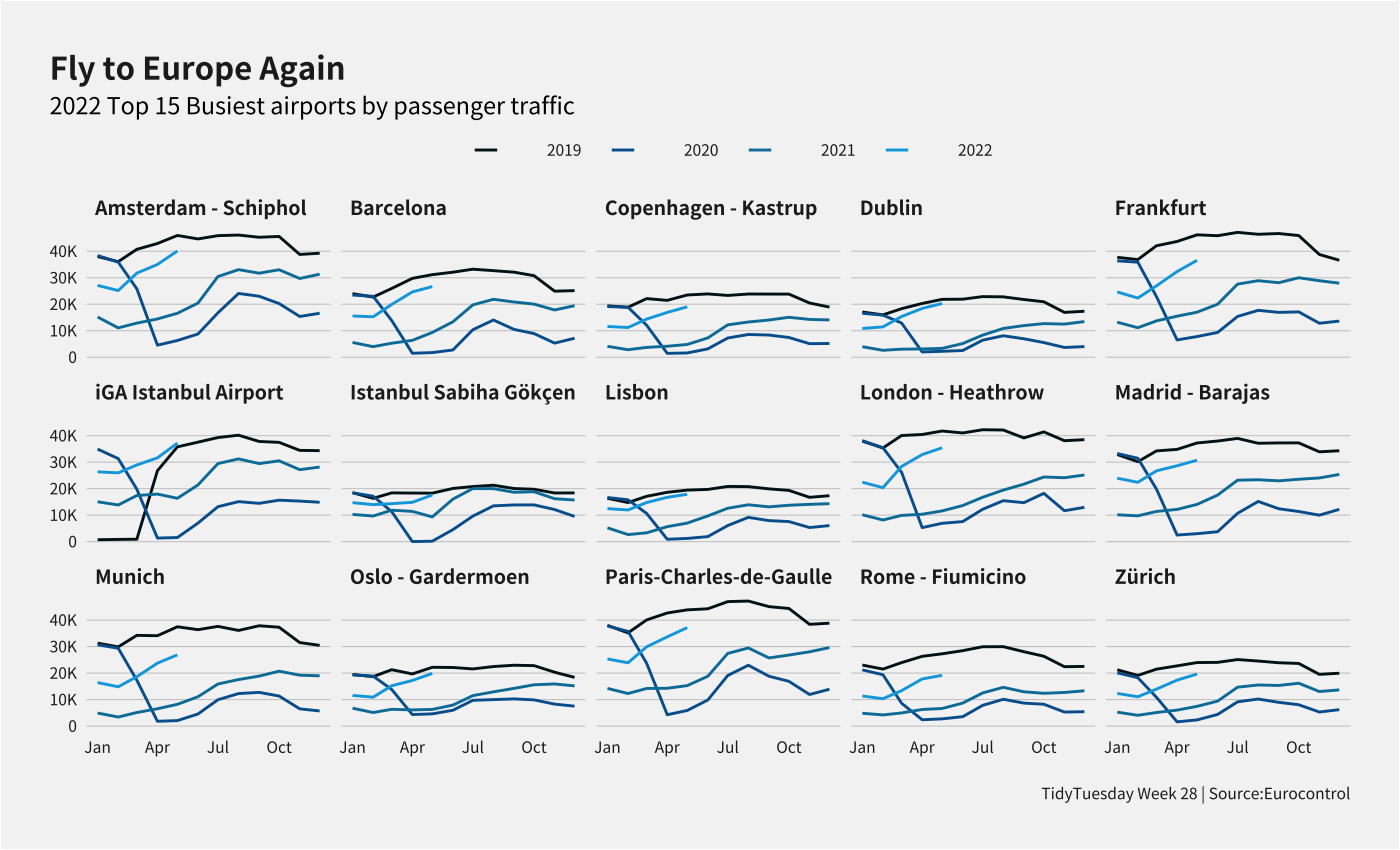} &
        \includegraphics[width=0.17\textwidth]{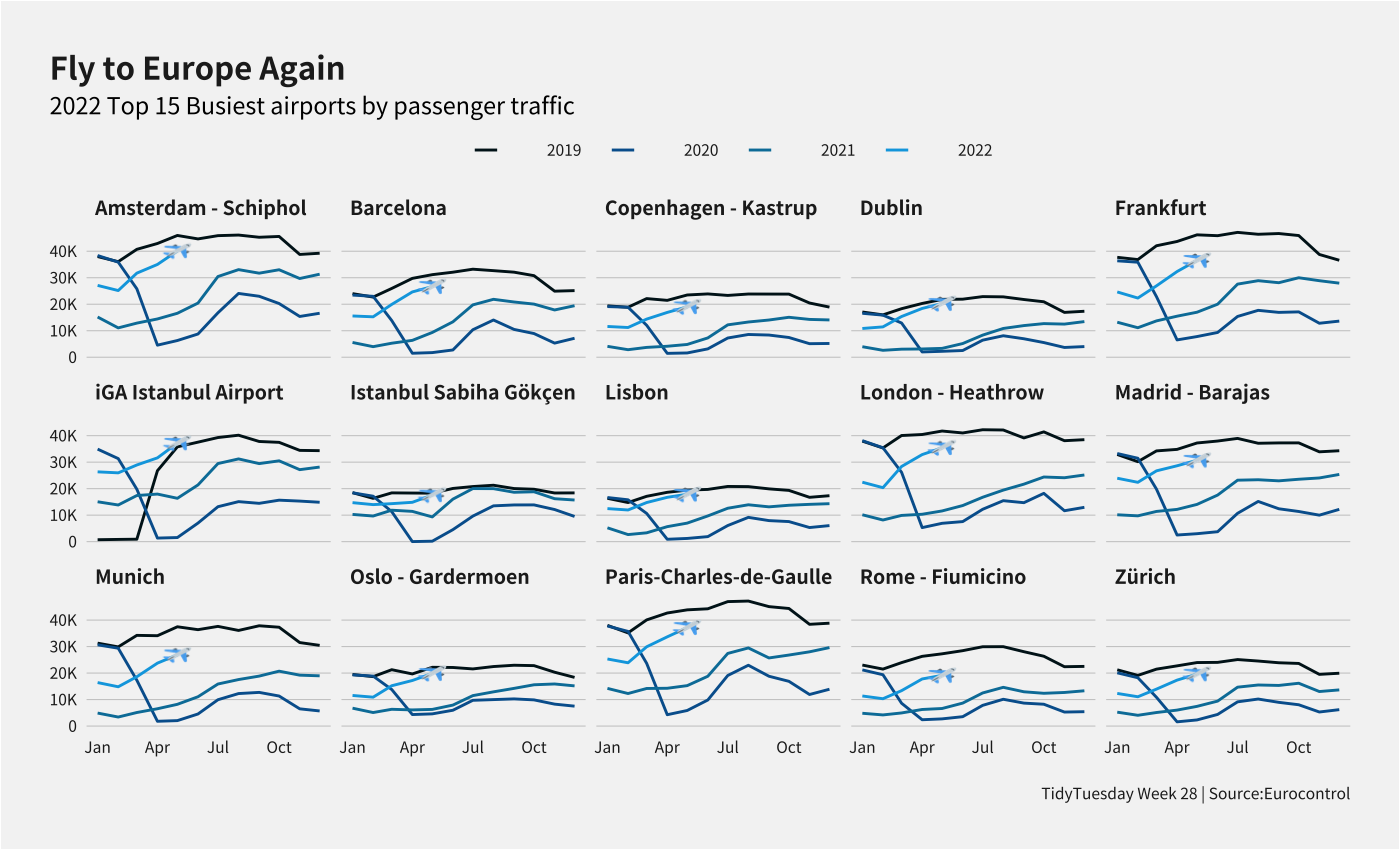} &
        \includegraphics[width=0.17\textwidth]{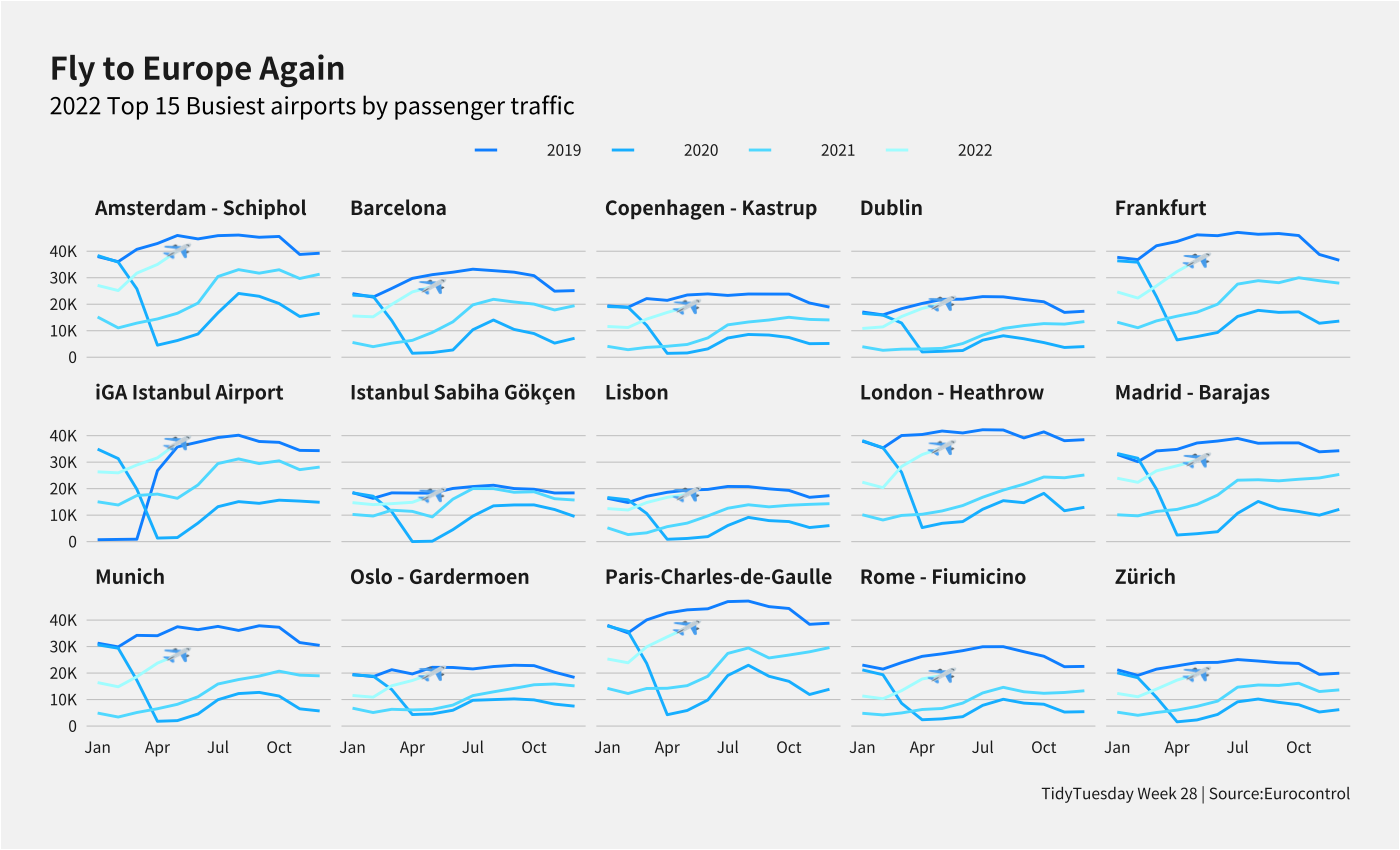}
        \\
        \bottomrule
        
    \end{tabular}

    \caption{\textbf{Overview of collected designs.}
    These images represent the five most significant versions (as identified by the participants) during the design process of creating a new visualization artifact from scratch using the Pat Design Lab and the Perceptual Pat Suite.
    All four participants from the study are represented (five were recruited, but one person abandoned the study after a week).}
    \Description{This figure shows how 4 participants' chart designs have evolved over 4 iterative processes using our tool, the Pat Design Lab. For each participant, we provide 5 updated versions.}
    \label{fig:all-versions}
\end{figure*}

\subsection{Evolution of Visualization Snapshots}

Fig.~\ref{fig:all-versions} shows how visualizations of the four participants evolved over 5 iterations of designs. 
Here we describe how each participant changed their design based entirely on the uploaded images.
In the following subsection we explore the rationale behind each change.

\paragraph{Participant 1.}

Originally, P1 created his chart using Matplotlib, a Python library for drawing charts. 
The first version of the chart was the initial version that Matplotlib creates, without any manual constraints involved. 
In the second and third versions, he enlarged the fonts of the letters within in the heatmap, the title, and axis labels. 
In the fourth version, he changed the heatmap from a bipolar to a unipolar color scheme. 
At last, he removed the potentially disturbing icons next to the labels, to keep from disrupting the readability of the chart. 

\paragraph{Participant 2.} 

P2 selected a research dataset classifying the interactions commonly used for specific chart types. 
Initially, she started off with a 2-axis grid scatterplot (version 1), where the horizontal axis represents the types of interactions (e.g., hover, brush, click, etc.), and the vertical axis represents the types of charts (e.g., bar, line, etc.).
The size of the blue-outlined circle in each point \revt{refers} to the frequency of charts. 
In the next version, she changed her chart into a heatmap (version 2), and then into a grayscale (version 3).
Then, in the fourth version, she created both a heatmap and a grid scatterplot.  
This time, she filled the circle with blue. 
Finally, in the fifth version, she created a stacked bar chart, where each column represents the proportion of different interactions and interactions are represented as different colors of the bar. 

\paragraph{Participant 3.}

P3 used a dataset on characters in a video game. 
She used D3 to create a visualization.
She started by creating a bar chart with 2 columns using grayscale (version 1). 
Afterwards, she tried, in an iterative manner, different designs by changing the number of columns, sizes of bars, and color (versions 2, 3, 4).
She finalized her visualization using green-colored bars on a gray background (version 5).

\paragraph{Participant 4.}

P4 wanted to visualize air passenger traffic data of 15 different cities.
She decided immediately to use line charts displayed in a 3 by 5 grid format. 
The initial version had gray charts with a white background and a legend at the right side of the visualization. 
In the second version, she moved the legend to the top and removed grids that are drawn in each chart. 
In the next version, she added a title in black, subtitle in dark gray, and source information about the chart a the right end of the chart.
In the fourth version, she added a small airplane embellishment to the image. 
In her final version, she performed finishing touches by changing the color of the chart to a more saturated blue.

\subsection{Evolution Rationale}

Recall that in the report we asked the participants to record what influenced them to refine their design. 
These descriptions, intentions, and reasons for the change are based on both these annotations as well as the interviews conducted with each participant.

\paragraph{Participant 1.}

P1 was conducting feature analysis of a deep neural network, and wanted to create a chart to share with his peers in the company.
He initially looked at the OCR to see if the \rev{component} could detect letters in the chart.
He consistently checked the size of the font in the chart, and left notes below the report on OCR. 
For example, he left a note below the report on OCR in the third version, as ``\textit{Increased the font size, after knowing that the label size of the chart is too small.}''
Then he noted that the pink-colored areas in the heatmap were not visible for people with deuteranopia, and that the icons next to the label were detected by the potentially distracting object.
This led him in the last version to change from pink and green color palettes to green palettes, and to removed small icons next to the labels. 

\paragraph{Participant 2.}

P2 wanted to construct a visualization that is capable of effectively contrasting the different types of interfaces used for each chart.
She was interested in searching for the chart that can best show the types of interactions per each type of chart. 
In the first version, she noted that she was not satisfied with using size as the only encoding type.
Then, in her second version, she changed her scatterplot into a heatmap. 
She noted two points: (1) increasing the brightness past a certain threshold can lead to information loss, and (2) the visual entropy was overly focused towards the labels. 
In her third version, she converted the chart into a grayscale version and found that the grayscale version did not have colorblind issues. 
In her fourth version, she added two different types of charts to the design.
She was also concerned about the results of the Scanner Deeply, noting that gazemaps focused on textual information such as labels and titles when she wanted people to focus more on the data representation. 
However, she noted that while this concerned her, she was unsure whether this was a good sign or not. 
Finally, in the last version, she replaced the two charts with one stacked bar chart with each bar representing the proportion of different interaction per each chart. 
Then she noted that (1) it did not make sense to have the data encoded as either points or a square in a heatmap, and that (2) the Scanner Deeply finally focused on the data representation more than the labels and legends. 

\paragraph{Participant 3.}

P3's goal in creating the visualization was to find the proper size of columns for the bar chart and the proper color.
In the initial version, she left notes on all filters except the color vision deficiency component.
In the second version, she changed the color using red palette, and created 4 columns to present bars from 2 columns. 
Here, she discussed the visual entropy and low-level salience results, noting that there is more visual entropy because of the icons in the navigation bar.
She was also concerned that the low-level salience concentrated on specific parts of the bars, but was unsure how to change the design. 
In the third and fourth versions, she tried to change the colors and the number of columns in the main view. 
In the third version, she noted that the new low saturated colors helped reduce user attention and instead focus on the view with bars. 
In the fourth version, she noted that this version looked like the worst design, considering the results form visual entropy, Scanner Deeply, and low-level salience.
In the final version, she used these findings to change to a highly saturated green to represent bars and to darken the navigation bar. 

\paragraph{Participant 4.} 

After receiving a report on his initial version, P4 left remarks about all \rev{components} in the report. 
In particular, she noted that the Scanner Deeply focuses only on the center of the chart, and that the grids installed in each chart raised the complexity of the charts. 
Consequently, she removed grids in each chart, and placed the legend at the top of the charts in the next version. 
Again, she left a note that the attention from the Scanner Deeply is still focused too much on the center of the visualization.
So in the third version, she added a title, subtitle, and caption to the chart to see if they help divert the attention to other parts of the chart. 
On the report from the third version, she made comments on OCR and Scanner Deeply.
Using OCR, she identified that title, subtitle and captions are noticed by the OCR \rev{component}, and also that the attention of the chart diverged after putting in title and subtitle. 
In the fourth version, she slightly modified the heights of each chart and added a small airplane icon in the chart.
Here, she mentioned that image with higher saturation improved the visibility of line charts.
At last, in the final version, she changed the colors of the lines with more saturated blue. 
From this, she observed that the increased saturation also helped diverge the attention of gaze heatmaps from the center to charts in various locations.

\subsection[External Assessment]{\rev{External Assessment}}

\rev{Table~\ref{tab:external-assessment} summarizes the quality change ratings of the external evaluators in the final phase (Phase IV) of our experiment.
As can be seen from the results, the evaluators mostly felt that all four participants had successfully improved on their visualization designs over the course of the design process.
Only P2's design received an average rating below the neutral; evaluators were split in their assessment of this design process.}

\rev{The evaluators also provided qualitative feedback on the changes they saw.
Even if the evaluators only saw the visualization designs (and not the Perceptual Pat reports), their written feedback often called out specific design improvements, such as more salient color scales, more legible fonts, and better visual layouts that seemed inspired by Perceptual Pat feedback.
For example, they noted several instances of increased legibility, larger font sizes for labels, improved color scales, better visual layout, and better visual encodings.
We were able to match the majority of these qualitative observations to the feedback that the individual participants received in the longitudinal user study.}

\begin{table*}[htb]
    \centering
    \begin{tabular}{l|cccc|c}
        \toprule
        \rowcolor{gray!10}        
        \textbf{\textsc{Participant}} & \textbf{v1 $\rightarrow$ v2} & \textbf{v2 $\rightarrow$ v3} & \textbf{v3 $\rightarrow$ v4} & \textbf{v4 $\rightarrow$ v5} & \textbf{\textsc{Overall}}\\
        \midrule
        P1 & 4.50 (0.71) & 3.50 (0.71) & 3.50 (2.12) & 4.50 (0.71) & 4.50 (0.71)\\
        \rowcolor{gray!10}
        P2 & 4.50 (0.71) & 2.00 (1.41) & 2.00 (0.00) & 3.00 (2.83) & 2.50 (2.12)\\
        P3 & 3.00 (0.00) & 3.00 (0.00) & 3.00 (0.00) & 4.00 (1.41) & 3.50 (0.71)\\
        \rowcolor{gray!10}
        P4 & 4.00 (0.00) & 4.33 (0.58) & 3.33 (0.58) & 2.33 (0.58) & 3.67 (0.58)\\
        \midrule
        All & 4.00 (0.71) & 3.21 (0.98) & 2.96 (0.67) & 3.46 (0.98) & 3.54 (0.82)\\
        \bottomrule
    \end{tabular}
    \caption{\rev{\textbf{External assessments.} Average 1-5 Likert scale ratings (1 = significant decline, 3 = neutral, 5 = significant improvement) by external evaluators assessing the quality of changes from one version to another (v$x \rightarrow$ v$y$) as well as overall from the initial to the final version. 
    Standard deviations are given within parentheses.}}
    \Description{ This table measures the improvement of the design per each update from participants.
    The assessment is done by external evaluators. All evaluators were senior visualization faculty or researchers with experience in teaching data visualization and/or designing their own visualizations. 
    To sum up, in Likert Scale, from 1 to 5, the overall improvement of P1's design is 4.33, P2's is 3.33, P3's is 3.33 and P4's is 3.67.
    }
    \label{tab:external-assessment}
\end{table*}

\subsection{Post-experimental Interview}

After the experiment, we asked the participants six questions with regards to their experience in Perceptual Pat.
Q1, Q4, Q5, and Q6 are shown in Table~\ref{tab:post_study}, and Q2 and Q3 are described in Table~\ref{tab:adv_disadv}.

\begin{table*}[t!]
    \centering
    \begin{tabular}{lm{8em}m{12em}m{10em}m{10em}m{10em}}
        \toprule
        \rowcolor{gray!10}
        \textbf{P\#} &
        \textbf{\textsc{Q1}:} How useful was the Pat Design Lab? &
        \textbf{\textsc{Q4}:} What tool(s) was the most useful? &
        \textbf{\textsc{Q5}:} What tool(s) was the least useful? &
        \textbf{\textsc{Q6}:} What tool would you most like to add?\\
        \midrule
        
        \textbf{\textsc{P1}} & very useful & OCR, color vision deficiency & chart junk, color suggestions & suggestion tool \\
        
        \rowcolor{gray!10}
        \textbf{\textsc{P2}} & somewhat useful & Scanner Deeply, color vision deficiency, visual entropy  & color suggestions (except contrast)  & suggestion tool, result interpreter\\

        \textbf{\textsc{P3}} & somewhat useful & Scanner Deeply, low-level salience & color suggestions & suggestion tool\\
        
        \rowcolor{gray!10}
        \textbf{\textsc{P4}} & somewhat useful & Scanner Deeply, OCR & color suggestions (except saturation) & suggestion tool, result interpreter\\
        
        \bottomrule
    \end{tabular}
    \caption{\textbf{Post-experimental interview with participants.}
    After each participant finished updating their versions of visualization, in the post-experimental interview we asked 6 following questions to understand how they get feedback and whether they are familiar with a tool that provides design feedback.
    In Q1, the answers are written in the order each participant mentioned.}
    \Description{This place has information about 4 post-experimental questions as well as answers from the participants. These questions are how useful the tool was, what tool was the most or least useful, and what tool would you most like to add. 
    About the first question, P1 answered as very useful, while the others responded as somewhat useful.
    About the second question, P1 answered as OCR and color vision deficiency. P2 answered as Scanner Deeply, color vision deficiency, and visual entropy. P3 answered as Scanner Deeply and low-level salience. P4 answered as OCR and the Scanner Deeply. 
    About the third question, most participants answered as color suggestions. 
    About the fourth question, they are all interested in adding a suggestion tool that could guide them for additional design, and additional information to interpret the filters.}
    \label{tab:post_study}
\end{table*}

\begin{table*}[t!]
    \centering
    \begin{tabular}{m{22em}m{22em}}
        \toprule
        \rowcolor{gray!10}
        \textbf{\textsc{Q2}: Advantages of \techname{}} &
        \textbf{\textsc{Q3}: Disadvantages of \techname{}} \\
        \midrule
        
        Detects problems within a chart (P1, P2, P3, P4) & No guidance for improvement (P1, P2, P3, P4) \\
    
        \rowcolor{gray!10}
        Convenience in asking for feedback (P1, P2, P3, P4) & Difficulty in interpreting results (P2, P3) \\
        
        Feedback on general audience reaction (P2, P3, P4) & Unclear target audience (P3)  \\
        
        \rowcolor{gray!10}
        Helps overcome biases in visualizations (P2, P3) & \\
        
        Acts as a checklist for improving visualizations (P1, P4) &  \\
        
        \bottomrule
    \end{tabular}
    \caption{\label{tab:adv_disadv} \textbf{Advantages and disadvantages of Perceptual Pat.}
    After the experiment, we asked participants about the advantages/disadvantages of a virtual human visual system providing visual feedback for visualization designers.
    The table shows a summary of the answers from 4 visualization designers that participated in our experiment.}
    \Description{This table presents the advantages and disadvantages of the Perceptual Pat that the participants mentioned in the post-experimental interview. 
    As for the advantages, all participants mentioned of the capability of detecting problems within a chart, and of the convenience in asking for feedback. Other advantages include providing feedback about how a general audience would react, helping overcome biases in visualizations, and acting as a checklist for improving visualizations.
    As for the disadvantage, all participants mentioned that it does not provide guidance for improvement. Other disadvantages include the difficulty in interpreting the result and unclear target audience. }
\end{table*}

When asked again about whether the tool was useful, all of them answered positively about the system, answering either `somewhat useful' or `very useful.'
Then we asked for the advantages as well as the disadvantages of Perceptual Pat. 
As for advantages, all participants agreed that the tool is capable of detecting problems within the chart.
We could identify from their notes in the report that \rev{many changes in the design were} based on these reports. 
Another advantage shared among all participants was that it is convenient to obtain feedback compared to when asking to get feedback from clients, peers, and supervisors.  
P2 and P4 liked that it provides feedback quickly without having to ask peers for feedback. 
P3 thought it could thus save time in getting feedback. 
Three participants liked the fact that tools such as Scanner Deeply provide feedback on how general audience would react to their visualizations. 
Also, both P2 and P3 thought that the tool help overcome their biases in visualizations.
For example, P3 originally thought that high levels of saturation would make the chart look bad aesthetically, but when changing the saturation value higher using Pat Design Lab, she felt that adding saturation to the visualization was a choice worth considering.
Finally, another advantage P1 and P4 pointed out was that the tools work as a checklist in evaluating the design, including factors that are otherwise easily overlooked (e.g., inclusive designs for the color blind).

In total, participants listed three disadvantages about the Perceptual Pat.
All participants were looking for not just detecting problems, but actual guidance on how to improve a chart.
While the filters provided can help detect problems in the design, it is up to the designers to find the right solution to the problem. 
The participants thought that a guidance could facilitate the process of their design processes.  
Also, P2 and P3 expressed concerns about the difficulty in interpreting the results. 
P2 said that while some feedback is easy to address, some of the filters are not easy to interpret. 
She said, ``\textit{some of the feedback, such as changing contrast or brightness is easy, but some aren't.
Consider the Scanner Deeply, for example.
It is sometimes difficult to judge whether the heatmap is a good or bad. }''
P3 also talked about the importance of interpreting the tool. 
To use \techname{} in industry, she said, ``\textit{one must be fully comfortable with the mechanisms behind how the tool works, so that when something is wrong, we know how to fix it.}''
Last but not least, P3 asserted that visualizations are designed differently by different target audience, and the tool will become more useful if it suggested a specific group of \rev{components} according to the target audience. 

In Q4 and Q5, when asked about the most/least useful \rev{component}, three participants thought that the Scanner Deeply was the most useful \rev{component}, being followed by OCR and color vision deficiency.
All participants thought that the color analysis and suggestions were the least useful. 
As for which new \rev{component} people would most like to add (Q6), all participants asked for a \rev{functionality} that could provide design suggestions.
In addition, P2 and P4 thought that a \rev{component} that can provide interpretation about the result could also help in improving a visualization design. 

%% -------------------------------------------------------------
%% DISCUSSION 
%% -------------------------------------------------------------
\section{Discussion}
\label{sec:discussion}

We here discuss the findings and implications from our user study, followed by the limitations of our work, as well as our plans for future work.

\subsection{Benefits of the Perceptual Pat Suite}
\label{subsec:benefits}

\paragraph{\rev{Shorten the feedback loop.}}

All the participants commented on the convenience of getting immediate feedback from Perceptual Pat.
\rev{In their regular workflows, participants reported their} common practice \rev{for receiving design feedback} was to ask their peers or supervisors, which \rev{is time-consuming and resource-intensive.}
\rev{In contrast, Pat gives them to receive feedback within a minute or two, thus saving time and resources for actual design activities.
This confirms our rationale for embarking on this project in the first place.}

\paragraph{\rev{Provide design guidance.}}

\rev{Results from our user study confirmed that the Pat Suite indeed helped our user study participants to improve their visualization design.
The quality assessments by the external evaluators show a positive improvement trajectory for each of the four participants even if there were inevitable setbacks during the process.}

\rev{The Perceptual Pat feedback was seen as direct and actionable.}
P3, for example, mentioned that \rev{the gaze map predicted by Scanner Deeply caused them to avoid users mostly focusing on the center of their visualization}.
To tackle this issue, they raised the saturation of the chart.
After changing the colors to those with high saturation, they re-ran the analysis and found that the Scanner Deeply \rev{predicted that the user's attention would be more equally distributed across the whole chart}.
\rev{Based on notes left in their reports, we also found that} that the majority of improvements---such as increasing font size, \rev{changing the color scale, or reorganizing the spatial layout}---\rev{were} made based on the feedback received from Pat.

\subsection{Limitations and Future Work}
\label{subsec:possibility}

\paragraph{Interpretability.}

Perceptual Pat's feedback system contains components that utilize deep neural networks for generating feedback, such as Scanner Deeply or chart junk detection.
The decision-making process of those models, while accurate, is opaque, and therefore developing \emph{interpretable} models and tools is an active area of research~\cite{selvaraju17gradcam, sdn19kaya, Dissect, DBLP:journals/corr/abs-2011-05429}.
\rev{P3 pointed out that for designers to be able to trust tools such as Perceptual Pat requires significant knowledge of how those models work.}
Despite our best effort in providing the technical details of Scanner Deeply, P3 wanted to know more about the factors that led to the model's feedback responses.
Said P3, ``\textit{In the end, it is the designer who decides whether to take advantage of this tool or judge whether what I am looking at is a real problem.
I must be able to know the extent I can trust when I see a tool so that I can properly judge.}''
Thus, \rev{we hope to improve the Perceptual Pat suite in the future by explaining the design feedback it generates.}

\paragraph{Recommendations.}

Perceptual Pat \rev{can be seen as an early-warning sanity checker for visualization designs (i.e., a perception-based visualization linter~\cite{chen22vizlinter}) similar to the unit-tests} that \rev{software engineers} use to ensure the functionality of the software \rev{they} develop.
However, \rev{just like unit-tests}, while Pat provides a list of potential issues for a specific visualization, it does not offer \rev{recommendations for how} to fix them.
P1 said ``\emph{I know my design has several issues, but Pat does not show suggestions for fixes.}"
P2 pointed out by saying that ``\emph{I love this approach, but it would have been better if Pat could tell me 'these are the issues and here is how you can fix them.'}''
\rev{Participants} also mentioned that, for \rev{less experienced designers}, giving concrete guidance would reduce the duration of the design process.
However, \rev{experienced visualization designers may not need such guidance, as they generally know how to refine their designs from Pat's feedback.}
Thus, \rev{a future research direction is to add design recommendations to the Perceptual Pat suite.}

\paragraph{\rev{Accuracy}}

\rev{As with any automated method, there is always a risk for erroneous results.
False positives can be particularly problematic, since it may give the designer the impression that everything is fine, and thus not investigate further.
A false negative, on the other hand, means that the designer will inspect the problem manually.
This should only be a problem if the number of false negatives flagged is excessive.}

\rev{As a case in point, applying an OCR filter to extract all of the textual labels in a chart may yield some recognition failures that a human would not make.
That is acceptable: the OCR filter is acting as a canary in a coal mine by indicating possible concerns that may not be a problem in practice, or which may only be a problem for people with diminished visual acuity.
For this reason, Perceptual Pat has been designed with the philosophy of favoring false negatives over false positives; it is better to highlight many potential problems than to run the risk of missing a real one.}

\paragraph{\rev{Bias}}

\rev{Similarly, automation may also give rise to bias.
Paradoxically, the fact that Perceptual Pat currently does not provide design recommendations may actually reduce this effect.
On the other hand, even the mere reporting of perceptual aspects cannot be said to be entirely unbiased; for example,highlighting one type of issue (visual attention, text legibility, or color saturation) and not other types (such as animation, use of visual channels, or Gestalt Law groupings; all examples of components not currently supported by Perceptual Pat) means that the former issues will tend to get highlighted and thus fixed, whereas the latter ones won't. 
In other words, Perceptual Pat clearly has blind spots (no pun) that will have an impact on visualization designs iterated using the tool. 
The only remedy---besides progressively adding new components to the suite to eliminate each of these gaps---is to at least make users aware of their existence.
Thus, in our future deployments of the tool, we will inform users not only which perceptual errors the suite checks for, but also which ones it does \textit{not}.}

\paragraph{Generality vs.\ Specificity.}

While all the participants in our study felt that \rev{Perceptual Pat's} feedback helped improve their designs, some pointed out that this ``one-size-fits-all'' strategy may not be \rev{effective for all  audiences}.
Said P3, ``\emph{it would be helpful if Pat would suggest not only which chart to use, but also which tool to use depending on the audience and domain}.''
\rev{In fact, we see the potential for taking this idea to its limit by providing customized ``flavors'' of Pat---or Steve, Susan, and Xian---that embody specific audiences, domains, and visual design philosophies.
We leave such integration and customization as future work, however.}

% \begin{comment}
% \techname{} is still only a rudimentary virtual human visual system, and its scope is on supporting iterative visual design. 
% Significant functionality is missing from large swathes of both vision science, perceptual psychology, and design practice. 
% This is to be expected for a new research tool that provides an entirely new affordance to the visualization design process.
% \techname{} is an open and extensible architecture, and the research team commits to releasing and maintaining the toolkit as Open Source upon acceptance.

% Scientific models of the human visual system are currently outside the scope of our work.
% Nevertheless, a lofty future goal will be to use the ideas proposed in this paper to build increasingly accurate models of the human visual system that can also be used for more scientific applications.

% Finally, the Pat Design Lab has several shortcomings that we will address in the future.
% One such drawback is the rudimentary comparison feature for seeing differences between two different design reports.
% We will explore visual difference tools for this purpose.
% In general, our future work revolves on further supporting designers through the use of human-centered AI~\cite{Shneiderman2022} such as computer vision and machine learning.
% We want to continue building supertools that can support the human in this process.
% \end{comment}

%% -------------------------------------------------------------
%% CONCLUSION AND FUTURE WORK
%% -------------------------------------------------------------
\section{Conclusion}

We have presented Perceptual Pat, a virtual human visual system designed for iterative visualization design.
Pat comprises a suite of image filters that are built using computer vision and related technologies to provide design feedback akin to what a designer may receive from both peer designers and supervisors as well as perceptual evaluation results collected from an empirical usability evaluation.
The Pat Design Lab provides a web-based interface to Perceptual Pat, enabling a user to track the evolution of their design work by repeatedly uploading new screenshots of their work, analyzing it with Pat, and viewing the results in a design report.
\revt{To assess the utility of Pat and his Design Lab, we conducted a longitudinal evaluation involving four professional visualization designers who used the tool to support the design of a new visualization artifact.}
Our findings showcase the utility of receiving quick turnaround feedback from the Pat suite in the design process.

%% -------------------------------------------------------------
%% ACKNOWLEDGMENTS
%% -------------------------------------------------------------

\begin{acks}
    We thank the anonymous reviewers for their feedback on this paper.
    We also thank Young-Ho Kim and Sunghyo Chung for their comments on our work.
    This work was partly supported by grant IIS-1901485 from the U.S.\ National Science Foundation.
    Any opinions, findings, and conclusions or recommendations expressed here are those of the authors and do not necessarily reflect the views of the funding agency.
\end{acks}

%% ---------------------------------------------------------------------
%% References
%% ---------------------------------------------------------------------
\bibliographystyle{ACM-Reference-Format}
\bibliography{perceptual-pat}


\begin{thebibliography}{83}

%%% the Web address.

\ifx \showCODEN    \undefined \def \showCODEN     #1{\unskip}     \fi
\ifx \showDOI      \undefined \def \showDOI       #1{#1}\fi
\ifx \showISBNx    \undefined \def \showISBNx     #1{\unskip}     \fi
\ifx \showISBNxiii \undefined \def \showISBNxiii  #1{\unskip}     \fi
\ifx \showISSN     \undefined \def \showISSN      #1{\unskip}     \fi
\ifx \showLCCN     \undefined \def \showLCCN      #1{\unskip}     \fi
\ifx \shownote     \undefined \def \shownote      #1{#1}          \fi
\ifx \showarticletitle \undefined \def \showarticletitle #1{#1}   \fi
\ifx \showURL      \undefined \def \showURL       {\relax}        \fi
% invisible to TeX
\providecommand\bibfield[2]{#2}
\providecommand\bibinfo[2]{#2}
\providecommand\natexlab[1]{#1}
\providecommand\showeprint[2][]{arXiv:#2}

\bibitem[Adebayo et~al\mbox{.}(2020)]%
        {DBLP:journals/corr/abs-2011-05429}
\bibfield{author}{\bibinfo{person}{Julius Adebayo}, \bibinfo{person}{Michael
  Muelly}, \bibinfo{person}{Ilaria Liccardi}, {and} \bibinfo{person}{Been
  Kim}.} \bibinfo{year}{2020}\natexlab{}.
\newblock \showarticletitle{Debugging Tests for Model Explanations}.
\newblock \bibinfo{journal}{\emph{CoRR}}  \bibinfo{volume}{abs/2011.05429}
  (\bibinfo{year}{2020}), \bibinfo{numpages}{64}~pages.
\newblock
\showeprint[arXiv]{2011.05429}
\urldef\tempurl%
\url{https://arxiv.org/abs/2011.05429}
\showURL{%
\tempurl}


\bibitem[Alvarez(2011)]%
        {Alvarez2011}
\bibfield{author}{\bibinfo{person}{George~A. Alvarez}.}
  \bibinfo{year}{2011}\natexlab{}.
\newblock \showarticletitle{Representing multiple objects as an ensemble
  enhances visual cognition}.
\newblock \bibinfo{journal}{\emph{Trends in Cognitive Sciences}}
  \bibinfo{volume}{15}, \bibinfo{number}{3} (\bibinfo{year}{2011}),
  \bibinfo{pages}{122--131}.
\newblock
\urldef\tempurl%
\url{https://doi.org/10.1016/j.tics.2011.01.003}
\showDOI{\tempurl}


\bibitem[Angerbauer et~al\mbox{.}(2022)]%
        {DBLP:conf/chi/AngerbauerRCOPM22}
\bibfield{author}{\bibinfo{person}{Katrin Angerbauer}, \bibinfo{person}{Nils
  Rodrigues}, \bibinfo{person}{Ren{\'{e}} Cutura}, \bibinfo{person}{Seyda
  {\"{O}}ney}, \bibinfo{person}{Nelusa Pathmanathan}, \bibinfo{person}{Cristina
  Morariu}, \bibinfo{person}{Daniel Weiskopf}, {and} \bibinfo{person}{Michael
  Sedlmair}.} \bibinfo{year}{2022}\natexlab{}.
\newblock \showarticletitle{Accessibility for Color Vision Deficiencies:
  Challenges and Findings of a Large Scale Study on Paper Figures}. In
  \bibinfo{booktitle}{\emph{Proceedings of the {ACM} Conference on Human
  Factors in Computing Systems}}. \bibinfo{publisher}{{ACM}},
  \bibinfo{address}{{New York, NY, USA}}, \bibinfo{pages}{134:1--134:23}.
\newblock
\urldef\tempurl%
\url{https://doi.org/10.1145/3491102.3502133}
\showDOI{\tempurl}


\bibitem[Badler et~al\mbox{.}(1993)]%
        {badler93simulating}
\bibfield{author}{\bibinfo{person}{Norman~I. Badler}, \bibinfo{person}{Cary~B.
  Phillips}, {and} \bibinfo{person}{Bonnie~Lynn Webber}.}
  \bibinfo{year}{1993}\natexlab{}.
\newblock \bibinfo{booktitle}{\emph{Simulating Humans: Computer Graphics,
  Animation, and Control}}.
\newblock \bibinfo{publisher}{Oxford University Press},
  \bibinfo{address}{Oxford, United Kingdom}.
\newblock
\urldef\tempurl%
\url{https://doi.org/10.1093/oso/9780195073591.001.0001}
\showDOI{\tempurl}


\bibitem[Bardzell(2011)]%
        {DBLP:journals/iwc/Bardzell11}
\bibfield{author}{\bibinfo{person}{Jeffrey Bardzell}.}
  \bibinfo{year}{2011}\natexlab{}.
\newblock \showarticletitle{Interaction criticism: An introduction to the
  practice}.
\newblock \bibinfo{journal}{\emph{Interacting with Computers}}
  \bibinfo{volume}{23}, \bibinfo{number}{6} (\bibinfo{year}{2011}),
  \bibinfo{pages}{604--621}.
\newblock
\urldef\tempurl%
\url{https://doi.org/10.1016/j.intcom.2011.07.001}
\showDOI{\tempurl}


\bibitem[Bardzell et~al\mbox{.}(2010)]%
        {DBLP:journals/interactions/BardzellBL10}
\bibfield{author}{\bibinfo{person}{Jeffrey Bardzell}, \bibinfo{person}{Jay~D.
  Bolter}, {and} \bibinfo{person}{Jonas L{\"{o}}wgren}.}
  \bibinfo{year}{2010}\natexlab{}.
\newblock \showarticletitle{Interaction criticism: three readings of an
  interaction design, and what they get us}.
\newblock \bibinfo{journal}{\emph{Interactions}} \bibinfo{volume}{17},
  \bibinfo{number}{2} (\bibinfo{year}{2010}), \bibinfo{pages}{32--37}.
\newblock
\urldef\tempurl%
\url{https://doi.org/10.1145/1699775.1699783}
\showDOI{\tempurl}


\bibitem[Bateman et~al\mbox{.}(2010)]%
        {DBLP:conf/chi/BatemanMGGMB10}
\bibfield{author}{\bibinfo{person}{Scott Bateman}, \bibinfo{person}{Regan~L.
  Mandryk}, \bibinfo{person}{Carl Gutwin}, \bibinfo{person}{Aaron Genest},
  \bibinfo{person}{David McDine}, {and} \bibinfo{person}{Christopher~A.
  Brooks}.} \bibinfo{year}{2010}\natexlab{}.
\newblock \showarticletitle{Useful junk?: the effects of visual embellishment
  on comprehension and memorability of charts}. In
  \bibinfo{booktitle}{\emph{Proceedings of the {ACM} Conference on Human
  Factors in Computing Systems}}. \bibinfo{publisher}{{ACM}},
  \bibinfo{address}{{New York, NY, USA}}, \bibinfo{pages}{2573--2582}.
\newblock
\urldef\tempurl%
\url{https://doi.org/10.1145/1753326.1753716}
\showDOI{\tempurl}


\bibitem[Borkin et~al\mbox{.}(2016)]%
        {borkin16recall}
\bibfield{author}{\bibinfo{person}{Michelle~A. Borkin}, \bibinfo{person}{Zoya
  Bylinskii}, \bibinfo{person}{Nam~Wook Kim}, \bibinfo{person}{Constance~May
  Bainbridge}, \bibinfo{person}{Chelsea~S. Yeh}, \bibinfo{person}{Daniel
  Borkin}, \bibinfo{person}{Hanspeter Pfister}, {and} \bibinfo{person}{Aude
  Oliva}.} \bibinfo{year}{2016}\natexlab{}.
\newblock \showarticletitle{Beyond Memorability: Visualization Recognition and
  Recall}.
\newblock \bibinfo{journal}{\emph{{{IEEE} Transactions on Visualization and
  Computer Graphics}}} \bibinfo{volume}{22}, \bibinfo{number}{1}
  (\bibinfo{year}{2016}), \bibinfo{pages}{519--528}.
\newblock
\urldef\tempurl%
\url{https://doi.org/10.1109/TVCG.2015.2467732}
\showDOI{\tempurl}


\bibitem[Bruce and Tsotsos(2005)]%
        {bruce04saliencyIM}
\bibfield{author}{\bibinfo{person}{Neil Bruce} {and} \bibinfo{person}{John
  Tsotsos}.} \bibinfo{year}{2005}\natexlab{}.
\newblock \showarticletitle{Saliency Based on Information Maximization}. In
  \bibinfo{booktitle}{\emph{Proceedings of the Advances in Neural Information
  Processing Systems}}, Vol.~\bibinfo{volume}{18}. \bibinfo{publisher}{Curran
  Associates, Inc.}, \bibinfo{address}{Red Hook, NY, USA},
  \bibinfo{numpages}{8}~pages.
\newblock


\bibitem[Bylinskii et~al\mbox{.}(2012)]%
        {mit-saliency-benchmark}
\bibfield{author}{\bibinfo{person}{Zoya Bylinskii}, \bibinfo{person}{Tilke
  Judd}, \bibinfo{person}{Ali Borji}, \bibinfo{person}{Laurent Itti},
  \bibinfo{person}{Fr{\'e}do Durand}, \bibinfo{person}{Aude Oliva}, {and}
  \bibinfo{person}{Antonio Torralba}.} \bibinfo{year}{2012}\natexlab{}.
\newblock \bibinfo{title}{{MIT} Saliency Benchmark}.
\newblock \bibinfo{howpublished}{http://saliency.mit.edu/}.
\newblock


\bibitem[Bylinskii et~al\mbox{.}(2019)]%
        {bylinskii19differentmodels}
\bibfield{author}{\bibinfo{person}{Zoya Bylinskii}, \bibinfo{person}{Tilke
  Judd}, \bibinfo{person}{Aude Oliva}, \bibinfo{person}{Antonio Torralba},
  {and} \bibinfo{person}{Frédo Durand}.} \bibinfo{year}{2019}\natexlab{}.
\newblock \showarticletitle{What Do Different Evaluation Metrics Tell Us About
  Saliency Models?}
\newblock \bibinfo{journal}{\emph{IEEE Transactions on Pattern Analysis and
  Machine Intelligence}} \bibinfo{volume}{41}, \bibinfo{number}{3}
  (\bibinfo{year}{2019}), \bibinfo{pages}{740--757}.
\newblock
\urldef\tempurl%
\url{https://doi.org/10.1109/TPAMI.2018.2815601}
\showDOI{\tempurl}


\bibitem[Cavanagh and Alvarez(2005)]%
        {Cavanagh2005}
\bibfield{author}{\bibinfo{person}{Patrick Cavanagh} {and}
  \bibinfo{person}{George~A. Alvarez}.} \bibinfo{year}{2005}\natexlab{}.
\newblock \showarticletitle{Tracking multiple targets with multifocal
  attention}.
\newblock \bibinfo{journal}{\emph{Trends in Cognitive Sciences}}
  \bibinfo{volume}{9}, \bibinfo{number}{7} (\bibinfo{year}{2005}),
  \bibinfo{pages}{349--354}.
\newblock
\urldef\tempurl%
\url{https://doi.org/10.1016/j.tics.2005.05.009}
\showDOI{\tempurl}


\bibitem[Chalbi et~al\mbox{.}(2020)]%
        {DBLP:journals/tvcg/ChalbiRPCREC20}
\bibfield{author}{\bibinfo{person}{Amira Chalbi}, \bibinfo{person}{Jacob
  Ritchie}, \bibinfo{person}{Deokgun Park}, \bibinfo{person}{Jungu Choi},
  \bibinfo{person}{Nicolas Roussel}, \bibinfo{person}{Niklas Elmqvist}, {and}
  \bibinfo{person}{Fanny Chevalier}.} \bibinfo{year}{2020}\natexlab{}.
\newblock \showarticletitle{Common Fate for Animated Transitions in
  Visualization}.
\newblock \bibinfo{journal}{\emph{{{IEEE} Transactions on Visualization and
  Computer Graphics}}} \bibinfo{volume}{26}, \bibinfo{number}{1}
  (\bibinfo{year}{2020}), \bibinfo{pages}{386--396}.
\newblock
\urldef\tempurl%
\url{https://doi.org/10.1109/TVCG.2019.2934288}
\showDOI{\tempurl}


\bibitem[Chen and J{\"{a}}nicke(2010)]%
        {DBLP:journals/tvcg/ChenJ10}
\bibfield{author}{\bibinfo{person}{Min Chen} {and} \bibinfo{person}{Heike
  J{\"{a}}nicke}.} \bibinfo{year}{2010}\natexlab{}.
\newblock \showarticletitle{An Information-theoretic Framework for
  Visualization}.
\newblock \bibinfo{journal}{\emph{{{IEEE} Transactions on Visualization and
  Computer Graphics}}} \bibinfo{volume}{16}, \bibinfo{number}{6}
  (\bibinfo{year}{2010}), \bibinfo{pages}{1206--1215}.
\newblock
\urldef\tempurl%
\url{https://doi.org/10.1109/TVCG.2010.132}
\showDOI{\tempurl}


\bibitem[Chen et~al\mbox{.}(2022)]%
        {chen22vizlinter}
\bibfield{author}{\bibinfo{person}{Qing Chen}, \bibinfo{person}{Fuling Sun},
  \bibinfo{person}{Xinyue Xu}, \bibinfo{person}{Zui Chen},
  \bibinfo{person}{Jiazhe Wang}, {and} \bibinfo{person}{Nan Cao}.}
  \bibinfo{year}{2022}\natexlab{}.
\newblock \showarticletitle{{VizLinter}: A Linter and Fixer Framework for Data
  Visualization}.
\newblock \bibinfo{journal}{\emph{{{IEEE} Transactions on Visualization and
  Computer Graphics}}} \bibinfo{volume}{28}, \bibinfo{number}{1}
  (\bibinfo{year}{2022}), \bibinfo{pages}{206--216}.
\newblock
\urldef\tempurl%
\url{https://doi.org/10.1109/TVCG.2021.3114804}
\showDOI{\tempurl}


\bibitem[Choi et~al\mbox{.}(2019)]%
        {DBLP:journals/cgf/ChoiJPCE19}
\bibfield{author}{\bibinfo{person}{Jinho Choi}, \bibinfo{person}{Sanghun Jung},
  \bibinfo{person}{Deok~Gun Park}, \bibinfo{person}{Jaegul Choo}, {and}
  \bibinfo{person}{Niklas Elmqvist}.} \bibinfo{year}{2019}\natexlab{}.
\newblock \showarticletitle{Visualizing for the Non-Visual: Enabling the
  Visually Impaired to Use Visualization}.
\newblock \bibinfo{journal}{\emph{Computer Graphics Forum}}
  \bibinfo{volume}{38}, \bibinfo{number}{3} (\bibinfo{year}{2019}),
  \bibinfo{pages}{249--260}.
\newblock
\urldef\tempurl%
\url{https://doi.org/10.1111/cgf.13686}
\showDOI{\tempurl}


\bibitem[Chundury et~al\mbox{.}(2022)]%
        {DBLP:journals/tvcg/ChunduryPRTLE22}
\bibfield{author}{\bibinfo{person}{Pramod Chundury}, \bibinfo{person}{Biswaksen
  Patnaik}, \bibinfo{person}{Yasmin Reyazuddin}, \bibinfo{person}{Christine
  Tang}, \bibinfo{person}{Jonathan Lazar}, {and} \bibinfo{person}{Niklas
  Elmqvist}.} \bibinfo{year}{2022}\natexlab{}.
\newblock \showarticletitle{Towards Understanding Sensory Substitution for
  Accessible Visualization: An Interview Study}.
\newblock \bibinfo{journal}{\emph{{{IEEE} Transactions on Visualization and
  Computer Graphics}}} \bibinfo{volume}{28}, \bibinfo{number}{1}
  (\bibinfo{year}{2022}), \bibinfo{pages}{1084--1094}.
\newblock
\urldef\tempurl%
\url{https://doi.org/10.1109/TVCG.2021.3114829}
\showDOI{\tempurl}


\bibitem[Cleveland(1985)]%
        {Cleveland1985}
\bibfield{author}{\bibinfo{person}{William~S. Cleveland}.}
  \bibinfo{year}{1985}\natexlab{}.
\newblock \bibinfo{booktitle}{\emph{The Elements of Graphing Data}}.
\newblock \bibinfo{publisher}{Wadsworth}, \bibinfo{address}{Monterey, CA, USA}.
\newblock


\bibitem[Cleveland and McGill(1984)]%
        {Cleveland1984}
\bibfield{author}{\bibinfo{person}{William~S. Cleveland} {and}
  \bibinfo{person}{Robert McGill}.} \bibinfo{year}{1984}\natexlab{}.
\newblock \showarticletitle{Graphical Perception: Theory, Experimentation and
  Application to the Development of Graphical Methods}.
\newblock \bibinfo{journal}{\emph{J. Amer. Statist. Assoc.}}
  \bibinfo{volume}{79}, \bibinfo{number}{387} (\bibinfo{year}{1984}),
  \bibinfo{pages}{531--554}.
\newblock
\urldef\tempurl%
\url{https://doi.org/10.2307/2288400}
\showDOI{\tempurl}


\bibitem[Croxton and Stein(1932)]%
        {Croxton1932}
\bibfield{author}{\bibinfo{person}{Frederick~E. Croxton} {and}
  \bibinfo{person}{Harold Stein}.} \bibinfo{year}{1932}\natexlab{}.
\newblock \showarticletitle{Graphic Comparisons by Bars, Squares, Circles, and
  Cubes}.
\newblock \bibinfo{journal}{\emph{J. Amer. Statist. Assoc.}}
  \bibinfo{volume}{27}, \bibinfo{number}{177} (\bibinfo{year}{1932}),
  \bibinfo{pages}{54--60}.
\newblock
\urldef\tempurl%
\url{https://doi.org/10.2307/2277880}
\showDOI{\tempurl}


\bibitem[Croxton and Stryker(1927)]%
        {Croxton1927}
\bibfield{author}{\bibinfo{person}{Frederick~E. Croxton} {and}
  \bibinfo{person}{Roy~E. Stryker}.} \bibinfo{year}{1927}\natexlab{}.
\newblock \showarticletitle{Bar charts versus circle diagrams}.
\newblock \bibinfo{journal}{\emph{J. Amer. Statist. Assoc.}}
  \bibinfo{volume}{22}, \bibinfo{number}{160} (\bibinfo{year}{1927}),
  \bibinfo{pages}{473--482}.
\newblock
\urldef\tempurl%
\url{https://doi.org/10.2307/2276829}
\showDOI{\tempurl}


\bibitem[Dragicevic et~al\mbox{.}(2011)]%
        {DBLP:conf/chi/DragicevicBJEF11}
\bibfield{author}{\bibinfo{person}{Pierre Dragicevic},
  \bibinfo{person}{Anastasia Bezerianos}, \bibinfo{person}{Waqas Javed},
  \bibinfo{person}{Niklas Elmqvist}, {and} \bibinfo{person}{Jean{-}Daniel
  Fekete}.} \bibinfo{year}{2011}\natexlab{}.
\newblock \showarticletitle{Temporal distortion for animated transitions}. In
  \bibinfo{booktitle}{\emph{Proceedings of the {ACM} Conference on Human
  Factors in Computing Systems}}. \bibinfo{publisher}{{ACM}},
  \bibinfo{address}{{New York, NY, USA}}, \bibinfo{pages}{2009--2018}.
\newblock
\urldef\tempurl%
\url{https://doi.org/10.1145/1978942.1979233}
\showDOI{\tempurl}


\bibitem[Eells(1926)]%
        {Eells1926}
\bibfield{author}{\bibinfo{person}{Walter~C. Eells}.}
  \bibinfo{year}{1926}\natexlab{}.
\newblock \showarticletitle{The relative merits of circles and bars for
  representing component parts}.
\newblock \bibinfo{journal}{\emph{J. Amer. Statist. Assoc.}}
  \bibinfo{volume}{21}, \bibinfo{number}{154} (\bibinfo{year}{1926}),
  \bibinfo{pages}{119--132}.
\newblock
\urldef\tempurl%
\url{https://doi.org/10.2307/2277140}
\showDOI{\tempurl}


\bibitem[Elavsky et~al\mbox{.}(2022)]%
        {DBLP:journals/cgf/ElavskyBM22}
\bibfield{author}{\bibinfo{person}{Frank Elavsky}, \bibinfo{person}{Cynthia~L.
  Bennett}, {and} \bibinfo{person}{Dominik Moritz}.}
  \bibinfo{year}{2022}\natexlab{}.
\newblock \showarticletitle{How accessible is my visualization? Evaluating
  visualization accessibility with Chartability}.
\newblock \bibinfo{journal}{\emph{Computer Graphics Forum}}
  \bibinfo{volume}{41}, \bibinfo{number}{3} (\bibinfo{year}{2022}),
  \bibinfo{pages}{57--70}.
\newblock
\urldef\tempurl%
\url{https://doi.org/10.1111/cgf.14522}
\showDOI{\tempurl}


\bibitem[Elrefaei(2018)]%
        {elrefaei18colorblind}
\bibfield{author}{\bibinfo{person}{Lamiaa~A. Elrefaei}.}
  \bibinfo{year}{2018}\natexlab{}.
\newblock \showarticletitle{Smartphone Based Image Color Correction for Color
  Blindness}.
\newblock \bibinfo{journal}{\emph{International Journal of Interactive Mobile
  Technologies}} \bibinfo{volume}{12}, \bibinfo{number}{3}
  (\bibinfo{year}{2018}), \bibinfo{pages}{104–119}.
\newblock
\urldef\tempurl%
\url{https://doi.org/10.3991/ijim.v12i3.8160}
\showDOI{\tempurl}


\bibitem[Fan et~al\mbox{.}(2022)]%
        {fan22linechartdeception}
\bibfield{author}{\bibinfo{person}{Arlen Fan}, \bibinfo{person}{Yuxin Ma},
  \bibinfo{person}{Michelle Mancenido}, {and} \bibinfo{person}{Ross
  Maciejewski}.} \bibinfo{year}{2022}\natexlab{}.
\newblock \showarticletitle{Annotating Line Charts for Addressing Deception}.
  In \bibinfo{booktitle}{\emph{Proceedings of the {ACM} Conference on Human
  Factors in Computing Systems}}. \bibinfo{publisher}{{ACM}},
  \bibinfo{address}{{New York, NY, USA}}, \bibinfo{pages}{80:1--80:12}.
\newblock
\urldef\tempurl%
\url{https://doi.org/10.1145/3491102.3502138}
\showDOI{\tempurl}


\bibitem[Franconeri et~al\mbox{.}(2021)]%
        {franconeri21visualdatacommunication}
\bibfield{author}{\bibinfo{person}{Steven~L. Franconeri},
  \bibinfo{person}{Lace~M. Padilla}, \bibinfo{person}{Priti Shah},
  \bibinfo{person}{Jeffrey~M. Zacks}, {and} \bibinfo{person}{Jessica Hullman}.}
  \bibinfo{year}{2021}\natexlab{}.
\newblock \showarticletitle{The Science of Visual Data Communication: What
  Works}.
\newblock \bibinfo{journal}{\emph{Psychological Science in the Public
  Interest}} \bibinfo{volume}{22}, \bibinfo{number}{3} (\bibinfo{year}{2021}),
  \bibinfo{pages}{110--161}.
\newblock
\urldef\tempurl%
\url{https://doi.org/10.1177/15291006211051956}
\showDOI{\tempurl}


\bibitem[Ghandeharioun et~al\mbox{.}(2021)]%
        {Dissect}
\bibfield{author}{\bibinfo{person}{Asma Ghandeharioun}, \bibinfo{person}{Been
  Kim}, \bibinfo{person}{Chun{-}Liang Li}, \bibinfo{person}{Brendan Jou},
  \bibinfo{person}{Brian Eoff}, {and} \bibinfo{person}{Rosalind~W. Picard}.}
  \bibinfo{year}{2021}\natexlab{}.
\newblock \bibinfo{title}{{DISSECT:} Disentangled Simultaneous Explanations via
  Concept Traversals}.
\newblock
\newblock
\urldef\tempurl%
\url{https://arxiv.org/abs/2105.15164}
\showURL{%
\tempurl}


\bibitem[Haehn et~al\mbox{.}(2019)]%
        {haehn19understandingpercwithcnns}
\bibfield{author}{\bibinfo{person}{Daniel Haehn}, \bibinfo{person}{James
  Tompkin}, {and} \bibinfo{person}{Hanspeter Pfister}.}
  \bibinfo{year}{2019}\natexlab{}.
\newblock \showarticletitle{Evaluating ‘Graphical Perception’ with CNNs}.
\newblock \bibinfo{journal}{\emph{IEEE Transactions on Visualization and
  Computer Graphics}} \bibinfo{volume}{25}, \bibinfo{number}{1}
  (\bibinfo{year}{2019}), \bibinfo{pages}{641--650}.
\newblock
\urldef\tempurl%
\url{https://doi.org/10.1109/TVCG.2018.2865138}
\showDOI{\tempurl}


\bibitem[Harel et~al\mbox{.}(2006)]%
        {harel06gbvs}
\bibfield{author}{\bibinfo{person}{Jonathan Harel}, \bibinfo{person}{Christof
  Koch}, {and} \bibinfo{person}{Pietro Perona}.}
  \bibinfo{year}{2006}\natexlab{}.
\newblock \showarticletitle{Graph-Based Visual Saliency}. In
  \bibinfo{booktitle}{\emph{Proceedings of the Advances in Neural Information
  Processing Systems}}, Vol.~\bibinfo{volume}{19}. \bibinfo{publisher}{Curran
  Associates, Inc.}, \bibinfo{address}{Red Hook, NY, USA},
  \bibinfo{numpages}{8}~pages.
\newblock
\urldef\tempurl%
\url{https://papers.nips.cc/paper/2006/file/4db0f8b0fc895da263fd77fc8aecabe4-Paper.pdf}
\showURL{%
\tempurl}


\bibitem[Healey(1996)]%
        {DBLP:conf/visualization/Healey96}
\bibfield{author}{\bibinfo{person}{Christopher~G. Healey}.}
  \bibinfo{year}{1996}\natexlab{}.
\newblock \showarticletitle{Choosing Effective Colours for Data Visualization}.
  In \bibinfo{booktitle}{\emph{Proceedings of the {IEEE} Visualization
  Conference}}. \bibinfo{publisher}{{IEEE}}, \bibinfo{address}{Piscataway, NJ,
  USA}, \bibinfo{pages}{263--270}.
\newblock
\urldef\tempurl%
\url{https://doi.org/10.1109/VISUAL.1996.568118}
\showDOI{\tempurl}


\bibitem[Hering(1964)]%
        {Hering1964}
\bibfield{author}{\bibinfo{person}{Ewald Hering}.}
  \bibinfo{year}{1964}\natexlab{}.
\newblock \bibinfo{booktitle}{\emph{Outlines of a Theory of the Light Sense}}.
\newblock \bibinfo{publisher}{Harvard University Press},
  \bibinfo{address}{Cambridge, MA, USA}.
\newblock


\bibitem[Hopkins et~al\mbox{.}(2020)]%
        {hopkins20visualint}
\bibfield{author}{\bibinfo{person}{Aspen~K. Hopkins}, \bibinfo{person}{Michael
  Correll}, {and} \bibinfo{person}{Arvind Satyanarayan}.}
  \bibinfo{year}{2020}\natexlab{}.
\newblock \showarticletitle{{VisuaLint}: Sketchy In Situ Annotations of Chart
  Construction Errors}.
\newblock \bibinfo{journal}{\emph{Computer Graphics Forum}}
  \bibinfo{volume}{39}, \bibinfo{number}{3} (\bibinfo{year}{2020}),
  \bibinfo{pages}{219--228}.
\newblock
\urldef\tempurl%
\url{https://doi.org/10.1111/cgf.13975}
\showDOI{\tempurl}


\bibitem[Hu et~al\mbox{.}(2019)]%
        {hu19vizml}
\bibfield{author}{\bibinfo{person}{Kevin Hu}, \bibinfo{person}{Michiel~A.
  Bakker}, \bibinfo{person}{Stephen Li}, \bibinfo{person}{Tim Kraska}, {and}
  \bibinfo{person}{C\'{e}sar Hidalgo}.} \bibinfo{year}{2019}\natexlab{}.
\newblock \showarticletitle{{VizML}: A Machine Learning Approach to
  Visualization Recommendation}. In \bibinfo{booktitle}{\emph{Proceedings of
  the {ACM} Conference on Human Factors in Computing Systems}}.
  \bibinfo{publisher}{{ACM}}, \bibinfo{address}{{New York, NY, USA}},
  \bibinfo{pages}{128:1–128:12}.
\newblock
\urldef\tempurl%
\url{https://doi.org/10.1145/3290605.3300358}
\showDOI{\tempurl}


\bibitem[Hurvich and Jameson(1957)]%
        {Hurvich1957}
\bibfield{author}{\bibinfo{person}{Leo~M. Hurvich} {and}
  \bibinfo{person}{Dorothea Jameson}.} \bibinfo{year}{1957}\natexlab{}.
\newblock \showarticletitle{An opponent-process theory of color vision}.
\newblock \bibinfo{journal}{\emph{Psychological Review}} \bibinfo{volume}{64},
  \bibinfo{number}{6, Pt.1} (\bibinfo{year}{1957}), \bibinfo{pages}{384--404}.
\newblock
\urldef\tempurl%
\url{https://doi.org/10.1037/h0041403}
\showDOI{\tempurl}


\bibitem[Isenberg et~al\mbox{.}(2013)]%
        {DBLP:journals/tvcg/Isenberg2013}
\bibfield{author}{\bibinfo{person}{Tobias Isenberg}, \bibinfo{person}{Petra
  Isenberg}, \bibinfo{person}{Jian Chen}, \bibinfo{person}{Michael Sedlmair},
  {and} \bibinfo{person}{Torsten M{\"{o}}ller}.}
  \bibinfo{year}{2013}\natexlab{}.
\newblock \showarticletitle{A Systematic Review on the Practice of Evaluating
  Visualization}.
\newblock \bibinfo{journal}{\emph{{{IEEE} Transactions on Visualization and
  Computer Graphics}}} \bibinfo{volume}{19}, \bibinfo{number}{12}
  (\bibinfo{year}{2013}), \bibinfo{pages}{2818--2827}.
\newblock
\urldef\tempurl%
\url{https://doi.org/10.1109/TVCG.2013.126}
\showDOI{\tempurl}


\bibitem[{Itti} et~al\mbox{.}(1998)]%
        {itti98saliency}
\bibfield{author}{\bibinfo{person}{Laurent {Itti}}, \bibinfo{person}{Christof
  {Koch}}, {and} \bibinfo{person}{Ernst {Niebur}}.}
  \bibinfo{year}{1998}\natexlab{}.
\newblock \showarticletitle{A model of saliency-based visual attention for
  rapid scene analysis}.
\newblock \bibinfo{journal}{\emph{IEEE Transactions on Pattern Analysis and
  Machine Intelligence}} \bibinfo{volume}{20}, \bibinfo{number}{11}
  (\bibinfo{year}{1998}), \bibinfo{pages}{1254--1259}.
\newblock
\urldef\tempurl%
\url{https://doi.org/10.1109/34.730558}
\showDOI{\tempurl}


\bibitem[{Jiang} et~al\mbox{.}(2015)]%
        {jiang15salicon}
\bibfield{author}{\bibinfo{person}{M. {Jiang}}, \bibinfo{person}{S. {Huang}},
  \bibinfo{person}{J. {Duan}}, {and} \bibinfo{person}{Q. {Zhao}}.}
  \bibinfo{year}{2015}\natexlab{}.
\newblock \showarticletitle{{SALICON}: Saliency in Context}. In
  \bibinfo{booktitle}{\emph{Proceedings of the IEEE Conference on Computer
  Vision and Pattern Recognition}}. \bibinfo{publisher}{{IEEE}},
  \bibinfo{address}{Piscataway, NJ, USA}, \bibinfo{pages}{1072--1080}.
\newblock
\urldef\tempurl%
\url{https://doi.org/10.1109/CVPR.2015.7298710}
\showDOI{\tempurl}


\bibitem[Kay(2007)]%
        {kay07tesseract}
\bibfield{author}{\bibinfo{person}{Anthony Kay}.}
  \bibinfo{year}{2007}\natexlab{}.
\newblock \showarticletitle{Tesseract: An Open-Source Optical Character
  Recognition Engine}.
\newblock \bibinfo{journal}{\emph{Linux J.}} \bibinfo{volume}{2007},
  \bibinfo{number}{159} (\bibinfo{year}{2007}), \bibinfo{pages}{2}.
\newblock


\bibitem[Kaya et~al\mbox{.}(2019)]%
        {sdn19kaya}
\bibfield{author}{\bibinfo{person}{Yigitcan Kaya}, \bibinfo{person}{Sanghyun
  Hong}, {and} \bibinfo{person}{Tudor Dumitras}.}
  \bibinfo{year}{2019}\natexlab{}.
\newblock \showarticletitle{Shallow-deep networks: Understanding and mitigating
  network overthinking}. In \bibinfo{booktitle}{\emph{Proceedings of the
  International Conference on Machine Learning}}. \bibinfo{publisher}{PMLR},
  \bibinfo{address}{Long Beach, CA, United States},
  \bibinfo{pages}{3301--3310}.
\newblock


\bibitem[Key et~al\mbox{.}(2012)]%
        {key12vizdeck}
\bibfield{author}{\bibinfo{person}{Alicia Key}, \bibinfo{person}{Bill Howe},
  \bibinfo{person}{Daniel Perry}, {and} \bibinfo{person}{Cecilia Aragon}.}
  \bibinfo{year}{2012}\natexlab{}.
\newblock \showarticletitle{{VizDeck}: Self-Organizing Dashboards for Visual
  Analytics}. In \bibinfo{booktitle}{\emph{Proceedings of the ACM Conference on
  Management of Data}}. \bibinfo{publisher}{{ACM}}, \bibinfo{address}{{New
  York, NY, USA}}, \bibinfo{pages}{681–684}.
\newblock
\urldef\tempurl%
\url{https://doi.org/10.1145/2213836.2213931}
\showDOI{\tempurl}


\bibitem[Kim et~al\mbox{.}(2012)]%
        {kim12eyetracker}
\bibfield{author}{\bibinfo{person}{Sung-Hee Kim}, \bibinfo{person}{Zhihua
  Dong}, \bibinfo{person}{Hanjun Xian}, \bibinfo{person}{Benjavan Upatising},
  {and} \bibinfo{person}{Ji~Soo Yi}.} \bibinfo{year}{2012}\natexlab{}.
\newblock \showarticletitle{Does an Eye Tracker Tell the Truth about
  Visualizations?: Findings while Investigating Visualizations for Decision
  Making}.
\newblock \bibinfo{journal}{\emph{{{IEEE} Transactions on Visualization and
  Computer Graphics}}} \bibinfo{volume}{18}, \bibinfo{number}{12}
  (\bibinfo{year}{2012}), \bibinfo{pages}{2421--2430}.
\newblock
\urldef\tempurl%
\url{https://doi.org/10.1109/TVCG.2012.215}
\showDOI{\tempurl}


\bibitem[Koffka(1922)]%
        {Koffka1922}
\bibfield{author}{\bibinfo{person}{Kurt Koffka}.}
  \bibinfo{year}{1922}\natexlab{}.
\newblock \showarticletitle{Perception: An introduction to the
  {G}estalt-theorie}.
\newblock \bibinfo{journal}{\emph{Psychological Bulletin}}
  \bibinfo{volume}{19} (\bibinfo{year}{1922}), \bibinfo{pages}{531--585}.
\newblock


\bibitem[Lall\'{e} et~al\mbox{.}(2016)]%
        {lalle16confusion}
\bibfield{author}{\bibinfo{person}{S\'{e}bastien Lall\'{e}},
  \bibinfo{person}{Cristina Conati}, {and} \bibinfo{person}{Giuseppe
  Carenini}.} \bibinfo{year}{2016}\natexlab{}.
\newblock \showarticletitle{Predicting Confusion in Information Visualization
  from Eye Tracking and Interaction Data}. In
  \bibinfo{booktitle}{\emph{Proceedings of the International Joint Conference
  on Artificial Intelligence}}. \bibinfo{publisher}{AAAI Press},
  \bibinfo{address}{New York, New York, USA}, \bibinfo{pages}{2529–2535}.
\newblock


\bibitem[Lin et~al\mbox{.}(2014)]%
        {lin14coco}
\bibfield{author}{\bibinfo{person}{Tsung-Yi Lin}, \bibinfo{person}{Michael
  Maire}, \bibinfo{person}{Serge Belongie}, \bibinfo{person}{James Hays},
  \bibinfo{person}{Pietro Perona}, \bibinfo{person}{Deva Ramanan},
  \bibinfo{person}{Piotr Doll{\'a}r}, {and} \bibinfo{person}{C.~Lawrence
  Zitnick}.} \bibinfo{year}{2014}\natexlab{}.
\newblock \showarticletitle{Microsoft COCO: Common Objects in Context}. In
  \bibinfo{booktitle}{\emph{Proceedings of the European Conference on Computer
  Vision}}. \bibinfo{publisher}{Springer International Publishing},
  \bibinfo{address}{Cham}, \bibinfo{pages}{740--755}.
\newblock


\bibitem[Lohse(1993)]%
        {Lohse1993}
\bibfield{author}{\bibinfo{person}{Gerald~L. Lohse}.}
  \bibinfo{year}{1993}\natexlab{}.
\newblock \showarticletitle{A cognitive model for understanding graphical
  perception}.
\newblock \bibinfo{journal}{\emph{Human-Computer Interaction}}
  \bibinfo{volume}{8}, \bibinfo{number}{4} (\bibinfo{year}{1993}),
  \bibinfo{pages}{353--388}.
\newblock
\urldef\tempurl%
\url{https://doi.org/10.1207/s15327051hci0804_3}
\showDOI{\tempurl}


\bibitem[Luo et~al\mbox{.}(2018)]%
        {luo18deepeye}
\bibfield{author}{\bibinfo{person}{Yuyu Luo}, \bibinfo{person}{Xuedi Qin},
  \bibinfo{person}{Nan Tang}, {and} \bibinfo{person}{Guoliang Li}.}
  \bibinfo{year}{2018}\natexlab{}.
\newblock \showarticletitle{{DeepEye}: Towards Automatic Data Visualization}.
  In \bibinfo{booktitle}{\emph{Proceedings of the IEEE International Conference
  on Data Engineering}}. \bibinfo{publisher}{{IEEE}},
  \bibinfo{address}{Piscataway, NJ, USA}, \bibinfo{pages}{101--112}.
\newblock
\urldef\tempurl%
\url{https://doi.org/10.1109/ICDE.2018.00019}
\showDOI{\tempurl}


\bibitem[Mackinlay(1986)]%
        {Mackinlay1986}
\bibfield{author}{\bibinfo{person}{Jock Mackinlay}.}
  \bibinfo{year}{1986}\natexlab{}.
\newblock \showarticletitle{Automating the Design of Graphical Presentations of
  Relational Information}.
\newblock \bibinfo{journal}{\emph{ACM Transactions on Graphics}}
  \bibinfo{volume}{5}, \bibinfo{number}{2} (\bibinfo{year}{1986}),
  \bibinfo{pages}{110--141}.
\newblock


\bibitem[Mackinlay et~al\mbox{.}(2007)]%
        {mackinlay07showme}
\bibfield{author}{\bibinfo{person}{Jock Mackinlay}, \bibinfo{person}{Pat
  Hanrahan}, {and} \bibinfo{person}{Chris Stolte}.}
  \bibinfo{year}{2007}\natexlab{}.
\newblock \showarticletitle{Show Me: Automatic Presentation for Visual
  Analysis}.
\newblock \bibinfo{journal}{\emph{IEEE Transactions on Visualization and
  Computer Graphics}} \bibinfo{volume}{13}, \bibinfo{number}{6}
  (\bibinfo{year}{2007}), \bibinfo{pages}{1137--1144}.
\newblock
\urldef\tempurl%
\url{https://doi.org/10.1109/TVCG.2007.70594}
\showDOI{\tempurl}


\bibitem[Matzen et~al\mbox{.}(2018)]%
        {matzen18visualsaliency}
\bibfield{author}{\bibinfo{person}{Laura~E. Matzen},
  \bibinfo{person}{Michael~J. Haass}, \bibinfo{person}{Kristin~M. Divis},
  \bibinfo{person}{Zhiyuan Wang}, {and} \bibinfo{person}{Andrew~T. Wilson}.}
  \bibinfo{year}{2018}\natexlab{}.
\newblock \showarticletitle{Data Visualization Saliency Model: A Tool for
  Evaluating Abstract Data Visualizations}.
\newblock \bibinfo{journal}{\emph{{{IEEE} Transactions on Visualization and
  Computer Graphics}}} \bibinfo{volume}{24}, \bibinfo{number}{1}
  (\bibinfo{year}{2018}), \bibinfo{pages}{563--573}.
\newblock
\urldef\tempurl%
\url{https://doi.org/10.1109/TVCG.2017.2743939}
\showDOI{\tempurl}


\bibitem[McNutt et~al\mbox{.}(2020)]%
        {mcnutt20mirage}
\bibfield{author}{\bibinfo{person}{Andrew McNutt}, \bibinfo{person}{Gordon
  Kindlmann}, {and} \bibinfo{person}{Michael Correll}.}
  \bibinfo{year}{2020}\natexlab{}.
\newblock \showarticletitle{Surfacing Visualization Mirages}. In
  \bibinfo{booktitle}{\emph{Proceedings of the {ACM} Conference on Human
  Factors in Computing Systems}}. \bibinfo{publisher}{{ACM}},
  \bibinfo{address}{{New York, NY, USA}}, \bibinfo{pages}{1–16}.
\newblock
\urldef\tempurl%
\url{https://doi.org/10.1145/3313831.3376420}
\showDOI{\tempurl}


\bibitem[Michal and Franconeri(2017)]%
        {michal17visual}
\bibfield{author}{\bibinfo{person}{Audrey~L. Michal} {and}
  \bibinfo{person}{Steven~L. Franconeri}.} \bibinfo{year}{2017}\natexlab{}.
\newblock \showarticletitle{Visual routines are associated with specific graph
  interpretations}.
\newblock \bibinfo{journal}{\emph{Cognitive Research: Principles and
  Implications}} \bibinfo{volume}{2}, \bibinfo{number}{1}
  (\bibinfo{year}{2017}), \bibinfo{pages}{1--10}.
\newblock
\urldef\tempurl%
\url{https://doi.org/10.1186/s41235-017-0059-2}
\showDOI{\tempurl}


\bibitem[Minaee et~al\mbox{.}(2022)]%
        {DBLP:journals/pami/MinaeeBPPKT22}
\bibfield{author}{\bibinfo{person}{Shervin Minaee}, \bibinfo{person}{Yuri
  Boykov}, \bibinfo{person}{Fatih Porikli}, \bibinfo{person}{Antonio Plaza},
  \bibinfo{person}{Nasser Kehtarnavaz}, {and} \bibinfo{person}{Demetri
  Terzopoulos}.} \bibinfo{year}{2022}\natexlab{}.
\newblock \showarticletitle{Image Segmentation Using Deep Learning: {A}
  Survey}.
\newblock \bibinfo{journal}{\emph{{IEEE} Transactions on Pattern Analysis and
  Machine Intelligence}} \bibinfo{volume}{44}, \bibinfo{number}{7}
  (\bibinfo{year}{2022}), \bibinfo{pages}{3523--3542}.
\newblock
\urldef\tempurl%
\url{https://doi.org/10.1109/TPAMI.2021.3059968}
\showDOI{\tempurl}


\bibitem[Moritz et~al\mbox{.}(2019)]%
        {moritz19draco}
\bibfield{author}{\bibinfo{person}{Dominik Moritz}, \bibinfo{person}{Chenglong
  Wang}, \bibinfo{person}{Greg~L. Nelson}, \bibinfo{person}{Halden Lin},
  \bibinfo{person}{Adam~M. Smith}, \bibinfo{person}{Bill Howe}, {and}
  \bibinfo{person}{Jeffrey Heer}.} \bibinfo{year}{2019}\natexlab{}.
\newblock \showarticletitle{Formalizing Visualization Design Knowledge as
  Constraints: Actionable and Extensible Models in Draco}.
\newblock \bibinfo{journal}{\emph{IEEE Transactions on Visualization and
  Computer Graphics}} \bibinfo{volume}{25}, \bibinfo{number}{1}
  (\bibinfo{year}{2019}), \bibinfo{pages}{438--448}.
\newblock
\urldef\tempurl%
\url{https://doi.org/10.1109/TVCG.2018.2865240}
\showDOI{\tempurl}


\bibitem[Munzner(2009)]%
        {munzner2009}
\bibfield{author}{\bibinfo{person}{Tamara Munzner}.}
  \bibinfo{year}{2009}\natexlab{}.
\newblock \showarticletitle{A Nested Model for Visualization Design and
  Validation}.
\newblock \bibinfo{journal}{\emph{IEEE Transactions on Visualization and
  Computer Graphics}} \bibinfo{volume}{15}, \bibinfo{number}{6}
  (\bibinfo{date}{Nov.} \bibinfo{year}{2009}), \bibinfo{pages}{921--928}.
\newblock
\showISSN{1077-2626}
\urldef\tempurl%
\url{https://doi.org/10.1109/TVCG.2009.111}
\showDOI{\tempurl}


\bibitem[Munzner(2014)]%
        {munzner14visualization}
\bibfield{author}{\bibinfo{person}{Tamara Munzner}.}
  \bibinfo{year}{2014}\natexlab{}.
\newblock \bibinfo{booktitle}{\emph{Visualization Analysis and Design}}.
\newblock \bibinfo{publisher}{CRC Press}, \bibinfo{address}{Boca Raton, FL,
  USA}.
\newblock


\bibitem[Peterson and Schramm(1954)]%
        {Peterson1954}
\bibfield{author}{\bibinfo{person}{Lewis~V. Peterson} {and}
  \bibinfo{person}{Wilbur Schramm}.} \bibinfo{year}{1954}\natexlab{}.
\newblock \showarticletitle{How accurately are different kinds of graphs read?}
\newblock \bibinfo{journal}{\emph{Educational Technology Research and
  Development}} \bibinfo{volume}{2}, \bibinfo{number}{3} (\bibinfo{date}{June}
  \bibinfo{year}{1954}), \bibinfo{pages}{178--189}.
\newblock
\urldef\tempurl%
\url{https://doi.org/10.1007/BF02713334}
\showDOI{\tempurl}


\bibitem[Plaisant(2004)]%
        {DBLP:conf/avi/Plaisant04}
\bibfield{author}{\bibinfo{person}{Catherine Plaisant}.}
  \bibinfo{year}{2004}\natexlab{}.
\newblock \showarticletitle{The challenge of information visualization
  evaluation}. In \bibinfo{booktitle}{\emph{Proceedings of the ACM Conference
  on Advanced Visual Interfaces}}. \bibinfo{publisher}{{ACM}},
  \bibinfo{address}{{New York, NY, USA}}, \bibinfo{pages}{109--116}.
\newblock
\urldef\tempurl%
\url{https://doi.org/10.1145/989863.989880}
\showDOI{\tempurl}


\bibitem[Poco and Heer(2017)]%
        {Poco2017}
\bibfield{author}{\bibinfo{person}{Jorge Poco} {and} \bibinfo{person}{Jeffrey
  Heer}.} \bibinfo{year}{2017}\natexlab{}.
\newblock \showarticletitle{Reverse-Engineering Visualizations: Recovering
  Visual Encodings from Chart Images}.
\newblock \bibinfo{journal}{\emph{Computer Graphics Forum}}
  \bibinfo{volume}{36}, \bibinfo{number}{3} (\bibinfo{date}{June}
  \bibinfo{year}{2017}), \bibinfo{pages}{353--363}.
\newblock
\urldef\tempurl%
\url{https://doi.org/10.1111/cgf.13193}
\showDOI{\tempurl}


\bibitem[Pylyshyn and Storm(1988)]%
        {Pylyshyn1988}
\bibfield{author}{\bibinfo{person}{Zenon~W. Pylyshyn} {and}
  \bibinfo{person}{R.~W. Storm}.} \bibinfo{year}{1988}\natexlab{}.
\newblock \showarticletitle{Tracking multiple independent targets: Evidence for
  a parallel tracking mechanism}.
\newblock \bibinfo{journal}{\emph{Spatial Vision}}  \bibinfo{volume}{3}
  (\bibinfo{year}{1988}), \bibinfo{pages}{179--197}.
\newblock
\urldef\tempurl%
\url{https://doi.org/10.1163/156856888x00122}
\showDOI{\tempurl}


\bibitem[Reddy et~al\mbox{.}(2020)]%
        {reddy20deepsaliencyprediction}
\bibfield{author}{\bibinfo{person}{Navyasri Reddy}, \bibinfo{person}{Samyak
  Jain}, \bibinfo{person}{Pradeep Yarlagadda}, {and} \bibinfo{person}{Vineet
  Gandhi}.} \bibinfo{year}{2020}\natexlab{}.
\newblock \showarticletitle{Tidying Deep Saliency Prediction Architectures}. In
  \bibinfo{booktitle}{\emph{Proceedings of the IEEE/RSJ International
  Conference on Intelligent Robots and Systems}}. \bibinfo{publisher}{{IEEE}},
  \bibinfo{address}{Piscataway, NJ, USA}, \bibinfo{pages}{10241--10247}.
\newblock
\urldef\tempurl%
\url{https://doi.org/10.1109/IROS45743.2020.9341574}
\showDOI{\tempurl}


\bibitem[Rhyne(2016)]%
        {Rhyne2016}
\bibfield{author}{\bibinfo{person}{Theresa-Marie Rhyne}.}
  \bibinfo{year}{2016}\natexlab{}.
\newblock \bibinfo{booktitle}{\emph{Applying Color Theory to Digital Media and
  Visualization}}.
\newblock \bibinfo{publisher}{CRC Press}, \bibinfo{address}{Boca Raton, FL,
  USA}.
\newblock


\bibitem[Riche et~al\mbox{.}(2013)]%
        {riche13metricssaliency}
\bibfield{author}{\bibinfo{person}{Nicolas Riche}, \bibinfo{person}{Matthieu
  Duvinage}, \bibinfo{person}{Matei Mancas}, \bibinfo{person}{Bernard
  Gosselin}, {and} \bibinfo{person}{Thierry Dutoit}.}
  \bibinfo{year}{2013}\natexlab{}.
\newblock \showarticletitle{Saliency and Human Fixations: State-of-the-Art and
  Study of Comparison Metrics}. In \bibinfo{booktitle}{\emph{Proceedings of the
  IEEE International Conference on Computer Vision}}.
  \bibinfo{publisher}{{IEEE}}, \bibinfo{address}{Piscataway, NJ, USA},
  \bibinfo{pages}{1153--1160}.
\newblock
\urldef\tempurl%
\url{https://doi.org/10.1109/ICCV.2013.147}
\showDOI{\tempurl}


\bibitem[Savva et~al\mbox{.}(2011)]%
        {Savva2011}
\bibfield{author}{\bibinfo{person}{Manolis Savva}, \bibinfo{person}{Nicholas
  Kong}, \bibinfo{person}{Arti Chhajta}, \bibinfo{person}{Li Fei-Fei},
  \bibinfo{person}{Maneesh Agrawala}, {and} \bibinfo{person}{Jeffrey Heer}.}
  \bibinfo{year}{2011}\natexlab{}.
\newblock \showarticletitle{{ReVision}: Automated Classification, Analysis and
  Redesign of Chart Images}. In \bibinfo{booktitle}{\emph{Proceedings of the
  {ACM} Symposium on User Interface Software and Technology}}.
  \bibinfo{publisher}{{ACM}}, \bibinfo{address}{{New York, NY, USA}},
  \bibinfo{pages}{393--402}.
\newblock
\urldef\tempurl%
\url{https://doi.org/10.1145/2047196.2047247}
\showDOI{\tempurl}


\bibitem[Sedlmair et~al\mbox{.}(2011)]%
        {DBLP:journals/ivs/SedlmairIBB11}
\bibfield{author}{\bibinfo{person}{Michael Sedlmair}, \bibinfo{person}{Petra
  Isenberg}, \bibinfo{person}{Dominikus Baur}, {and} \bibinfo{person}{Andreas
  Butz}.} \bibinfo{year}{2011}\natexlab{}.
\newblock \showarticletitle{Information visualization evaluation in large
  companies: Challenges, experiences and recommendations}.
\newblock \bibinfo{journal}{\emph{Information Visualization}}
  \bibinfo{volume}{10}, \bibinfo{number}{3} (\bibinfo{year}{2011}),
  \bibinfo{pages}{248--266}.
\newblock
\urldef\tempurl%
\url{https://doi.org/10.1177/1473871611413099}
\showDOI{\tempurl}


\bibitem[Sedlmair et~al\mbox{.}(2012)]%
        {Sedlmair2012}
\bibfield{author}{\bibinfo{person}{Michael Sedlmair},
  \bibinfo{person}{Miriah~D. Meyer}, {and} \bibinfo{person}{Tamara Munzner}.}
  \bibinfo{year}{2012}\natexlab{}.
\newblock \showarticletitle{Design Study Methodology: Reflections from the
  Trenches and the Stacks}.
\newblock \bibinfo{journal}{\emph{{{IEEE} Transactions on Visualization and
  Computer Graphics}}} \bibinfo{volume}{18}, \bibinfo{number}{12}
  (\bibinfo{year}{2012}), \bibinfo{pages}{2431--2440}.
\newblock
\urldef\tempurl%
\url{https://doi.org/10.1109/TVCG.2012.213}
\showDOI{\tempurl}


\bibitem[Selvaraju et~al\mbox{.}(2017)]%
        {selvaraju17gradcam}
\bibfield{author}{\bibinfo{person}{Ramprasaath~R. Selvaraju},
  \bibinfo{person}{Michael Cogswell}, \bibinfo{person}{Abhishek Das},
  \bibinfo{person}{Ramakrishna Vedantam}, \bibinfo{person}{Devi Parikh}, {and}
  \bibinfo{person}{Dhruv Batra}.} \bibinfo{year}{2017}\natexlab{}.
\newblock \showarticletitle{Grad-CAM: Visual Explanations from Deep Networks
  via Gradient-Based Localization}. In \bibinfo{booktitle}{\emph{Proceedings of
  the IEEE International Conference on Computer Vision}}.
  \bibinfo{publisher}{{IEEE}}, \bibinfo{address}{Piscataway, NJ, USA},
  \bibinfo{pages}{618--626}.
\newblock
\urldef\tempurl%
\url{https://doi.org/10.1109/ICCV.2017.74}
\showDOI{\tempurl}


\bibitem[Shapiro and Stockman(2001)]%
        {Shapiro2001}
\bibfield{author}{\bibinfo{person}{Linda~G. Shapiro} {and}
  \bibinfo{person}{George~C. Stockman}.} \bibinfo{year}{2001}\natexlab{}.
\newblock \bibinfo{booktitle}{\emph{Computer Vision}}.
\newblock \bibinfo{publisher}{Prentice-Hall}, \bibinfo{address}{Hoboken, NJ,
  USA}.
\newblock


\bibitem[Shi et~al\mbox{.}(2017)]%
        {shi17ocrnn}
\bibfield{author}{\bibinfo{person}{Baoguang Shi}, \bibinfo{person}{Xiang Bai},
  {and} \bibinfo{person}{Cong Yao}.} \bibinfo{year}{2017}\natexlab{}.
\newblock \showarticletitle{An End-to-End Trainable Neural Network for
  Image-Based Sequence Recognition and Its Application to Scene Text
  Recognition}.
\newblock \bibinfo{journal}{\emph{IEEE Transactions on Pattern Analysis and
  Machine Intelligence}} \bibinfo{volume}{39}, \bibinfo{number}{11}
  (\bibinfo{year}{2017}), \bibinfo{pages}{2298--2304}.
\newblock
\urldef\tempurl%
\url{https://doi.org/10.1109/TPAMI.2016.2646371}
\showDOI{\tempurl}


\bibitem[Shin et~al\mbox{.}(2023)]%
        {shin23scannerdeeply}
\bibfield{author}{\bibinfo{person}{Sungbok Shin}, \bibinfo{person}{Sunghyo
  Chung}, \bibinfo{person}{Sanghyun Hong}, {and} \bibinfo{person}{Niklas
  Elmqvist}.} \bibinfo{year}{2023}\natexlab{}.
\newblock \showarticletitle{A Scanner Deeply: Predicting Gaze Heatmaps on
  Visualizations Using Crowdsourced Eye Movement Data}.
\newblock \bibinfo{journal}{\emph{IEEE Transactions on Visualization and
  Computer Graphics}} \bibinfo{volume}{29}, \bibinfo{number}{1}
  (\bibinfo{year}{2023}), \bibinfo{numpages}{11}~pages.
\newblock


\bibitem[Shneiderman(2022)]%
        {Shneiderman2022}
\bibfield{author}{\bibinfo{person}{Ben Shneiderman}.}
  \bibinfo{year}{2022}\natexlab{}.
\newblock \bibinfo{booktitle}{\emph{Human-Centered AI}}.
\newblock \bibinfo{publisher}{Oxford University Press},
  \bibinfo{address}{Oxford, United Kingdom}.
\newblock


\bibitem[Szafir(2018)]%
        {szafir18modelingcolordifference}
\bibfield{author}{\bibinfo{person}{Danielle~Albers Szafir}.}
  \bibinfo{year}{2018}\natexlab{}.
\newblock \showarticletitle{Modeling Color Difference for Visualization
  Design}.
\newblock \bibinfo{journal}{\emph{{{IEEE} Transactions on Visualization and
  Computer Graphics}}} \bibinfo{volume}{24}, \bibinfo{number}{1}
  (\bibinfo{year}{2018}), \bibinfo{pages}{392--401}.
\newblock
\urldef\tempurl%
\url{https://doi.org/10.1109/TVCG.2017.2744359}
\showDOI{\tempurl}


\bibitem[Tesseract-OCR(1999)]%
        {PyTesseract_perf}
\bibfield{author}{\bibinfo{person}{Tesseract-OCR}.}
  \bibinfo{year}{1999}\natexlab{}.
\newblock \bibinfo{title}{Performance of Tessarct-OCR v.4.0}.
\newblock
  \bibinfo{howpublished}{\url{https://github.com/tesseract-ocr/tessdoc/blob/main/tess4/4.0-Accuracy-and-Performance.md}}.
\newblock


\bibitem[Toker et~al\mbox{.}(2013)]%
        {toker13infoviseyetrackingchar}
\bibfield{author}{\bibinfo{person}{Dereck Toker}, \bibinfo{person}{Cristina
  Conati}, \bibinfo{person}{Ben Steichen}, {and} \bibinfo{person}{Giuseppe
  Carenini}.} \bibinfo{year}{2013}\natexlab{}.
\newblock \showarticletitle{Individual User Characteristics and Information
  Visualization: Connecting the Dots through Eye Tracking}. In
  \bibinfo{booktitle}{\emph{Proceedings of the {ACM} Conference on Human
  Factors in Computing Systems}}. \bibinfo{publisher}{{ACM}},
  \bibinfo{address}{{New York, NY, USA}}, \bibinfo{pages}{295–304}.
\newblock
\urldef\tempurl%
\url{https://doi.org/10.1145/2470654.2470696}
\showDOI{\tempurl}


\bibitem[Toker et~al\mbox{.}(2014)]%
        {toker14untrainedvisusers}
\bibfield{author}{\bibinfo{person}{Dereck Toker}, \bibinfo{person}{Ben
  Steichen}, \bibinfo{person}{Matthew Gingerich}, \bibinfo{person}{Cristina
  Conati}, {and} \bibinfo{person}{Giuseppe Carenini}.}
  \bibinfo{year}{2014}\natexlab{}.
\newblock \showarticletitle{Towards Facilitating User Skill Acquisition:
  Identifying Untrained Visualization Users through Eye Tracking}. In
  \bibinfo{booktitle}{\emph{Proceedings of the International Conference on
  Intelligent User Interfaces}}. \bibinfo{publisher}{{ACM}},
  \bibinfo{address}{{New York, NY, USA}}, \bibinfo{pages}{105–114}.
\newblock
\urldef\tempurl%
\url{https://doi.org/10.1145/2557500.2557524}
\showDOI{\tempurl}


\bibitem[Treisman(1985)]%
        {Treisman1985}
\bibfield{author}{\bibinfo{person}{Anne Treisman}.}
  \bibinfo{year}{1985}\natexlab{}.
\newblock \showarticletitle{Preattentive Processing in Vision}.
\newblock \bibinfo{journal}{\emph{Computer Vision, Graphics and Image
  Processing}}  \bibinfo{volume}{31} (\bibinfo{year}{1985}),
  \bibinfo{pages}{156--177}.
\newblock
\urldef\tempurl%
\url{https://doi.org/10.1016/S0734-189X(85)80004-9}
\showDOI{\tempurl}


\bibitem[Treisman and Gelade(1980)]%
        {Treisman1980}
\bibfield{author}{\bibinfo{person}{Anne~M. Treisman} {and}
  \bibinfo{person}{Garry Gelade}.} \bibinfo{year}{1980}\natexlab{}.
\newblock \showarticletitle{A feature-integration theory of attention}.
\newblock \bibinfo{journal}{\emph{Cognitive Psychology}} \bibinfo{volume}{12},
  \bibinfo{number}{1} (\bibinfo{year}{1980}), \bibinfo{pages}{97--136}.
\newblock
\urldef\tempurl%
\url{https://doi.org/10.1016/0010-0285(80)90005-5}
\showDOI{\tempurl}


\bibitem[Tufte(1983)]%
        {Tufte1983}
\bibfield{author}{\bibinfo{person}{Edward~R. Tufte}.}
  \bibinfo{year}{1983}\natexlab{}.
\newblock \bibinfo{booktitle}{\emph{The Visual Display of Quantitative
  Information}}.
\newblock \bibinfo{publisher}{Graphics Press}, \bibinfo{address}{Cheshire, CT,
  USA}.
\newblock


\bibitem[Virtanen et~al\mbox{.}(2020)]%
        {virtanen20scipy}
\bibfield{author}{\bibinfo{person}{Pauli Virtanen}, \bibinfo{person}{Ralf
  Gommers}, \bibinfo{person}{Travis~E Oliphant}, \bibinfo{person}{Matt
  Haberland}, \bibinfo{person}{Tyler Reddy}, \bibinfo{person}{David
  Cournapeau}, \bibinfo{person}{Evgeni Burovski}, \bibinfo{person}{Pearu
  Peterson}, \bibinfo{person}{Warren Weckesser}, \bibinfo{person}{Jonathan
  Bright}, \bibinfo{person}{Stéfan~J. van~der Walt}, \bibinfo{person}{Matthew
  Brett}, \bibinfo{person}{Joshua Wilson}, \bibinfo{person}{K.~Jarrod Millman},
  \bibinfo{person}{Nikolay Mayorov}, \bibinfo{person}{Andrew R.~J. Nelson},
  \bibinfo{person}{Eric Jones}, \bibinfo{person}{Robert Kern},
  \bibinfo{person}{Eric Larson}, \bibinfo{person}{C.~J. Carey},
  \bibinfo{person}{İlhan Polat}, \bibinfo{person}{Yu Feng},
  \bibinfo{person}{Eric~W. Moore}, {and} \bibinfo{person}{Jake VanderPlas}.}
  \bibinfo{year}{2020}\natexlab{}.
\newblock \showarticletitle{{SciPy 1.0}: fundamental algorithms for scientific
  computing in {Python}}.
\newblock \bibinfo{journal}{\emph{Nature Methods}} \bibinfo{volume}{17},
  \bibinfo{number}{3} (\bibinfo{year}{2020}), \bibinfo{pages}{261--272}.
\newblock
\urldef\tempurl%
\url{https://doi.org/10.1038/s41592-019-0686-2}
\showDOI{\tempurl}


\bibitem[Wang et~al\mbox{.}(2021a)]%
        {wang21yolor}
\bibfield{author}{\bibinfo{person}{Chien{-}Yao Wang}, \bibinfo{person}{I{-}Hau
  Yeh}, {and} \bibinfo{person}{Hong{-}Yuan~Mark Liao}.}
  \bibinfo{year}{2021}\natexlab{a}.
\newblock \showarticletitle{You Only Learn One Representation: Unified Network
  for Multiple Tasks}.
\newblock \bibinfo{journal}{\emph{CoRR}}  \bibinfo{volume}{abs/2105.04206}
  (\bibinfo{year}{2021}), \bibinfo{numpages}{11}~pages.
\newblock
\showeprint[arXiv]{2105.04206}
\urldef\tempurl%
\url{https://arxiv.org/abs/2105.04206}
\showURL{%
\tempurl}


\bibitem[Wang et~al\mbox{.}(2021b)]%
        {yolor_perf}
\bibfield{author}{\bibinfo{person}{Chien-Yao Wang}, \bibinfo{person}{I-Hau
  Yeh}, {and} \bibinfo{person}{Hong-Yuan~Mark Liao}.}
  \bibinfo{year}{2021}\natexlab{b}.
\newblock \bibinfo{title}{Performance of YoloR}.
\newblock \bibinfo{howpublished}{\url{https://github.com/WongKinYiu/yolor}}.
\newblock


\bibitem[Wolfe and Utochkin(2019)]%
        {Wolfe2019}
\bibfield{author}{\bibinfo{person}{Jeremy~M. Wolfe} {and}
  \bibinfo{person}{Igor~S. Utochkin}.} \bibinfo{year}{2019}\natexlab{}.
\newblock \showarticletitle{What is a preattentive feature?}
\newblock \bibinfo{journal}{\emph{Current Opinion in Psychology}}
  \bibinfo{volume}{29} (\bibinfo{year}{2019}), \bibinfo{pages}{19--26}.
\newblock
\urldef\tempurl%
\url{https://doi.org/10.1016/j.copsyc.2018.11.005}
\showDOI{\tempurl}


\bibitem[Wongsuphasawat et~al\mbox{.}(2016)]%
        {wongsuphasawat16voyager}
\bibfield{author}{\bibinfo{person}{Kanit Wongsuphasawat},
  \bibinfo{person}{Dominik Moritz}, \bibinfo{person}{Anushka Anand},
  \bibinfo{person}{Jock Mackinlay}, \bibinfo{person}{Bill Howe}, {and}
  \bibinfo{person}{Jeffrey Heer}.} \bibinfo{year}{2016}\natexlab{}.
\newblock \showarticletitle{Voyager: Exploratory Analysis via Faceted Browsing
  of Visualization Recommendations}.
\newblock \bibinfo{journal}{\emph{{{IEEE} Transactions on Visualization and
  Computer Graphics}}} \bibinfo{volume}{22}, \bibinfo{number}{1}
  (\bibinfo{year}{2016}), \bibinfo{pages}{649--658}.
\newblock
\urldef\tempurl%
\url{https://doi.org/10.1109/TVCG.2015.2467191}
\showDOI{\tempurl}


\end{thebibliography}

%\clearpage
%\newpage
%% ---------------------------------------------------------------------
%% APPENDIX
%% ---------------------------------------------------------------------

%\input{content/09-appendix}

\end{document}